\documentclass[12pt]{article}

\renewcommand{\footnotesize}{\small}
\usepackage{setspace,caption}
\usepackage{lipsum}
\usepackage{xr}
\usepackage{amssymb,amsmath}
\usepackage{natbib}
\usepackage{hyperref}
\usepackage{xtab}
\usepackage{float}
\usepackage[T1]{fontenc}
\usepackage{comment}
\usepackage{array}

\usepackage{rotating}
\usepackage{lscape,graphicx}
\usepackage{multicol}
\usepackage{threeparttable}
\usepackage{url}
\usepackage{bookmark}
\usepackage{xcolor}
\usepackage[12pt]{moresize}
\usepackage{subfigure}
\usepackage{afterpage}
\usepackage{subcaption}
\usepackage{subfigure}
\newcommand{\stars}[1]{#1}

\hypersetup{
	colorlinks,
	linkcolor={blue!50!blue},
	citecolor={blue!50!blue},
	urlcolor={blue!80!blue}
}
\usepackage{booktabs,caption}
\usepackage{longtable}
\usepackage{mwe}

\renewcommand{\b}{\stars{**}}
\renewcommand{\c}{\stars{***}}

\usepackage{comment}
\usepackage{listings}
\usepackage[english]{babel}
\usepackage[normalem]{ulem}
\usepackage{multirow}
\usepackage[section]{placeins}

\usepackage[multiple]{footmisc}
\usepackage{bigfoot}

\DeclareNewFootnote{AAffil}[arabic]
\DeclareNewFootnote{ANote}[fnsymbol]

\usepackage[margin=1in]{geometry}

\renewcommand{\abstractname}{\vspace{-\baselineskip}}

\providecommand\abstractname{Abstract}
\def\abstract{}

\renewenvironment{abstract}{%
	\centering\small
	\textbf\abstractname
	\list{}{\leftmargin0.9cm \rightmargin\leftmargin}
	\item\relax
}{%
	\endlist \par\bigskip
}
\addto{\captionsenglish}{\renewcommand{\abstractname}{}}
\graphicspath{{./charts/}}

\begin{document}

\title{\Large \vspace{-2cm}Measuring Consumption with Credit Card Data: \\ Benchmarking and Beyond\thanks{We thank Colin Campbell and Erica Gonzales for outstanding research assistance. For helpful feedback and comments, we thank Vitaly Bord, Sinem Hacioglu Hoke, and Patrick Moran. We are also grateful to the conference and seminar participants of the Consumer Finance Round Robin and the Federal Reserve Board. The views expressed in this paper solely reflect those of the authors and not necessarily those of the Federal Reserve Board, the Federal Reserve System as a whole, nor of anyone else associated with the Federal Reserve System. 
}}

\author{
	Aditya Aladangady\footnote{Federal Reserve Board. E-mail: \href{mailto:aditya.aladangady@frb.gov}{aditya.aladangady@frb.gov}}
    \and
	Ricardo Duque Gabriel\footnote{Federal Reserve Board. E-mail: \href{mailto:ricardo.f.duquegabriel@frb.gov}{ricardo.f.duquegabriel@frb.gov}}
	\and
	Carlo Wix\footnote{Federal Reserve Board. E-mail: \href{mailto:carlo.e.wix@frb.gov}{carlo.e.wix@frb.gov}}
}

\date{July 2026}

\normalem
\maketitle
\thispagestyle{empty}
 
\begin{abstract}
	\vspace{-0.55cm}
	\begin{singlespacing} 
		\noindent  \normalsize \textbf{Abstract:} We introduce a novel monthly county-level consumption dataset constructed from spending data on over 350 million credit cards in the Federal Reserve's Y-14M reports, covering over 3,000 U.S. counties since 2014. We first show that the data closely approximate traditional consumption measures, explaining 92 percent of the variation in monthly adjusted personal consumption expenditures (PCE) growth at the national level and capturing meaningful cross-sectional variation in annual adjusted PCE growth at the state level. As a proof of concept, we use the county-month panel to estimate heterogeneous consumption responses to monetary policy shocks across the county-level income distribution, an analysis infeasible with traditional consumption data. We find that low-income counties exhibit larger spending declines than high-income counties, consistent with heterogeneous agent New Keynesian models. Finally, we provide practical guidance for researchers working with similar data, discussing coverage, sample composition, and the approximation of credit card spending from credit bureau data.\\
		\vspace{0.25cm}
		\noindent {\bf Keywords:} Measurement and Data, Consumption, Credit Card Spending \\
		\noindent {\bf JEL Codes:} D12, E01, E21	
	\end{singlespacing}
	\vspace{0.5cm}
\end{abstract}

\clearpage	
\section{Introduction}\label{sec:introduction}

National accounts, published by government statistical agencies, provide the central measurement system for macroeconomic analysis. Consumption, for example, is measured comprehensively in the National Income and Product Accounts (NIPA), yet official consumption statistics are available only with a delay and, at sub-national levels, at relatively low frequency or coarse geographic detail. These limitations are especially consequential when researchers and policymakers want to assess rapidly evolving shocks or heterogeneous responses across local areas and household groups. At the same time, the widespread use of credit cards generates large volumes of data on household spending. A growing literature shows that, when organized and benchmarked carefully, such data can be transformed into economic statistics that complement official accounts, offering more timely, higher-frequency, and more geographically detailed measures of economic activity \citep{Vavra2021,AladangadyEtAl2022,BudaCarvalhoHansenOrtizRodrigoMora2023,AndersenHuberJohannesenStraubVestergaard2026}.

In this paper, we use credit card data from the Federal Reserve’s Y-14M reports to construct high-frequency, geographically granular measures of household spending and evaluate their usefulness as consumption measures. Our paper makes three contributions. First, we systematically benchmark Y-14M credit card spending against traditional consumption measures, establishing its reliability (and limitations) as a proxy measure for consumption both in the national time series and in the geographic cross-section. Second, we introduce a novel monthly county-level consumption dataset and illustrate its research value through a proof-of-concept application to heterogeneous monetary policy transmission across the county-level income distribution, an analysis that is infeasible with traditional consumption data. Finally, we provide a guide for researchers using credit card data to measure consumption, discussing key caveats related to consumption-category coverage, sample representativeness, changes in card adoption and sample composition, and the use of credit bureau data to approximate credit card spending.

The Y-14M reports, collected for regulatory stress-testing purposes, require large bank holding companies with at least \$100 billion in total assets to provide detailed, de-identified monthly data on individual credit card accounts. Covering over 350 million U.S. credit cards over the period from 2014 to 2025, the data represent a significant share of the market, accounting for approximately 66 percent of outstanding consumer credit card balances \citep*{CFPB2025}. We obtain a card’s monthly net purchase dollar volume, which serves as our measure for credit card spending, as well as a card's zip code, which allows for a geographically granular attribution of spending.\footnote{Compared to transaction-level credit card datasets, the Y-14M data do not report individual purchases but only monthly account-level spending totals. We therefore cannot distinguish between spending categories.} Throughout the paper, our baseline spending measure is growth in average purchase volume among cards with positive purchase volume. This measure abstracts from changes in the total number of cards in the Y-14M sample. We discuss the construction of our spending measure in Section~\ref{sec:data} and examine how changes in card counts and the entry and exit of active cards affect measured spending growth in Section~\ref{sec:researcher_guide}.

We start by benchmarking monthly year-over-year growth in average credit card spending against PCE growth and retail sales growth at the national level. For our baseline PCE measure, we exclude a number of components that are not true out-of-pocket expenditures by households, such as spending by the non-profit sector, healthcare expenditures paid by insurance, and imputed rental equivalent consumption by households who own their homes. We refer to this baseline measure as \emph{adjusted PCE}.

We find that credit card spending growth approximates both adjusted PCE growth and retail sales growth very well, both during normal times and during the extraordinary circumstances of the COVID-19 pandemic. In a simple univariate time-series regression, variation in credit card spending growth explains 92 percent of the variation in monthly adjusted PCE growth and 71 percent of the variation in monthly retail sales growth over our sample period. We then study the cross-section and benchmark annual credit card spending growth against annual PCE growth at the state level. Credit card spending growth is strongly associated with adjusted PCE growth in pooled state-year regressions and captures meaningful variation in state-level consumption growth, although the within-year cross-sectional relationship is more uneven. Finally, we compare Y-14M credit card spending to several private-sector high-frequency spending measures and show that, over comparable sample periods, the Y-14M series more closely approximates adjusted PCE growth than the alternative transaction-based datasets we examine.  

Having shown that credit card spending closely tracks traditional consumption measures, we introduce a novel monthly county-level consumption dataset constructed from the Y-14M credit card data. Traditional consumption measures are available either at high frequency but with limited geographic detail or at finer geographic disaggregation but lower frequency. Our dataset helps break this trade-off by allowing us to track consumption at the county level at a monthly frequency, a combination of frequency and geographic detail unavailable in traditional public consumption data. We validate the county-level series in several ways. First, we show that counties’ shares of aggregate Y-14M credit card spending are strongly related to their shares of spending in the Economic Census. Second, we document that county-level spending dynamics align closely with well-documented economic developments: counties with higher COVID-19 intensity in the early stages of the pandemic experienced markedly sharper consumption declines in April 2020, while geographically proximate counties within the same metropolitan area exhibit similar spending dynamics throughout the sample period. Finally, county-level consumption as measured by credit card spending is positively associated with local income growth and negatively associated with local unemployment growth. Covering more than 3,000 counties since 2014, this dataset opens the door to research designs that require both high-frequency and granular geographic variation in consumption.

We present a proof-of-concept research application that illustrates the value of monthly county-level consumption data for studying heterogeneous monetary policy transmission across the county-level income distribution, an analysis that is infeasible with traditional consumption measures. Using local projections, we first estimate the impulse responses of credit card spending and traditional consumption measures to contractionary monetary policy shocks at the national and state level. Credit card spending exhibits responses that are qualitatively consistent with those of PCE and retail sales---a gradual decline, a trough at medium horizons, and partial mean reversion---supporting its use as a consumption proxy in dynamic settings. We then exploit the geographic granularity of our dataset to estimate heterogeneous consumption responses across counties with different income levels, an analysis that is infeasible with traditional public consumption data at monthly frequency. A large theoretical literature, anchored in heterogeneous-agent New Keynesian models, predicts that monetary policy transmits more strongly to lower-income and liquidity-constrained households, yet direct empirical evidence at fine geographic resolution has been limited by data availability. Using our county-month panel, we show that lower-income counties exhibit substantially larger consumption declines following a contractionary monetary policy shock than higher-income counties, consistent with these predictions.

We conclude with a researcher’s guide for using credit card data to measure consumption. This section highlights the strengths of the Y-14M data, discusses the main caveats involved in interpreting credit card spending as a consumption measure, and provides practical guidance for researchers working with similar credit card or credit bureau data. We first show that the relationship between credit card spending and PCE varies meaningfully across PCE consumption categories, underscoring the importance of choosing an appropriate benchmark when evaluating card-based spending measures. Next, we decompose credit card spending growth in two ways: first into average-spending and card-count components, and then into intensive- and extensive-margin components of average spending growth. These decompositions show that both the number and composition of active cards are important for interpreting measured spending growth over time. We further document substantial heterogeneity across banks, showing that individual bank-level spending series can be noisy even when the aggregate Y-14M series closely tracks consumption. Finally, we evaluate whether researchers can approximate credit card spending using the types of variables commonly available in credit bureau data. We find that spending can be approximated reasonably well using changes in balances and payments, but that approximation quality varies across borrower groups and deteriorates for lower-FICO accounts. Together, these results provide a set of caveats and best practices for researchers seeking to use credit card data as a consumption measure.

\textbf{Related literature and contribution.} First, we contribute to the growing literature on how naturally occurring private-sector data can be harnessed to construct high-frequency, geographically detailed measures of economic activity that complement official statistics. Most closely related to our paper are studies that develop transaction-based consumption datasets and benchmark them against established consumption measures.  \citet{AladangadyEtAl2022} construct real-time consumer spending indexes from Fiserv payment-processor data that closely track Census retail sales at the national, state, and MSA levels, while \citet{BudaCarvalhoHansenOrtizRodrigoMora2023} use BBVA retail bank-account transactions to construct a representative consumption panel that reproduces national-account consumption aggregates in Spain. More broadly, related work uses private-sector and administrative micro data to build real-time measures of economic activity \citep{ChettyFriedmanHendrenStepner2024} and disaggregated economic accounts \citep{AndersenHuberJohannesenStraubVestergaard2026}. We contribute to this literature by constructing a novel monthly county-level dataset based on U.S. credit card spending and by showing how these data can be used to approximate official consumption statistics. Unlike many existing transaction-based datasets, our data are not drawn from a single bank, fintech, or payment processor, but from comprehensive supervisory data covering a large share of the U.S. credit card market, allowing us to benchmark card spending against official consumption across time, categories, and geographies.

Our paper further relates to work that uses transaction-based spending data to study household consumption dynamics and responses to shocks. The COVID-19 pandemic highlighted the usefulness of these data, with researchers drawing on spending data from JPMorgan Chase \citep{CoxGanongNoelVavraWongFarrellGreig2020}, BBVA \citep{CarvalhoGarciaHansenOrtizRodrigoMoraRuiz2021}, and the fintech company SaverLife \citep{BakerFarrokhniaMeyerPagelYannelis2020} to study high-frequency, disaggregated spending dynamics. Earlier work also used payment-processor and credit card data to study the consumption consequences of housing, income, wealth, and monetary policy shocks \citep{MianRaoSufi2013,GanongNoel2019,GanongNoel2020,BudaEtAl_Short_Lags_of_MP2025}. \citet{Vavra2021} provides an overview of this literature and highlights the broader role of administrative micro data in real-time business-cycle analysis. We evaluate the measurement assumptions underlying these applications. Specifically, we provide the first systematic assessment of the validity and limitations of credit card spending as a proxy for consumption dynamics and offer practical guidance for researchers working with similar data.

Finally, we contribute to research on heterogeneous-agent macroeconomic models and their implications for household consumption. In HANK models, the transmission of monetary policy and other aggregate shocks depends on the distribution of household income, liquidity, balance sheets, and marginal propensities to consume \citep{KaplanMollViolante2018,Auclert2019}. Related empirical work shows that monetary policy can have heterogeneous effects across households and regions, including through balance-sheet positions and mortgage refinancing channels \citep{BerajaFusterHurstVavra2019}. Our data provide a way to study these mechanisms at high frequency and fine geographic detail using a broad measure of household spending. As an illustration, we use our credit-card-based consumption measures to estimate local consumption responses to contractionary monetary policy shocks and show that low-income counties exhibit substantially larger spending declines.  

The remainder of the paper is structured as follows. Section~\ref{sec:data} discusses our data and Section~\ref{sec:descriptive_statistics} presents summary statistics. In Section~\ref{sec:benchmarking}, we benchmark credit card spending against traditional data sources of consumption and in Section~\ref{sec:beyond_benchmarking_a_new_dataset} we present our novel monthly county-level consumption dataset. Section~\ref{sec:research_application} presents our research application and studies the transmission of monetary policy to consumption. In Section~\ref{sec:researcher_guide}, we discuss caveats and best practices and provide a guide for researchers working with credit card spending data. Section~\ref{sec:conclusion} concludes.

\section{Data}\label{sec:data}

\subsection{Y-14M Credit Card Data}\label{sec:data_y14m}

We obtain monthly account-level data on U.S. consumer credit cards from the Federal Reserve's supervisory FR Y-14M reports. These reports require large bank holding companies, which are subject to DFAST/CCAR stress tests, to report detailed information on individual credit card accounts on a monthly basis. The data are available from June 2013 and include, depending on the month and on prevailing reporting requirements, between 16 and 20 banks, covering a large portion of the credit card market and accounting for 66 percent of aggregate outstanding balances on consumer credit cards \citep*{CFPB2025}.

Our main variables of interest are a card's purchase volume, defined as ``the net purchase dollar volume during the current month's cycle'' and the account zip code, defined as the nine-digit or five-digit zip code ``the cardholder reported as their billing address'' \citep*{FRB2024}. These variables allow us to track monthly credit card spending at the zip-code level. We map account zip codes to counties using standard zip-code crosswalks. Thus, our data record spending by the place of residence of the consumer (household-based data), rather than by the location of expenditure (point-of-sale data).

We focus on general-purpose and private-label consumer credit cards that, in any given month, had any account activity within the previous twelve months. To ensure consistency over time, we further restrict the sample to 13 banks that are present throughout the entire sample period from 2014 to 2025. Finally, in each month, we only consider cards with nonzero purchase volume. This restriction focuses the analysis on cards used for purchases in a given month and avoids mechanically diluting average spending with dormant or unused accounts. This leaves us with a dataset of 142 to 247 million credit cards per month.

From these data, we construct three datasets: (A) monthly data at the national level, (B) annual data at the state level, and (C) monthly data at the county level. For each dataset, we calculate credit card spending growth as growth in average credit card spending among cards with positive purchase volume. Let
\[
\overline{CCSpending}_{r,t}
=
\frac{1}{N_{r,t}}
\sum_{i=1}^{N_{r,t}} CCSpending_{i,r,t}
\]
denote average credit card spending in region \(r\) and period \(t\), where \(i\) indexes individual cards and \(N_{r,t}\) is the number of cards with positive purchase volume in region \(r\) and period \(t\). We calculate credit card spending growth as
    \begin{equation}
    g^{CC}_{r,t}
    =
    \left(
    \frac{
    \overline{CCSpending}_{r,t}
    -
    \overline{CCSpending}_{r,t-\tau}
    }{
    \overline{CCSpending}_{r,t-\tau}
    }
    \right)
    \times 100,
    \label{eq:cc_spending_growth}
    \end{equation}
where \(\tau=12\) for the monthly datasets and \(\tau=1\) for the annual dataset.

Our baseline measure is growth in average spending among cards with positive purchase volume, rather than growth in total credit card spending. This choice is intended to abstract from changes in the number of cards observed in the Y-14M data, which can reflect secular growth in card adoption and bank portfolio growth in addition to changes in household consumption. However, average spending growth can still be affected by changes in the composition of cards with purchase activity, since the set of active cards is not constant over time. We investigate both of these issues in Section~\ref{sec:researcher_guide_payment_shifts} by decomposing total credit card spending growth into changes in average spending and changes in the number of cards, and by decomposing average credit card spending growth into intensive-margin spending changes among continuing cards and extensive-margin changes due to card entry and exit.

\subsection{Benchmarking Data}\label{sec:data_benchmarking}

We use two types of benchmarking data. First, we compare Y-14M credit card spending to official consumption and retail sales statistics produced by federal statistical agencies. These data provide the most widely used benchmarks for aggregate consumption, but they are available at limited geographic granularity or relatively low frequency and are released with some delay. Second, we also compare Y-14M credit card spending to several alternative private-sector high-frequency transaction datasets. These datasets are closer in spirit to our credit card measure as they are constructed from card transactions, electronic payments, or household purchase records rather than from surveys or national-accounting source data.

\subsubsection{Official Consumption Statistics}

\textbf{Personal Consumption Expenditures (PCE):} Our primary official consumption benchmarks are national Personal Consumption Expenditures (PCE) from the National Income and Product Accounts (NIPA) and annual state-level PCE from the BEA Regional Accounts, both published by the Bureau of Economic Analysis (BEA). National PCE comprises about 68 percent of U.S. GDP and includes all expenditures by households and non-profit institutions serving households (NPISH).  BEA constructs PCE using a variety of source data, which often arrive with a lag and are available only at an aggregated level. As a result, initial national PCE estimates are typically released about four weeks after the end of the reference month and are subject to revision as additional data become available. BEA also publishes estimates of PCE at the state level, reflecting spending by households and non-profits residing in each U.S. state.\footnote{State-level PCE is available annually rather than monthly and is released with a longer delay than national PCE. Details on the construction of National PCE are available \hyperlink{https://www.bea.gov/resources/methodologies/nipa-handbook/pdf/chapter-05.pdf}{here} and details on State PCE \hyperlink{https://www.bea.gov/sites/default/files/2024-11/BEA-PCE-by-State-Concepts-and-Methods.pdf}{here}.}  

Both national and state-level PCE include components that are not true out-of-pocket expenditures by households, such as spending by the non-profit sector, healthcare expenditures paid by insurance, and imputed rental equivalent consumption by owner-occupied households. They also include categories of spending that are often not paid for using credit cards, such as motor vehicle purchases and rents. For our main analysis, we construct an adjusted PCE measure designed to be more comparable to credit-card-based household spending. Specifically, we start from total PCE and subtract six broad components: motor vehicles and parts; housing and utilities; health care; financial services and insurance; other services; and final consumption expenditures of NPISH:
\begin{equation}
    \begin{aligned}
        \widetilde{\text{PCE}}_{t} = \text{PCE}_{t}   &- \text{Motor Vehicles}_{t} - \text{Housing and Utilities}_{t} - \text{Health Care}_{t} \\
                                        &- \text{Financial Services}_{t} - \text{Other Services}_{t} - \text{NPISH}_{t},
    \end{aligned}
\label{eq:adjusted_pce}
\end{equation} 

where \(\widetilde{PCE}_{t}\) denotes adjusted PCE. We exclude motor vehicles and parts as these purchases are large, infrequent expenditures that are rarely paid for primarily by credit card. We exclude housing and utilities, health care, and financial services as these categories include substantial expenditures that are imputed, paid by third parties, or otherwise not directly comparable to card-based
household purchases. We exclude final consumption expenditures by NPISH as our credit card data measure consumer spending by cardholders rather than expenditures by non-profit institutions. Finally, we exclude other services as this residual services category contains several components that are less directly comparable to the purchase-volume concept in the Y-14M data. We refer to the resulting aggregate as \emph{adjusted PCE}.

\textbf{Census Retail Sales.} We further benchmark our credit card spending data against Census retail and food services sales, one of the main source datasets used by the BEA in constructing PCE. These data capture spending on retail goods and food services, categories in which credit cards are a common method of payment. The data come from a monthly survey of establishments known as the Monthly Retail Trade Survey (MRTS). Since January 2019, the Census Bureau has also published experimental Monthly State Retail Sales (MSRS) estimates, which report modeled year-over-year percentage changes by state and retail subsector and provide an additional subnational benchmark for credit card spending. Sampled establishments report revenues, which are then weighted and aggregated to the industry level and published at a monthly frequency by the U.S. Census Bureau, typically about two weeks after the end of the reference month.

\subsubsection{Private Sector High-Frequency Data}

\textbf{Fiserv:} Fiserv is one of the largest electronic payment processors in the United States. The data are merchant-based and record the dollar amount, timing, and merchant location of card and electronic transactions at merchants that use Fiserv's payment-processing network. Following \citet{AladangadyEtAl2022}, the raw transaction data are filtered to construct constant-merchant spending indexes, mapped to Census retail categories using merchant category codes, and benchmarked to Economic Census levels. The resulting series provide high-frequency measures of retail spending by industry and geography. Relative to the Y-14M data, Fiserv is merchant-based rather than cardholder-based, making it conceptually close to Census retail sales but less directly comparable to residence-based consumption measures.

\textbf{Affinity Solutions.} Affinity Solutions aggregates anonymized credit and debit card transactions from financial institutions. We use the Affinity Solutions spending indexes distributed through the Opportunity Insights Economic Tracker \citep{ChettyFriedmanHendrenStepner2024}. The data are cardholder-based and can be aggregated by geography, industry, day, and ZIP-code income quartile, allowing researchers to track spending across local areas and income groups. Relative to Fiserv, Affinity is closer to the household-spending concept in the Y-14M data because transactions are assigned to cardholders rather than merchants; relative to the Y-14M data, however, it includes both credit and debit cards and is available only through processed spending indexes rather than account-level records.

\textbf{Numerator:} Numerator is a consumer panel dataset that records detailed household purchases from submitted paper receipts, online receipts, and linked loyalty or membership transactions \citep{HaciogluHokeFelerChylak2025}. The data include item-level information, retailer information, transaction timing, and household characteristics, and are weighted to match Census demographics and major retailer sales aggregates. Numerator therefore provides richer information on spending categories and household characteristics than the Y-14M data, but it is based on a selected consumer panel rather than administrative account-level card records.

\textbf{Verisk Analytics:} Verisk Analytics aggregates credit and debit card transactions at the ZIP-code-by-week level \citep{AladangadyEtAl2025StudentLoanRestart}. Transactions are assigned to the ZIP code of the card owner's billing address rather than to the location where the transaction occurred. This residence-based geographic assignment makes Verisk conceptually close to the Y-14M data, while its weekly frequency and ZIP-code-level demographic information provide additional flexibility. Relative to Y-14M, however, Verisk includes both credit and debit cards and is available only as an aggregated spending product.

\section{Descriptive Statistics}\label{sec:descriptive_statistics}

Table~\ref{tab:summary_stats} reports summary statistics for the datasets used in our main analysis: monthly data at the national level (Panel A), annual data at the state level (Panel B), and monthly data at the county level (Panel C). The 142 monthly observations in Panel A span nearly twelve years of data; the 510 state-year observations in Panel B correspond to 11 years across 51 states (including the District of Columbia); and the 444,886 county-month observations in Panel C cover more than 3,000 counties over approximately the same period. The national-level data include, on average, about 190 million credit cards from 13 banks per month; the state-level data include about 3.5 million credit cards from 13 banks per state-year; and the county-level data include about 59 thousand credit cards per county-month. While all 13 banks are represented in the national- and state-level datasets, bank coverage varies across counties, with the smallest counties covered by at least eight banks.

As shown in Panel A, monthly year-over-year credit card spending growth is, on average, slightly lower than both monthly year-over-year adjusted PCE growth and monthly year-over-year retail sales growth, a pattern that also holds for annual credit card spending growth relative to adjusted PCE growth at the state level (Panel B). Credit card spending growth also exhibits greater volatility than adjusted PCE growth: the first percentile of monthly credit card spending growth is (-20.4) percent, compared with (-13.2) percent for adjusted PCE, while the 99th percentile is 36.1 percent, compared with 29.4 percent. This wider range is consistent with the fact that credit card transactions disproportionately capture discretionary expenditures, which are more sensitive to economic conditions. Nonetheless, as we show in the next section, credit card spending growth closely tracks the time-series movement of both adjusted PCE growth and retail sales growth.

\begin{table}[!t]
    \caption{Descriptive Statistics}
    \footnotesize{This table reports summary statistics for the three main analysis samples. Panel A reports statistics for the monthly national sample from June 2014 to March 2026. Panel B reports statistics for the annual state-level sample from 2014 to 2024. Panel C reports statistics for the monthly county-level sample from June 2014 to March 2026.}
    \begin{center}
        \begin{tabular}{lccccc}
            \toprule
                                                    & Obs.      & Mean      & Median    & p(1)      & p(99)         \\
            \midrule
            \multicolumn{6}{l}{\textit{Panel A. Monthly National Level.}}                                       \\
            \addlinespace
            Number of Cards (in m)                  & 142       & 191.81    & 185.87    & 142.43	& 247.65    \\
            Number of Banks                         & 142       & 13        & 13        & 13        & 13        \\
            Monthly YoY CC Spending Growth (in \%)  & 142       & 3.63      & 2.90      & -20.44	& 36.12     \\
            Monthly YoY Adjusted PCE Growth (in \%) & 142       & 5.30      & 4.29      & -13.15	& 29.39     \\
            Monthly YoY Retail Sales Growth (in \%) & 142       & 5.08      & 3.87      & -6.20	    & 28.88     \\
            \addlinespace
            \multicolumn{6}{l}{\textit{Panel B. Annual State Level.}}                                           \\
            \addlinespace
            Number of Cards (in m)                  & 510       & 3.68      & 2.12      & 0.27      & 25.34     \\
            Number of Banks                         & 510       & 13        & 13        & 13        & 13        \\
            Annual CC Spending Growth (in \%)       & 510       & 3.77      & 2.59      & -8.31     & 18.26     \\
            Annual Adjusted PCE Growth (in \%)      & 510       & 5.06      & 3.92      & -7.29     & 21.07     \\
            \addlinespace
            \multicolumn{6}{l}{\textit{Panel C. Monthly County Level.}}                                         \\
            \addlinespace
            Number of Cards (in k)                  & 444{,}886 & 61.00     & 9.79      & 0.38      & 909.77    \\
            Number of Banks                         & 444{,}886 & 11.64     & 12        & 8         & 13        \\
            Monthly YoY CC Spending Growth (in \%)  & 444{,}886 & 3.90      & 3.13      & -15.30    & 30.55     \\
            \bottomrule
        \end{tabular}
    \end{center}
    \label{tab:summary_stats}
\end{table}

\section{Benchmarking}\label{sec:benchmarking}

We benchmark Y-14M credit card spending against two types of consumption measures. We first compare the data to official consumption and retail sales statistics produced by the BEA and Census Bureau, focusing on both monthly national time-series dynamics and annual state-level cross-sectional variation. We then compare Y-14M spending to several private-sector high-frequency spending measures, which are closer in spirit to credit card data but differ in coverage, transaction concept, payment instruments, and geographic attribution. These comparisons allow us to assess both how closely Y-14M credit card spending approximates traditional consumption benchmarks and whether our results reflect broader patterns also present in other transaction-based spending data.

\subsection{Monthly National Benchmarks}\label{sec:benchmarking_monthly_national}

\begin{figure}[!t]
    \caption{Monthly National Consumption Growth: Official Benchmarks} 
    \footnotesize{This figure plots monthly year-over-year growth rates in average credit card spending (solid blue line), adjusted personal consumption expenditures (adjusted PCE; dashed red line), and retail sales (dashed green line) from June 2014 to March 2026.}
    \begin{center}
        \centerline{\includegraphics[width=1\textwidth]{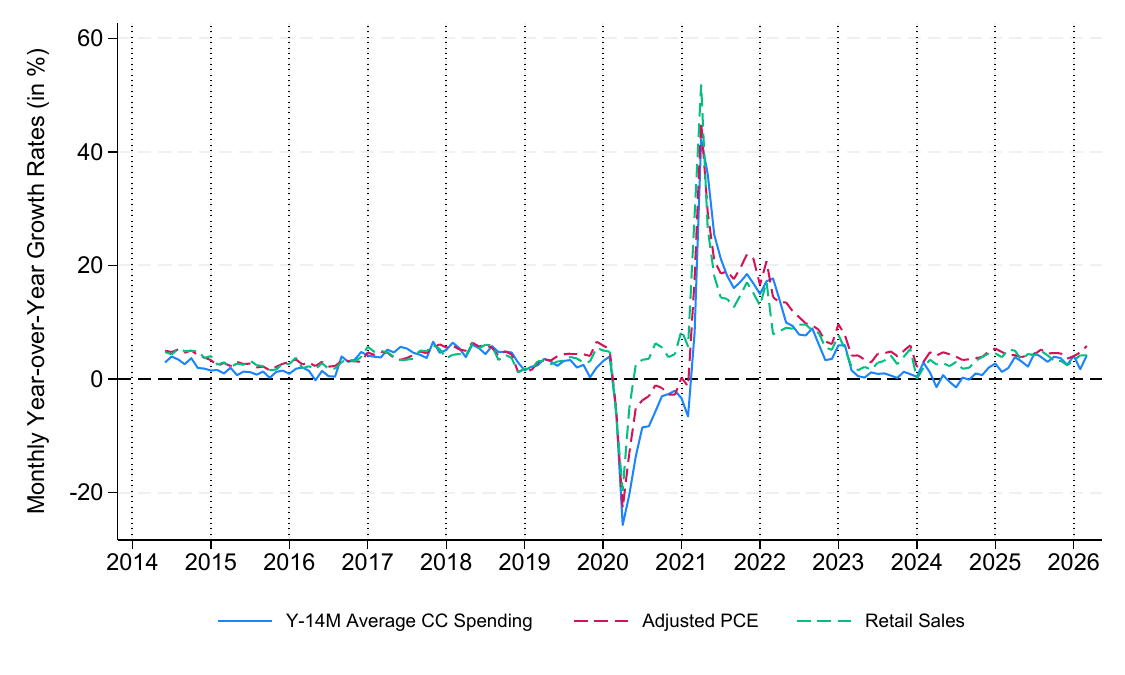}}
    \end{center}  	
    \label{fig:monthly_yoy_consumption_growth}
\end{figure}

Figure~\ref{fig:monthly_yoy_consumption_growth} plots monthly year-over-year growth in credit card spending (solid blue line) alongside monthly year-over-year growth in adjusted PCE (dashed red line) and retail sales (dashed green line). Credit card spending closely tracks both benchmark series over time, including during relatively stable periods and during the extraordinary circumstances of the COVID-19 pandemic. From 2014 to 2020, the three series fluctuate in a narrow range and exhibit very similar dynamics, with only modest differences in growth rates across the measures.    

Importantly, the sharp decline in credit card spending at the onset of the COVID-19 pandemic closely mirrors the declines in adjusted PCE and retail sales. Credit card spending fell by 25.7 percent at its trough in April 2020, compared with declines of 22.7 percent in adjusted PCE and 19.7 percent in retail sales. While retail sales recovered more rapidly in the months following the initial shock, credit card spending continued to provide a close approximation to adjusted PCE growth. From 2021 through 2023, credit card spending also tracks the normalization of consumption growth in the post-pandemic period. Beginning in 2023, credit card spending growth again falls slightly below both adjusted PCE and retail sales growth, before the three series converge once more by 2025.

Table~\ref{tab:monthly_yoy_consumption_growth} reports simple time-series regressions of monthly year-over-year growth in adjusted PCE and retail sales on credit card spending growth. As shown in Column (1), credit card spending growth is strongly correlated with PCE growth, yielding an $R^2$ of 0.92. The slope coefficient of 0.88 indicates that a one percentage point increase in credit card spending growth is associated with a 0.88 percentage point increase in PCE growth, consistent with credit card spending being a more volatile proxy for consumption. Notably, credit card spending growth explains more of the variation in PCE growth than retail sales growth does (Column 2, $R^2$ = 0.83). Column (3) shows that credit card spending growth is also strongly correlated with retail sales growth, although less so than with PCE growth. Taken together, Figure~\ref{fig:monthly_yoy_consumption_growth} and Table~\ref{tab:monthly_yoy_consumption_growth} indicate that credit card spending provides a close approximation to traditional consumption benchmarks at the monthly national level.

\begin{table}[!t]
    \caption{Benchmarking Monthly National Consumption} 
    \footnotesize{\flushleft This table reports monthly national time-series regressions from June 2014 to March 2026. All variables are measured as monthly year-over-year growth rates. Columns (1) and (2) use adjusted PCE growth as the dependent variable, and Column (3) uses retail sales growth as the dependent variable. The explanatory variables are average credit card spending growth in Columns (1) and (3) and retail sales growth in Column (2). Standard errors are reported in parentheses. $^{***}$, $^{**}$, and $^{*}$ denote statistical significance at the 1, 5, and 10 percent levels, respectively.}
    \begin{center}
        \begin{tabular}{lcccc}
        \toprule
         & \multicolumn{2}{c}{$\Delta$ Adjusted PCE}    & $\Delta$ Retail Sales \\
        \cmidrule(lr){2-3} \cmidrule(lr){4-4}
                                & (1)       & (2)       & (3)       \\
        \midrule
        $\Delta$ CC Spending    & 0.88\c    &           & 0.70\c    \\
                                & (0.02)    &           & (0.04)    \\
        \addlinespace
        $\Delta$ Retail Sales   &           & 1.00\c    &           \\
                                &           & (0.04)    &           \\
        \addlinespace
        Adjusted $R^2$          & 0.92      & 0.83      & 0.71      \\
        Observations            & 142       & 142       & 142       \\
        \bottomrule
        \end{tabular}
    \end{center}
    \label{tab:monthly_yoy_consumption_growth}
\end{table}

\subsection{State-Level Benchmarks}\label{sec:benchmarking_state}

We next examine whether credit card spending closely tracks consumption not only at the national level, but also across states. We benchmark our state-level spending measures against two official sources: annual PCE from the BEA, the broadest available measure of state-level consumption, and monthly Census retail sales, a higher-frequency but narrower measure of retail activity.

\subsubsection{Annual State-Level PCE}\label{sec:benchmarking_annual_state_pce}

\begin{figure}[!t]
    \caption{Annual State-Level Consumption Levels} 
    \footnotesize{This figure plots annual adjusted PCE per capita against average annual Y-14M credit card spending per card for state-year observations from 2014 to 2024. Each dot represents a state-year observation. The solid red line indicates the fitted regression line, and the dashed black line indicates the 45-degree line.}
    \begin{center}
        \centerline{\includegraphics[width=1\textwidth]{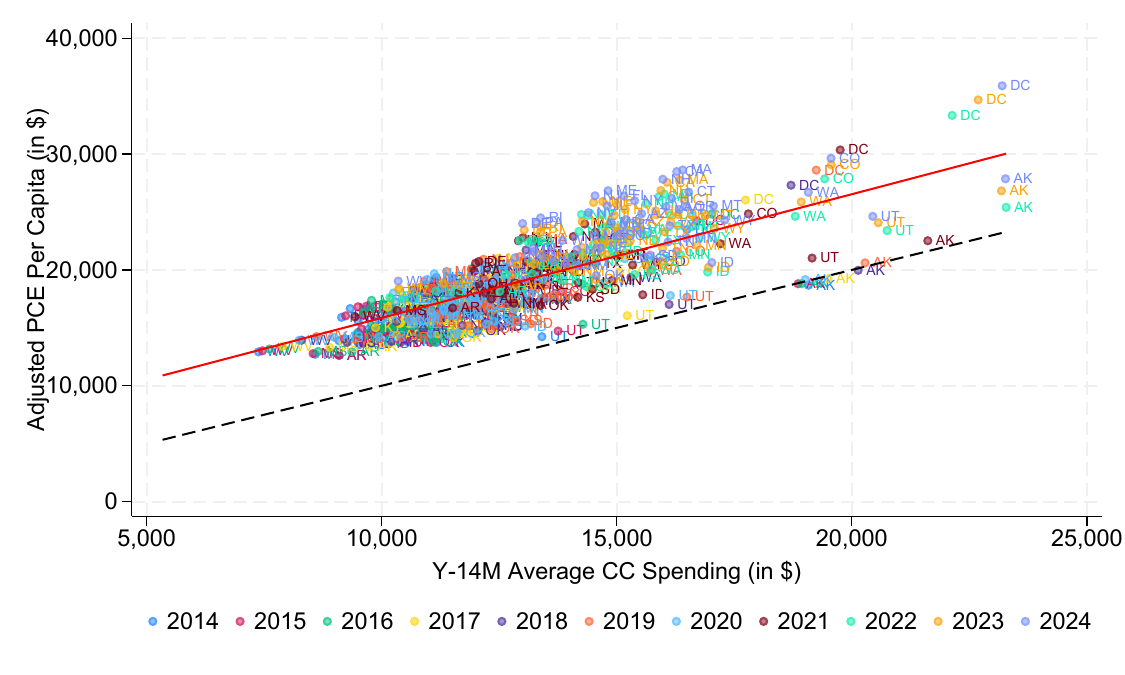}}
    \end{center}  	
    \label{fig:y14m_vs_pce_per_capita}
\end{figure}

We begin by examining whether cross-sectional differences in credit card spending levels are informative about cross-sectional differences in adjusted PCE levels. Figure~\ref{fig:y14m_vs_pce_per_capita} plots average annual credit card spending per card against adjusted PCE per capita across states for all years from 2014 to 2024. The figure shows a strong positive relationship: states with higher average credit card spending also tend to have higher per-capita consumption. A univariate regression yields a slope coefficient of 1.16 and an adjusted $R^2$ of 0.66, indicating that credit card spending captures a substantial share of cross-sectional variation in consumption levels across states.

\begin{figure}[!t]	
    \caption{Annual State-Level Consumption Growth: Cross-Sectional Evidence}
    \par
    \footnotesize{This figure plots annual growth rates in adjusted PCE against annual growth rates in average credit card spending at the state level for selected years. Panel (a) shows 2017, Panel (b) 2018, Panel (c) 2020, and Panel (d) 2021. Each dot represents a state. The solid red lines indicate fitted regression lines, and the dashed black lines indicate 45-degree lines.}
    \begin{center}
        \subfigure[2017]{\includegraphics[width=.495\textwidth]{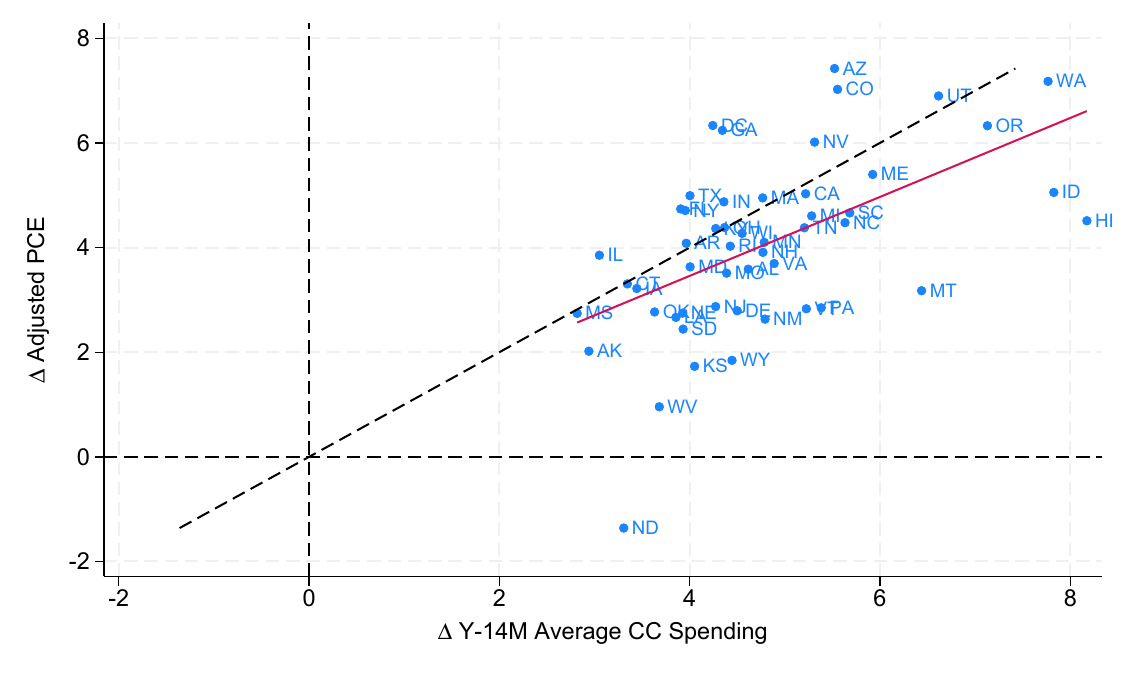}}\label{fig:Fig5a}
        \subfigure[2018]{\includegraphics[width=.495\textwidth]{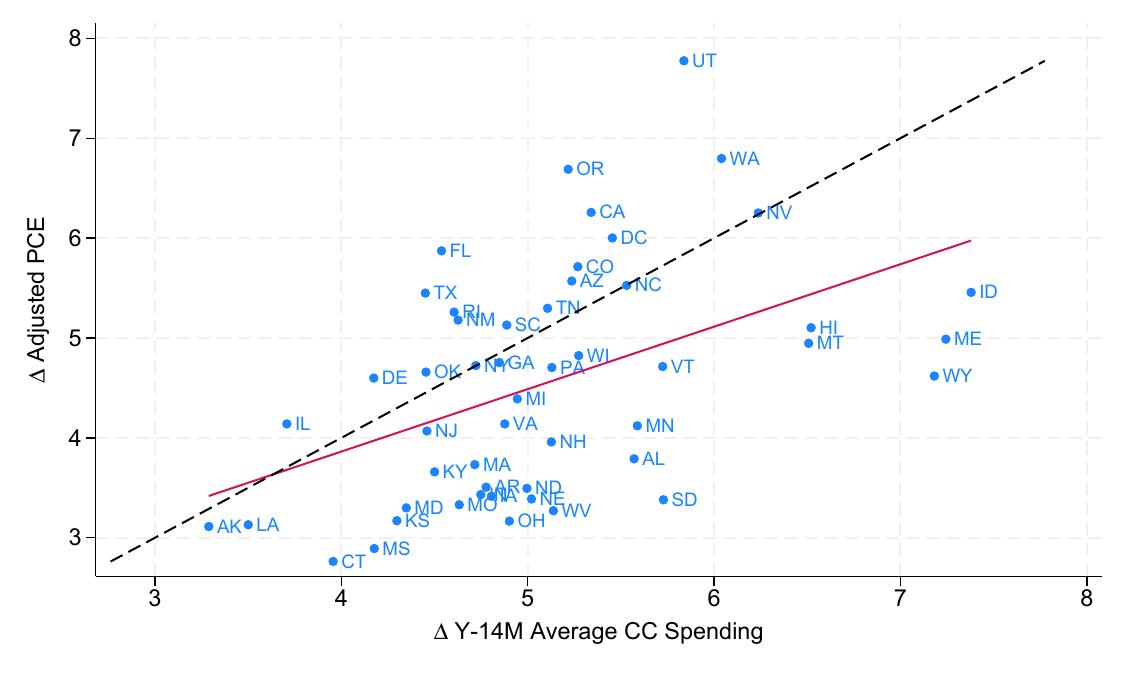}}\label{fig:Fig5b}
        \subfigure[2020]{\includegraphics[width=.495\textwidth]{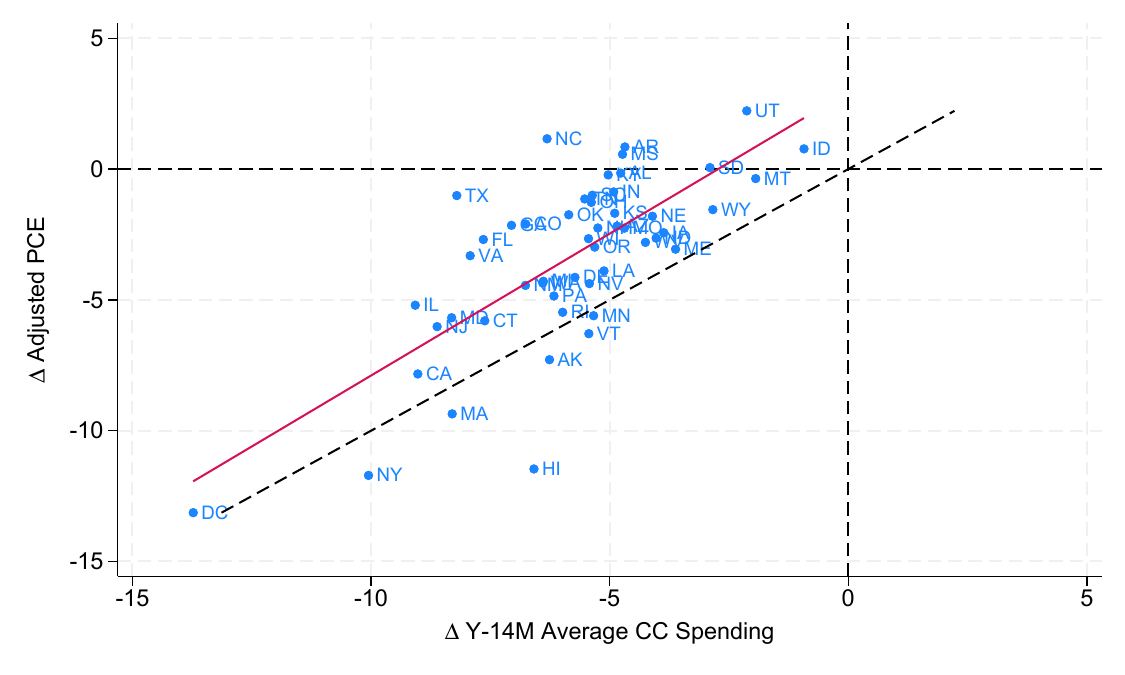}}\label{fig:Fig5c}
        \subfigure[2021]{\includegraphics[width=.495\textwidth]{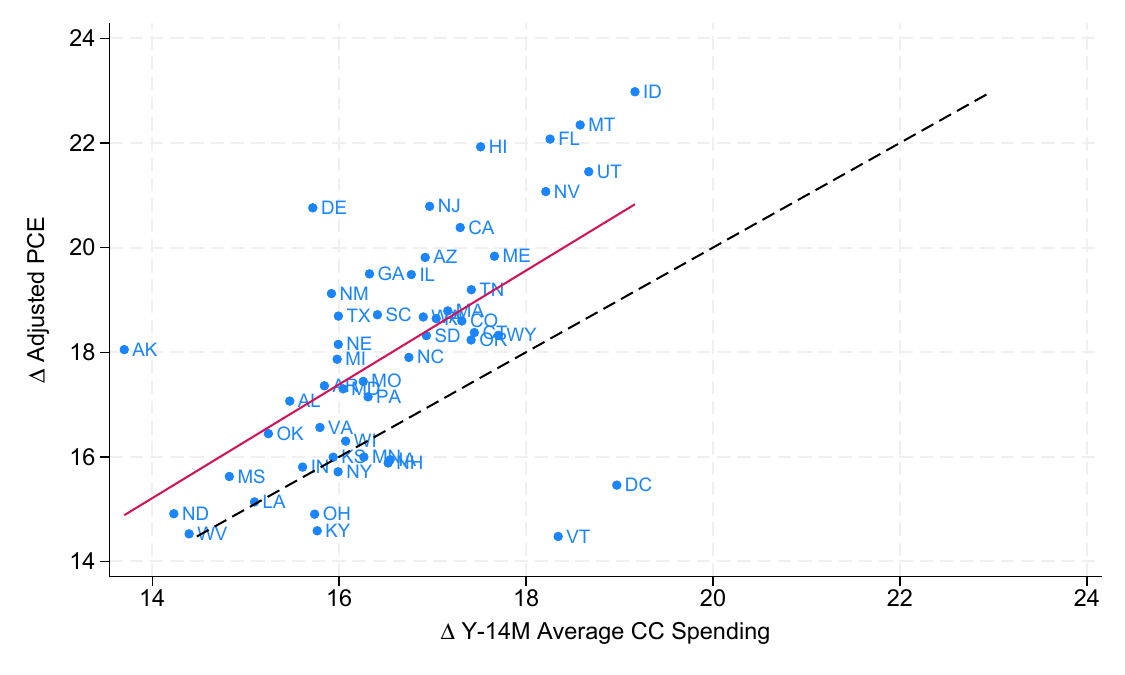}}\label{fig:Fig5d}
    \end{center}
    \label{fig:scatter_plots_annual_cc_versus_pce_state_year}
\end{figure}

We then turn from levels to growth rates. Because state-level PCE is available only at an annual frequency, we benchmark annual credit card spending growth against annual adjusted PCE growth. Figure~\ref{fig:scatter_plots_annual_cc_versus_pce_state_year} illustrates this relationship for selected years, while Table~\ref{tab:regression_annual_cc_versus_adjusted_pce_state_year} reports the corresponding regression evidence for the full sample and separately by year. Pooling all state-year observations, annual growth in average credit card spending is strongly associated with annual adjusted PCE growth, with a slope coefficient of 0.93 and an adjusted $R^2$ of 0.88. This pooled relationship, however, combines both time-series and cross-sectional variation and is influenced by years with large aggregate consumption movements, especially the pandemic contraction and rebound in 2020 and 2021. The within-year cross-sectional relationship is more modest. The fit is positive in most years, but weaker in 2019, 2022, and 2023, when annual consumption growth varied less across states and the estimated coefficients are not statistically significant.\footnote{Figure~\ref{fig:annual_state_level_consumption_growth} in the appendix further shows selected state-level time series and confirms that credit card spending growth also tracks adjusted PCE growth within individual states over time.} 

\begin{table}[!t]
    \caption{Benchmarking Annual State-Level Consumption}
    \footnotesize{\flushleft This table reports state-level regressions of annual adjusted PCE growth on annual growth in average credit card spending. The first column pools all state-year observations from 2015 to 2024, and the remaining columns estimate the regression separately by year. Robust standard errors are reported in parentheses.  $^{***}$, $^{**}$, and $^{*}$ denote statistical significance at the 1, 5, and 10 percent levels, respectively.}
    \begin{center}
        \begin{tabular}{lcccccc}
            \toprule
                                    & \multicolumn{6}{c}{$\Delta$ Adjusted PCE} \\ 
                                    \cmidrule(lr){2-7}
                                    & All Years & 2015      & 2016      & 2017      & 2018      & 2019      \\ 
            \midrule
            $\Delta$ CC Spending    & 0.93\c    & 0.53\c    & 0.67\b    & 0.75\c    & 0.63\c    & -0.07     \\
                                    & (0.02)    & (0.18)    & (0.27)    & (0.18)    & (0.15)    & (0.17)    \\ 
            \addlinespace
            Adjusted $R^2$          & 0.88      & 0.14      & 0.23      & 0.30      & 0.22      & 0.00      \\
            Observations            & 510       & 51        & 51        & 51        & 51        & 51        \\ 
            \midrule
                                    &           & 2020      & 2021      & 2022      & 2023      & 2024      \\
            \midrule
            $\Delta$ CC Spending    &           & 1.08\c    & 1.09\c    & 0.25      & 0.20      & 0.19\c    \\
                                    &           & (0.12)    & (0.30)    & (0.15)    & (0.13)    & (0.05)    \\ 
            \addlinespace
            Adjusted $R^2$          &           & 0.53      & 0.34      & 0.04      & 0.05      & 0.17      \\
            Observations            &           & 51        & 51        & 51        & 51        & 51        \\
            \bottomrule
        \end{tabular}
    \end{center}
    \label{tab:regression_annual_cc_versus_adjusted_pce_state_year}
\end{table}

Both Y-14M credit card spending and state-level PCE are conceptually household-based measures of consumption. However, the most comprehensive source data for state-level PCE statistics are Economic Census receipts by the state of the business establishment, which measure where purchases occur (point-of-sale data) rather than where purchasers live (household-based data). These point-of-sale-based data are converted into household-based measures using a residency adjustment procedure. For state-category cells in which point-of-sale spending appears distorted by purchases from nonresidents, expenditure shares are adjusted toward Consumer Expenditure Survey–based resident spending shares, with the remaining state shares rescaled to preserve national PCE totals \citep{bea2024}. By contrast, Y-14M credit card spending can be assigned directly to the cardholder's state of residence, without the need for an analogous residency adjustment. The reliance of state-level PCE on multiple source datasets, interpolation procedures, and selective residency adjustments reflects the well-known methodological challenges involved in estimating residence-based consumption at the state level. These challenges may contribute to weaker state-level correlations between Y-14M spending growth and adjusted PCE growth in some years.

\subsubsection{Monthly State-Level Retail Sales}\label{sec:benchmarking_monthly_state_census_retail_sales}

A limitation of the state-level PCE data is that they are available only at an annual frequency. To evaluate whether Y-14M spending also captures higher-frequency geographic variation in consumption, we compare monthly state-level credit card spending growth to monthly state-level retail sales growth from the Census Bureau.

\begin{figure}[!t]
    \caption{Monthly State-Level Retail Spending Growth: Cross-Sectional Evidence} 
    \footnotesize{This figure plots monthly year-over-year growth rates in Census retail sales against monthly year-over-year growth rates in average credit card spending at the state level from 2019 to 2023. Each dot represents a state-month observation. The solid red line indicates the fitted regression line, and the dashed black line indicates the 45-degree line.}
    \begin{center}
        \centerline{\includegraphics[width=1\textwidth]{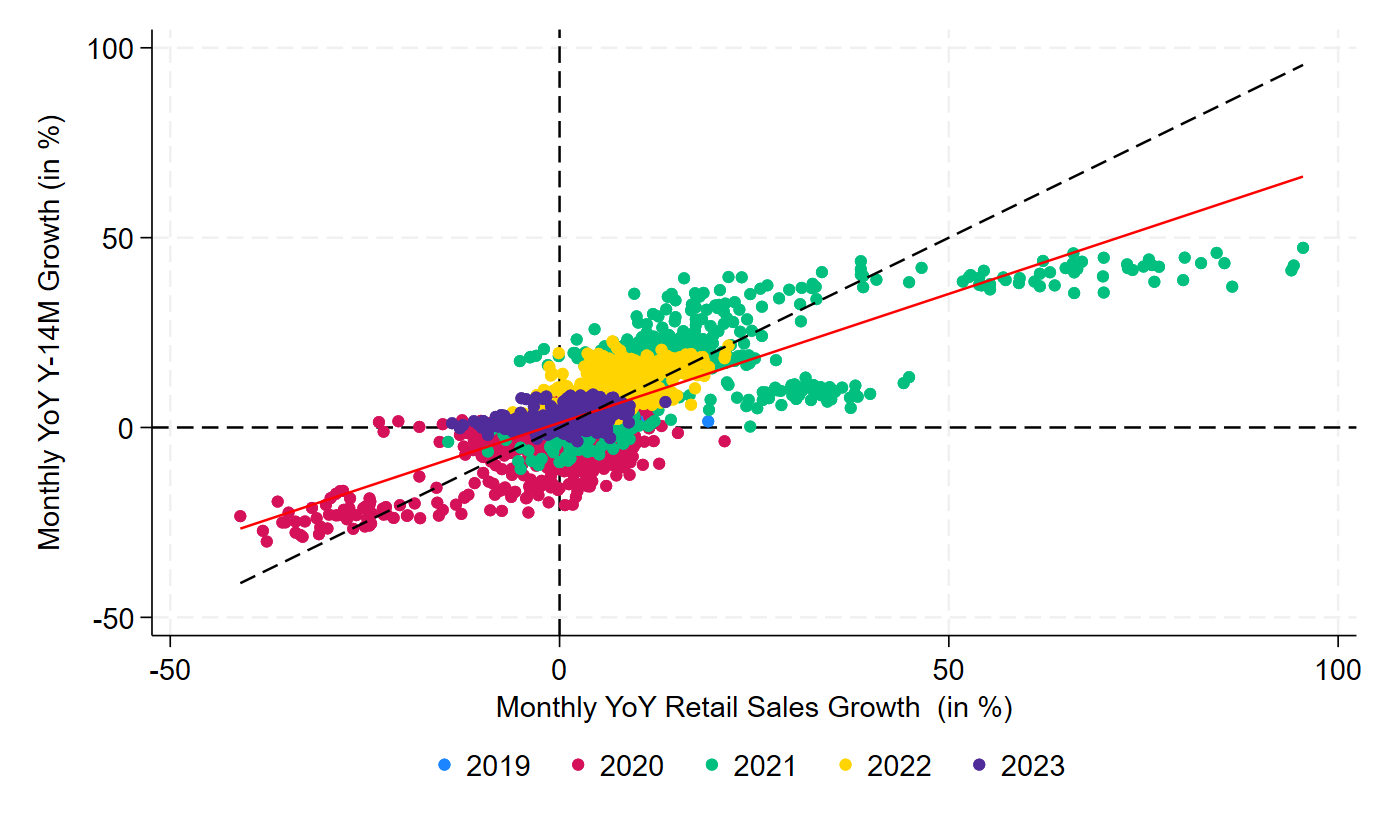}}
    \end{center}  	
    \label{fig:y14m_vs_retail_sales_monthly}
\end{figure}

Figure~\ref{fig:y14m_vs_retail_sales_monthly} plots monthly year-over-year growth in average Y-14M credit card spending against monthly year-over-year growth in Census retail sales, where each dot represents a state-month observation. The figure shows a positive relationship between the two series, with a slope coefficient of 0.81 and an $R^2$ of 0.40, indicating that Y-14M spending captures meaningful variation in monthly state-level retail spending growth. Nonetheless, the relationship is weaker than in the annual state-level PCE comparison.\footnote{Figure~\ref{fig:monthly_state_level_consumption_growth} in the appendix further shows selected state-level time series.} 

Unlike Y-14M credit card spending, monthly retail sales are a point-of-sale based measure of retail activity by the state in which sales occur. The Monthly State Retail Sales (MSRS) data are an experimental product by the Census Bureau published since January 2019. The estimates combine two types of information. First, when state-level sales data are not directly observed, national retail sales are allocated to states using each state's share of retail-establishment payroll. Second, where available, the estimates use state-level inputs, including third-party point-of-sale data, retailer survey responses, and modeled sales for establishments that do not report directly \citep{census_2025}. Thus, the weaker correlation relative to adjusted PCE likely reflects, at least in part, the fact that state-level Census retail sales are point-of-sale data, whereas Y-14M spending is assigned to cardholders' state of residence.

\subsection{Private-Sector High-Frequency Spending Measures}\label{sec:benchmarking_private_sector}

We next compare Y-14M credit card spending to several widely used private-sector high-frequency spending measures. These measures are mostly based on card transaction data and thus conceptually similar to the Y-14M data.

\begin{figure}[!t]
    \caption{Monthly National Consumption Growth: Private-Sector Benchmarks} 
    \footnotesize{This figure plots monthly year-over-year growth rates in average Y-14M credit card spending (solid blue line) and private-sector high-frequency indicators (HFIs) based on data from Fiserv (dashed red line), Numerator (dashed green line), and Verisk Analytics (dashed yellow line). The plotted sample varies across benchmark series based on data availability.}
    \begin{center}
        \centerline{\includegraphics[width=1\textwidth]{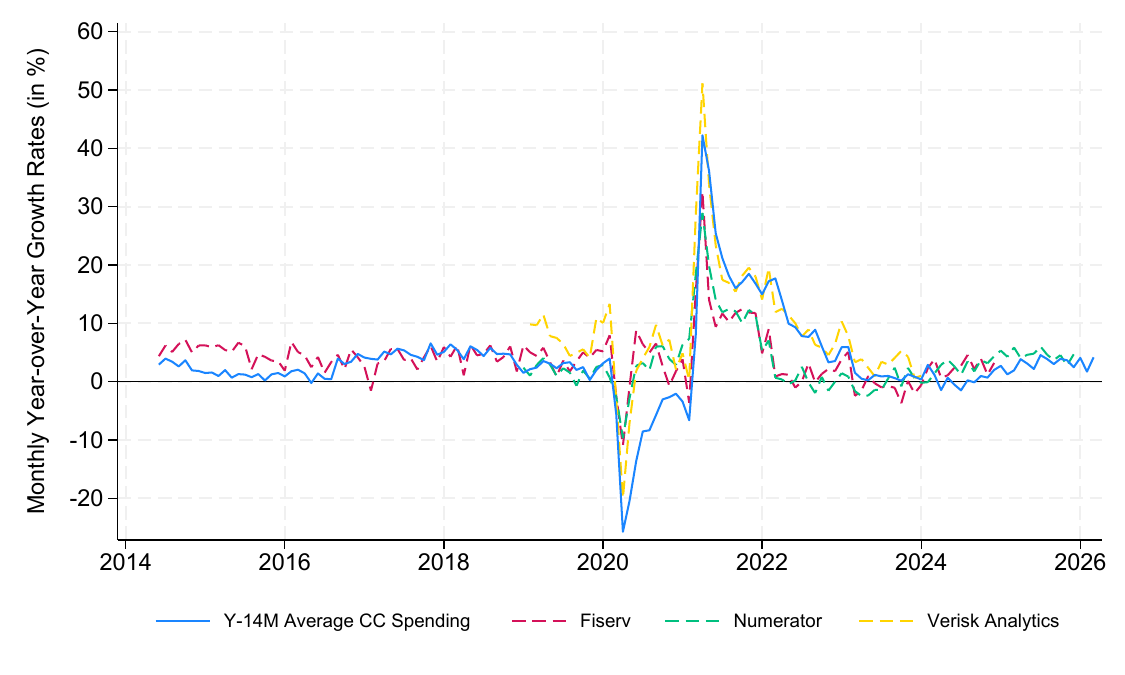}}
    \end{center}  	
    \label{fig:monthly_yoy_consumption_growth_hfi_measures}
\end{figure} 

Figure~\ref{fig:monthly_yoy_consumption_growth_hfi_measures} plots monthly year-over-year growth in Y-14M credit card spending alongside high-frequency spending measures from Fiserv, Numerator, and Verisk Analytics.\footnote{Figure~\ref{fig:monthly_yoy_consumption_growth_affinity} in the appendix reports the analogous comparison for Affinity Solutions, for which the publicly available data are reported as a seasonally adjusted change relative to January 2020 rather than as a monthly year-over-year growth rate.} Despite differences in coverage, transaction concept, and sample period, the private-sector series exhibit broadly similar dynamics to the Y-14M measure. In particular, all series capture the sharp decline in spending at the onset of the COVID-19 pandemic and the subsequent rebound in 2021. The series differ somewhat in levels and volatility, but the overall pattern suggests that the Y-14M data reflect broad high-frequency spending patterns also visible in other private-sector transaction datasets.\footnote{Table~\ref{tab:monthly_yoy_consumption_growth_hfi} in the appendix further evaluates the relationship between Y-14M credit card spending and private-sector high-frequency spending measures in a time-series regression framework. The results show a strong positive association across all four benchmarks.}

Table~\ref{tab:monthly_yoy_consumption_growth_hfi_y14m} evaluates the relationship between adjusted PCE growth and private-sector high-frequency spending measures and compares this relationship to the benchmarking performance of Y-14M credit card spending over the corresponding sample period. The table shows that credit card spending growth based on the Y-14M data explains a substantially larger share of the variation in adjusted PCE growth compared to various private-sector high-frequency spending measures. The adjusted $R^2$ rises from 0.41 to 0.92 in the Fiserv comparison, from 0.56 to 0.94 in the Numerator comparison, and from 0.84 to 0.95 in the Verisk comparison. These results indicate that, over the corresponding sample periods, Y-14M spending growth is at least as informative about national adjusted PCE growth as the private-sector high-frequency spending measures considered here.

\begin{table}[!t]
    \caption{Benchmarking Monthly National Consumption: Private-Sector HFIs}
    \footnotesize{\flushleft This table reports monthly national time-series regressions of adjusted PCE growth on private-sector high-frequency indicators (HFIs) and Y-14M credit card spending. All variables are measured as monthly year-over-year growth rates. Columns (1), (3), and (5) regress adjusted PCE growth on spending growth from Fiserv, Numerator, and Verisk Analytics, respectively. Columns (2), (4), and (6) regress adjusted PCE growth on average Y-14M credit card spending growth over the same sample periods as the corresponding HFI regressions. Standard errors are reported in parentheses. $^{***}$, $^{**}$, and $^{*}$ denote statistical significance at the 1, 5, and 10 percent levels, respectively.}
    \begin{center}
        \begin{tabular}{lcccccc}
            \toprule                & \multicolumn{6}{c}{$\Delta$ Adjusted PCE} \\ \cmidrule(lr){2-7} 
                                    & \multicolumn{2}{c}{Fiserv} 
                                    & \multicolumn{2}{c}{Numerator} 
                                    & \multicolumn{2}{c}{Verisk} \\
            \cmidrule(lr){2-3} \cmidrule(lr){4-5} \cmidrule(lr){6-7}
                                    & (1)               & (2)               & (3)               & (4)               & (5)               & (6)               \\  \midrule
            $\Delta$ HFI Spending   & 0.98$^{***}$      &                   & 1.22$^{***}$      &                   & 0.96$^{***}$      &                   \\
                                    & (0.10)            &                   & (0.12)            &                   & (0.06)            &                   \\  \addlinespace
            $\Delta$ CC Spending    &                   & 0.88$^{***}$      &                   & 0.88$^{***}$      &                   & 0.88$^{***}$      \\
                                    &                   & (0.02)            &                   & (0.02)            &                   & (0.03)            \\ \addlinespace
            Adjusted $R^2$          & 0.41              & 0.92              & 0.56              & 0.94              & 0.84              & 0.95              \\
            Observations            & 128               & 128               & 84                & 84                & 60                & 60                \\
            Sample                  & \multicolumn{2}{c}{2014m6--2025m1}   & \multicolumn{2}{c}{2019m1--2025m9}    & \multicolumn{2}{c}{2019m2--2024m1}    \\
            \bottomrule
        \end{tabular}
    \end{center}
    \label{tab:monthly_yoy_consumption_growth_hfi_y14m}
\end{table}

\section{Beyond Benchmarking}\label{sec:beyond_benchmarking_a_new_dataset}

Official consumption measures are available either at high frequency but with limited geographic detail or at finer geographic disaggregation but lower frequency. In this section, we move beyond these limitations by providing monthly spending measures that can be constructed at fine geographic disaggregation and across detailed cardholder characteristics. We illustrate two dimensions of this granularity. First, we introduce and validate a monthly county-level consumption dataset that allows us to track local consumption dynamics across more than 3,000 counties. Second, we show that the account-level structure of the data allows us to study high-frequency spending dynamics across cardholder characteristics such as credit scores and credit card utilization. These exercises highlight how credit card data can be used not only to approximate traditional consumption measures, but also to facilitate research designs that are infeasible with official consumption statistics.

\subsection{Validating the County-Level Data}\label{sec:validating_county_level_data}

To assess whether county-level credit card spending captures the geographic distribution of spending levels, we first benchmark it against the U.S. Economic Census, an independent source available in 2017 and 2022. Among other things, the Economic Census is used by BEA to benchmark retail sales at five-year intervals, and it provides industry-level measures of revenues at retail establishments in each county in the U.S. For each Census year, we compute each county's share of aggregate Y-14M credit card spending and its share of aggregate Economic Census spending in retail categories.\footnote{Because Economic Census collects data on revenues based on the location of the establishment, we focus on retail goods/service NAICS categories that are more likely to be purchased locally.  This includes 442 (furniture), 443 (electronics/appliances), 445 (food/beverage), 446 (health/personal care), 448 (clothing), 451 (sporting goods), 452 (general merchandise), 453 (miscellaneous), 722 (food services/restaurants), 811 (repair/maintenance), 812 (personal/laundry), along with equivalent 2020 codes. Figure~\ref{fig:y14m_versus_econ_census_spending_level_per_capita} in the appendix reports the analogous comparison in per-capita levels, which displays greater dispersion owing to cross-county variation in cards per resident but confirms the same positive association.}

\begin{figure}[!t]
    \caption{County-Level Spending Shares: Total Credit Card Spending and Economic Census} 
    \footnotesize{This figure plots county shares of aggregate Economic Census spending against county shares of aggregate Y-14M credit card spending for the years 2017 and 2022. Each dot represents a county-year observation. The solid red line indicates the fitted regression line, and the dashed black line indicates the 45-degree line.}
    \begin{center}
        \includegraphics[width=1\textwidth]{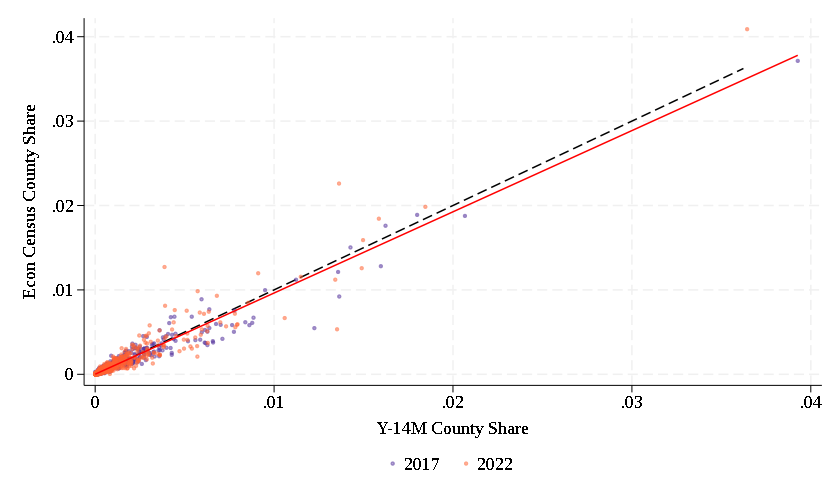}
    \end{center}
    \label{fig:y14m_versus_econ_census_spending_share}  	
\end{figure}

Figure~\ref{fig:y14m_versus_econ_census_spending_share} shows that this geographic allocation is very similar in the two datasets. Counties that account for a larger share of aggregate Economic Census spending also account for a larger share of aggregate Y-14M credit card spending, and this relationship is strongly positive in both benchmark years. The fit is especially tight for the bulk of counties, while a small number of very large counties generate some dispersion in the upper tail. A univariate regression yields a slope coefficient of 1.02 and an $R^2$ of 0.91.   

The close alignment with Economic Census spending shares indicates that Y-14M credit card spending captures not only aggregate time-series movements, but also meaningful cross-sectional variation in local spending activity. This provides support for using the county-level series to study high-frequency local consumption dynamics.

\subsection{Monthly County-Level Consumption Dynamics}\label{sec:monthly_county_level_consumption_dynamics}

This section showcases our novel monthly county-level consumption dataset, which allows us to track local consumption dynamics at high frequency and fine geographic detail. We present selected time-series evidence and cross-county patterns, and show that county-level consumption growth is closely associated with local economic developments.

\begin{figure}[!t]	
    \caption{Monthly County-Level Consumption Growth}
    \par
    \footnotesize{This figure plots monthly year-over-year growth rates in average credit card spending for selected counties from June 2014 to March 2026. Panel (a) compares New York County, NY, and Harris County, TX. Panel (b) compares the District of Columbia and Arlington County, VA.}
    \label{fig:monthly_county_level_credit_card_spending_growth}
    \begin{center}
        \subfigure[New York County, NY vs Harris County, TX]{\includegraphics[width=.495\textwidth]{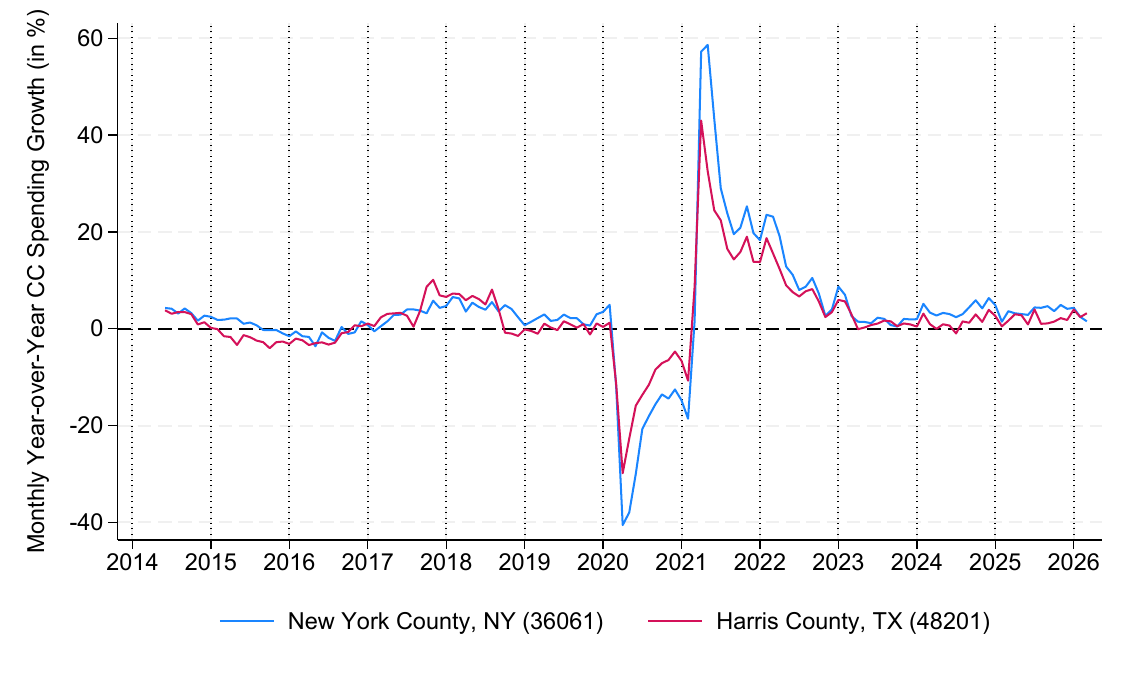}}\label{fig:Fig7a}
        \subfigure[Washington, DC vs Arlington County, VA]{\includegraphics[width=.495\textwidth]{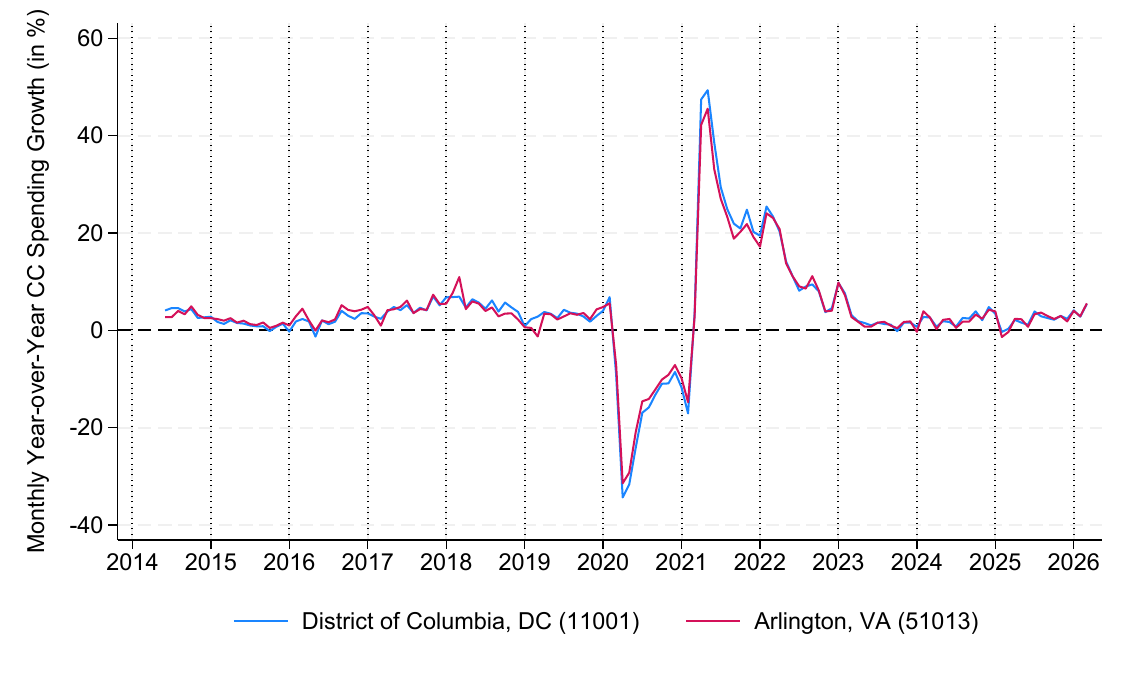}}\label{fig:Fig7b}
    \end{center}
\end{figure}

Figure~\ref{fig:monthly_county_level_credit_card_spending_growth} plots monthly growth rates of credit card spending for four selected counties. Panel (a) compares spending growth in New York County, NY, and Harris County, TX. New York County experienced a markedly sharper decline in consumption at the onset of the COVID-19 pandemic, consistent with the severe early impact of the pandemic in New York City and with evidence that local pandemic intensity strongly influenced consumer spending during this period \citep{ChettyFriedmanHendrenStepner2024}. Panel (b) compares spending growth in Washington, DC, and Arlington County, VA, two neighboring localities within the Washington metropolitan area. Reflecting their shared economic environment, the two counties exhibit nearly identical spending growth rates throughout the sample period, including during the pandemic contraction and subsequent rebound.  

\begin{figure}[!t]
    \caption{County-Level Consumption Growth and Pandemic Incidence} 
    \footnotesize{This figure maps county-level consumption growth and local pandemic incidence in April 2020, the first full month of the COVID-19 pandemic. Panel (a) shows monthly year-over-year growth in average credit card spending and darker colors indicate larger spending declines. Panel (b) shows new COVID-19 cases per 100,000 residents and darker colors indicate higher local pandemic incidence.}
    \label{fig:spending_vs_covid_april_2020}
    \begin{center}
        \subfigure[CC Spending Growth]{\includegraphics[width=.495\textwidth]{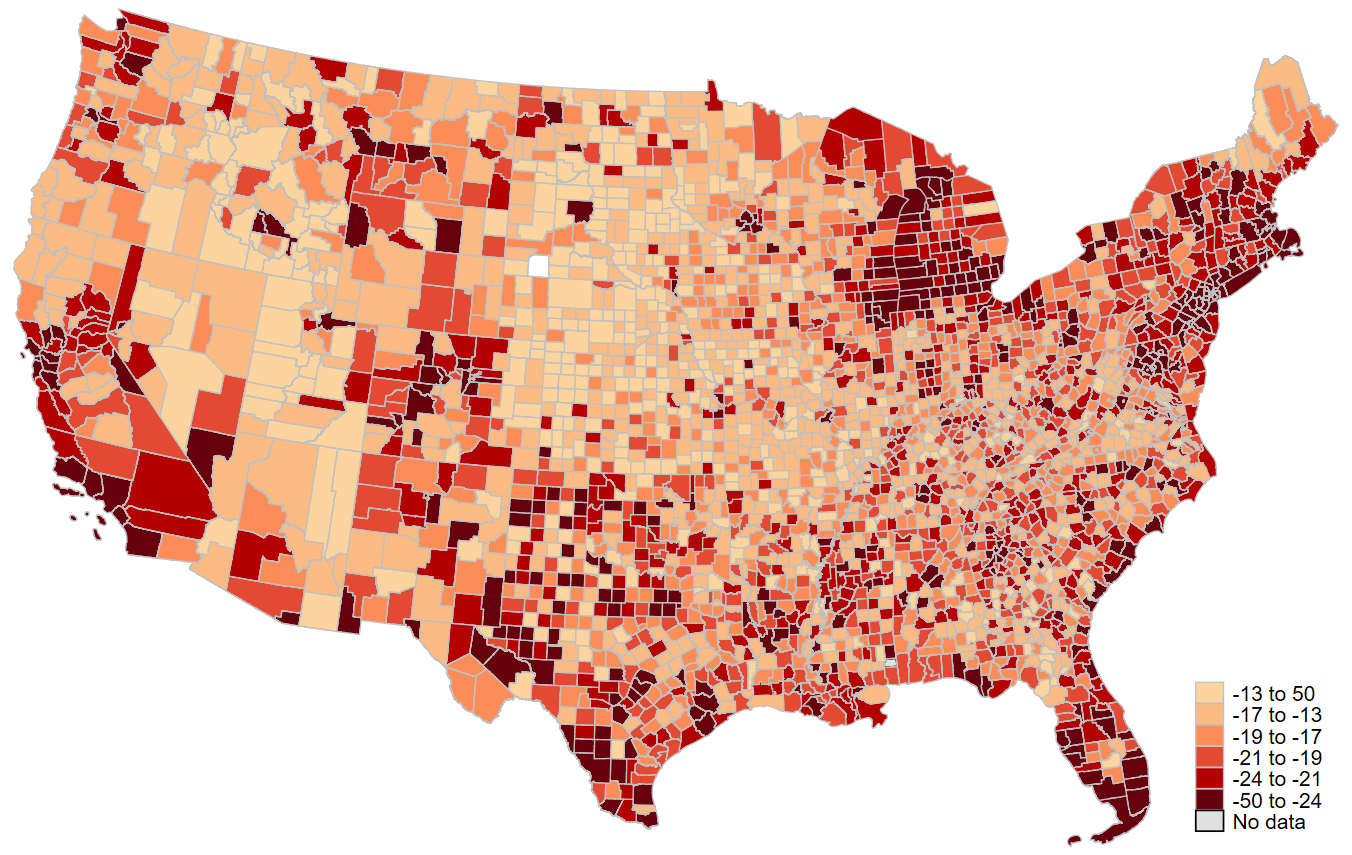}}\label{fig:Fig6a}
        \subfigure[New COVID Cases per 100k]{\includegraphics[width=.495\textwidth]{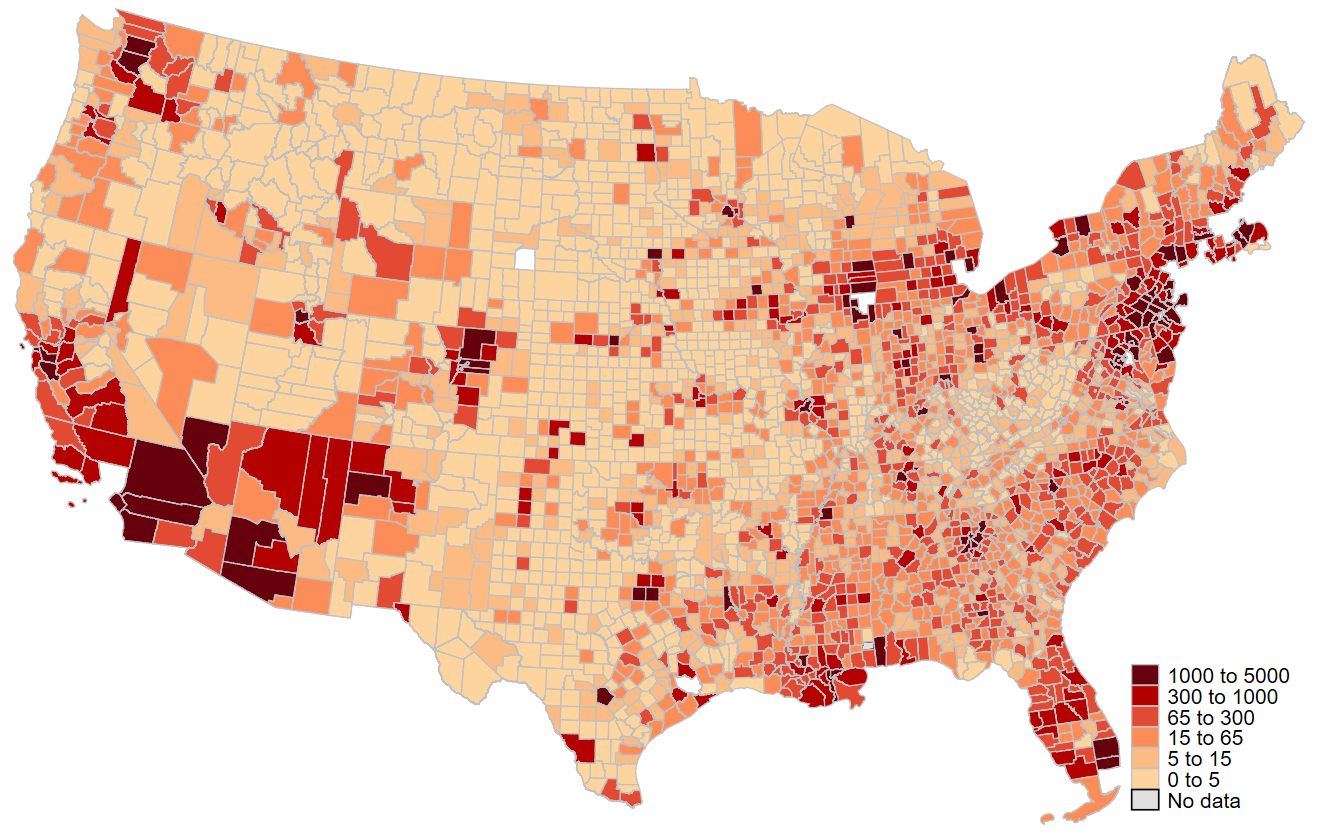}}\label{fig:Fig6b}
    \end{center}
\end{figure}

To illustrate the geographic breadth of the dataset, Figure~\ref{fig:spending_vs_covid_april_2020} presents county-level maps of credit card spending growth (Panel a) and local pandemic incidence (Panel b) as of April 2020, the first full month of the COVID-19 pandemic. The two maps exhibit strikingly similar spatial patterns. Counties with high COVID-19 incidence, particularly in the Northeast, the Bay Area, and South Florida, experienced the sharpest declines in credit card spending. By contrast, counties in the Mountain West and the rural Midwest, where case counts remained low in the early stages of the pandemic, exhibit substantially milder spending declines. To quantify this relationship, we regress county-level credit card spending growth on log new COVID-19 cases per 100,000 residents in April 2020. The estimated coefficient is $-1.33$ ($t = -31.77$), with an $R^2$ of 0.25, confirming that counties with higher pandemic intensity experienced significantly larger declines in consumption. This close spatial correspondence between pandemic intensity and consumption is consistent with evidence that local COVID-19 severity strongly influenced consumer spending during this period \citep{HorvathKayWix2023,ChettyFriedmanHendrenStepner2024}. 
 
\begin{figure}[!t]
    \caption{County-Level Consumption Growth and Local Economic Conditions} 
    \footnotesize{This figure presents binned scatterplots relating county-level credit card spending growth to local economic conditions. Panel (a) plots annual growth rates in average credit card spending against annual income growth. Panel (b) plots monthly year-over-year growth rates in average credit card spending against monthly year-over-year changes in the unemployment rate. Each dot represents a bin of county-level observations.}
    \begin{center}
        \subfigure[Income]{\includegraphics[width=.495\textwidth]{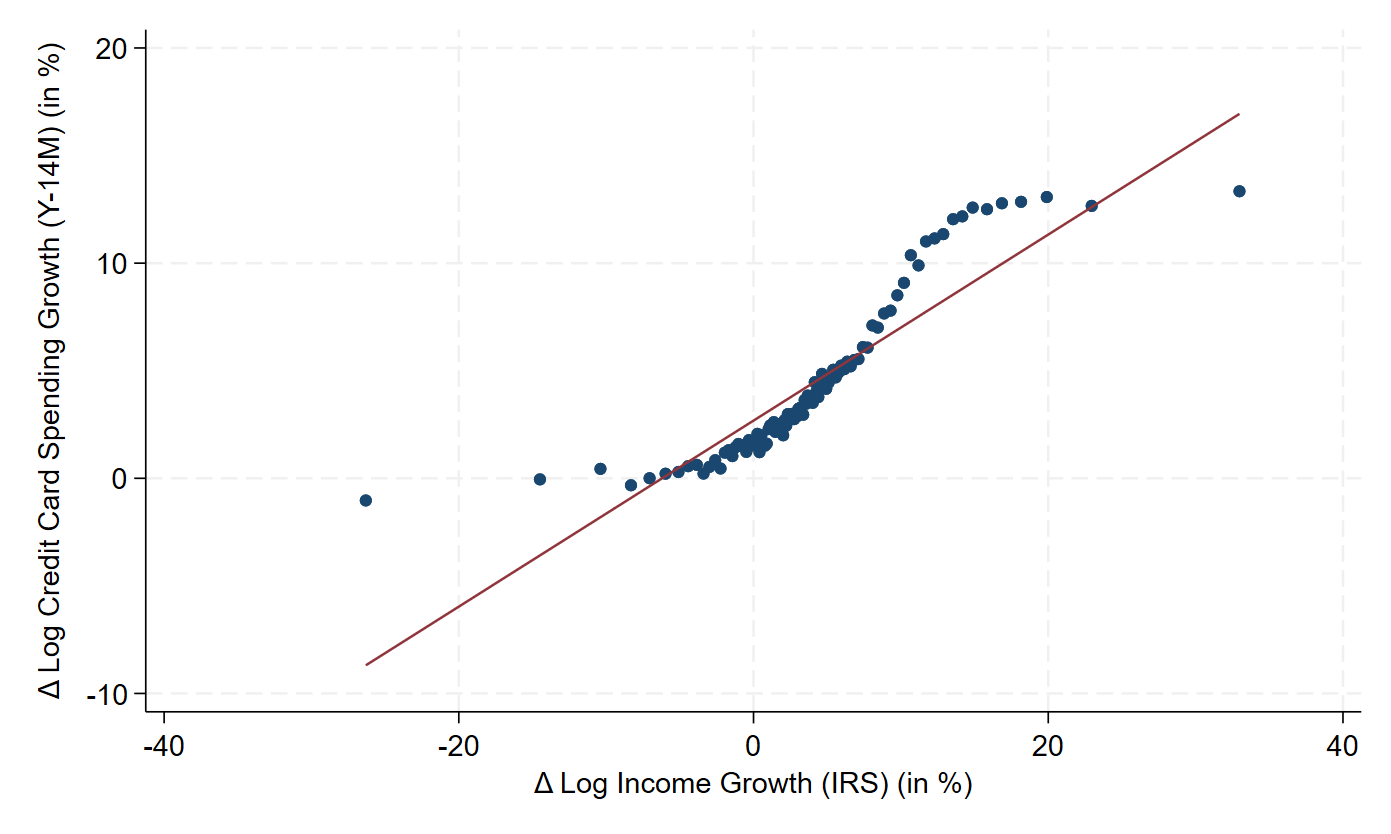}}\label{fig:Fig6a}
        \subfigure[Unemployment]{\includegraphics[width=.495\textwidth]{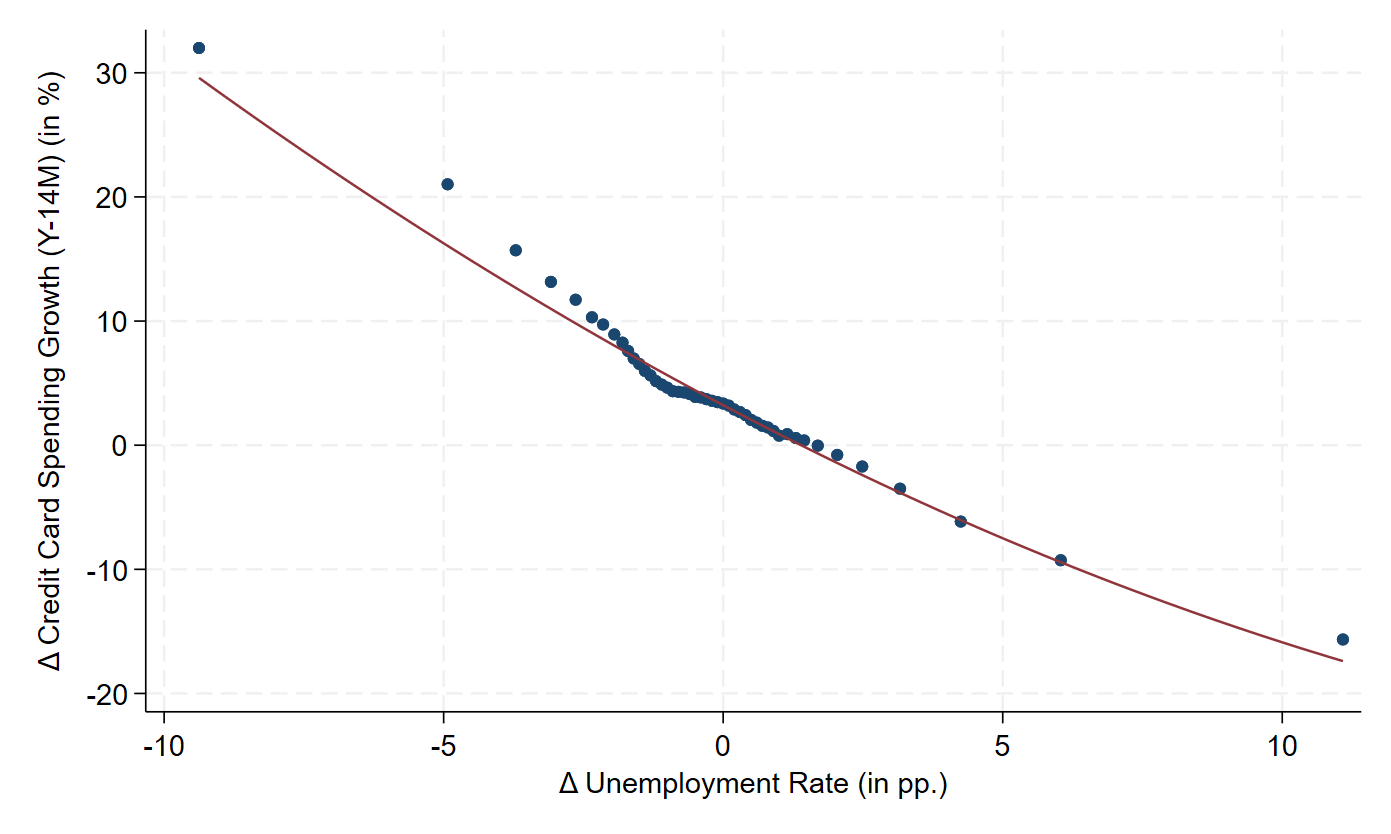}}\label{fig:Fig6b}
    \end{center}
    \label{fig:county_cc_spending_growth_vs_inc_unemp}
\end{figure}

We next show that our local consumption measure not only captures intuitive spending patterns in selected counties and periods, but exhibits the expected correlations with local economic conditions. Figure~\ref{fig:county_cc_spending_growth_vs_inc_unemp} presents a binned scatterplot that shows the correlation between annual credit card spending growth and annual income growth (Panel a) and monthly credit card spending growth and monthly unemployment growth (Panel b) at the county level. Annual consumption growth, as measured by Y-14M credit card spending, is positively associated with annual income growth, yielding a regression coefficient of $0.43$ ($t = 53.24$) with an $R^2$ of 0.32. Conversely, monthly consumption growth is negatively associated with monthly unemployment growth, yielding a regression coefficient of $-2.29$ ($t = -407.30$) with an $R^2$ of 0.36. These relationships are not intended as causal estimates, but they show that the Y-14M county series moves in the expected direction with local fundamentals.    

Together, the county time series, COVID maps, and correlations with local economic conditions show that Y-14M credit card spending captures meaningful local consumption dynamics in addition to aggregate national movements. Covering more than 3,000 counties at a monthly frequency since 2014, this dataset opens the door to research designs that require both high-frequency and granular geographic variation in consumption. In Section~\ref{sec:research_application}, we present one such application and study the heterogeneous transmission of monetary policy to consumption across the county-level income distribution.

\subsection{Cardholder-Level Spending Dynamics}\label{sec:cardholder_level_consumption_dynamics}

\begin{figure}[!t]
    \caption{Monthly Credit Card Spending Growth by Cardholder Characteristics} 
    \footnotesize{This figure plots monthly year-over-year growth rates in average credit card spending by cardholder characteristics from June 2014 to March 2026. Panel (a) compares cardholders with FICO scores above and below 720. Panel (b) compares cardholders with utilization rates above and below 30 percent.}
    \begin{center}
        \subfigure[By FICO Scores]{\includegraphics[width=.495\textwidth]{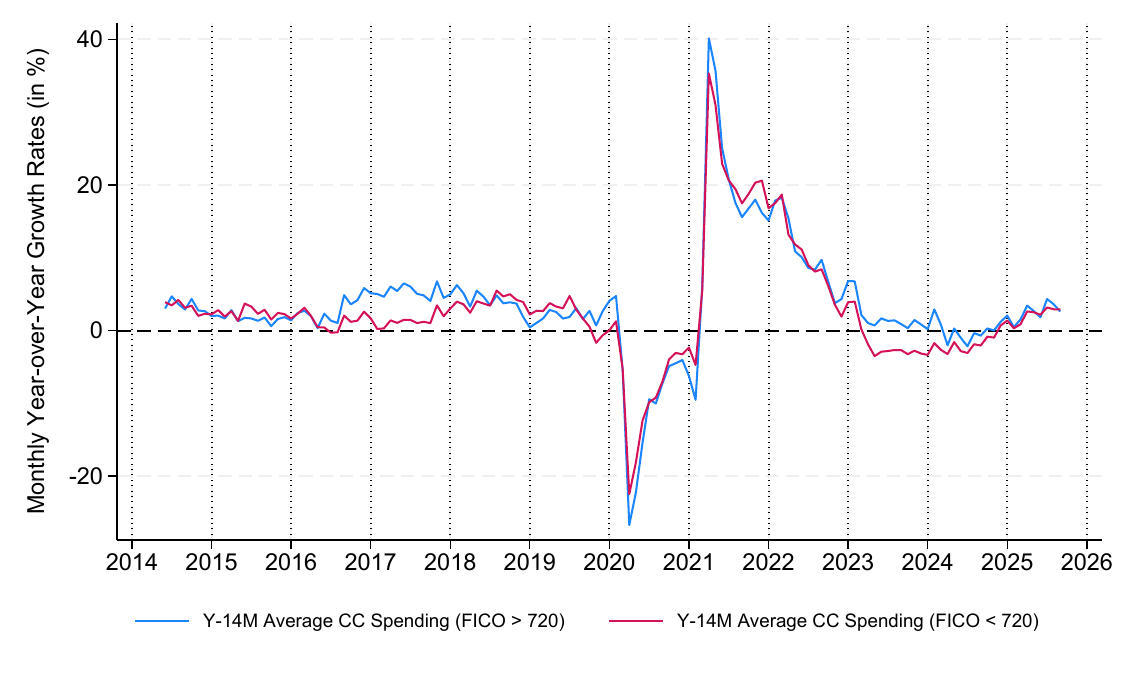}}\label{fig:Fig6a}
        \subfigure[By Utilization Rates]{\includegraphics[width=.495\textwidth]{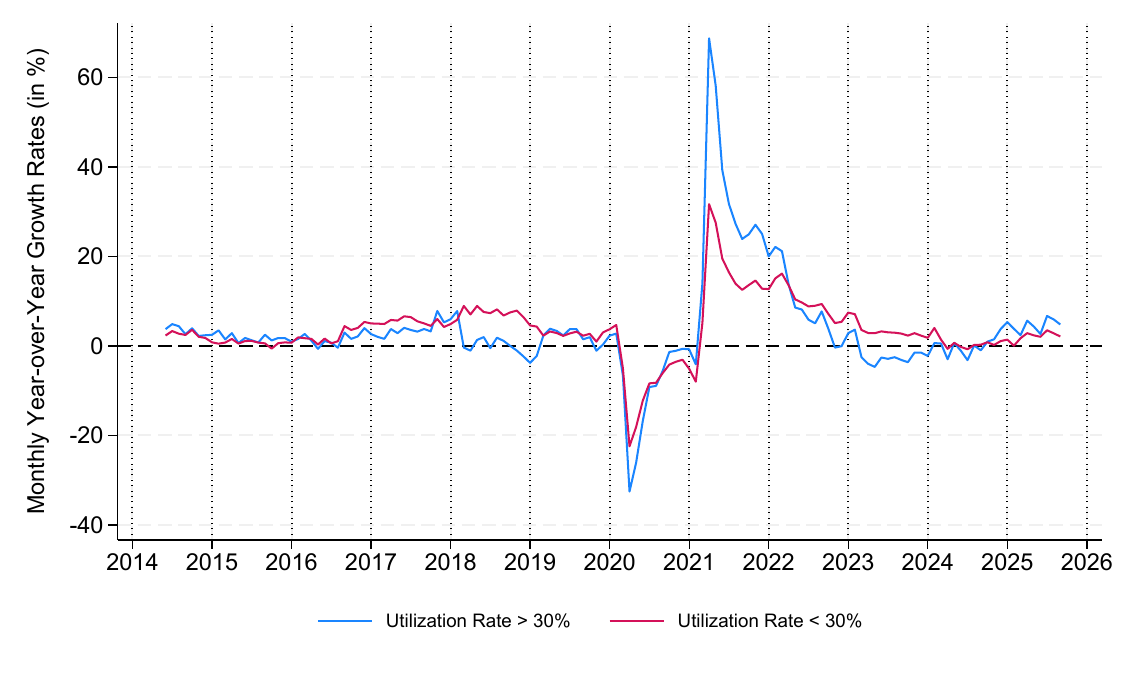}}\label{fig:Fig6b}
    \end{center}
    \label{fig:cc_spending_growth_by_fico_util}
\end{figure}

The preceding subsection highlights the geographic granularity of the Y-14M data. A second advantage of the data is that the underlying observations are at the account level, which allows us to study spending dynamics not only across places, but also across cardholder characteristics. To illustrate this dimension of the data, Figure~\ref{fig:cc_spending_growth_by_fico_util} plots monthly year-over-year growth in average credit card spending separately by FICO score and credit card utilization.  

Panel (a) compares the consumption dynamics of cardholders with FICO scores above and below 720. While spending growth of the two groups follows broadly similar dynamics over most of the sample, including the contraction at the onset of the COVID-19 pandemic and the subsequent rebound, low-FICO cardholders exhibited lower spending growth in 2017 and 2023. Analogously, Panel (b) compares the consumption dynamics of cardholders with utilization rates above and below 30 percent. Similarly, cardholders with high utilization rates exhibited lower spending growth in 2018 and 2023 and, additionally, a steeper contraction and stronger rebound in spending during and after the pandemic. Notably, these periods of weaker spending growth among low-FICO and high-utilization cardholders coincide with monetary policy tightening episodes in 2017--2018 and 2023. Since credit card interest rates adjust quickly to changes in short-term interest rates, these patterns are consistent with financially constrained cardholders being more exposed to rising borrowing costs. While this evidence is descriptive, it illustrates how the account-level structure of the Y-14M data can be used to study heterogeneity in spending dynamics across borrower characteristics.

\section{Research Application: The Consumption Response to Monetary Policy Shocks}\label{sec:research_application} 

In this section, we use the local projections methodology of \cite{jorda2005} to study the transmission of monetary policy to consumption. The goal of the section is to provide a proof of concept for the research value of the county-month consumption data. We focus on monetary policy because theory predicts heterogeneous consumption responses across the income and liquidity distribution, while traditional consumption data lack the frequency and geographic granularity needed to study these responses at a local level. We first estimate the impulse responses of credit card spending and traditional consumption measures to contractionary monetary policy shocks at the national and state level, providing a further validation that credit card spending captures economically meaningful consumption dynamics. We then exploit the high frequency and geographic granularity of the Y-14M data to estimate the heterogeneous effects of monetary policy on consumption across the income distribution at the county level.

\subsection{Empirical Specification}\label{sec:research_application_empirical_specification}

\paragraph{Consumption Measures:} We examine four consumption measures: (i) total Personal Consumption Expenditures (PCE), the standard aggregate measure; (ii) adjusted PCE, as discussed in Section~\ref{sec:data_benchmarking}; (iii) retail sales from the Census Bureau; and (iv) aggregated credit card spending from the Y-14M data. Comparing responses across these measures allows us to assess both the validity of credit card spending as a consumption proxy and differences in the estimated response across consumption measures. All consumption measures are deflated using the Personal Consumption Expenditures Price Index (PCEPI, base year 2017) to obtain real values:
\begin{equation}
    y_{i,t} = \ln\left(\frac{C_{i,t}}{\text{PCEPI}_t / 100}\right)
    \label{eq:deflation}
\end{equation}
where $C_{i,t}$ denotes nominal consumption in geographical unit $i$ in period $t$. We use PCEPI rather than CPI because it is purpose-built for deflating personal consumption expenditures, employs chain-weighting to better account for substitution across consumption categories, and exhibits less volatility than CPI.

\paragraph{Monetary Policy Shock:} We use high-frequency identification (HFI) to measure monetary policy shocks, following the approach of \cite{jarocinski2020}. Specifically, we use the median monetary policy surprise across Fed Funds futures contracts in a narrow window around FOMC announcements. To facilitate interpretation, we normalize the shock so that a one-unit increase corresponds to a 25 basis point rise in the Federal Funds Rate. This normalization is achieved by regressing changes in the Federal Funds Rate on the raw shock and rescaling accordingly.

\paragraph{Specification:} We estimate impulse responses using the following local projection specification:
\begin{equation}
    y_{i,t+h} - y_{i,t} = \beta_h \cdot \text{shock}_t + \Gamma'(L) X_{i,t} + \alpha_i + \varepsilon_{i,t+h}
    \label{eq:lp}
\end{equation}
where $y_{i,t+h} - y_{i,t}$ denotes the cumulative log change in real consumption from period $t$ to horizon $h$, $\text{shock}_t$ is the monetary policy shock, $\Gamma'(L) X_{i,t}$ represents a polynomial in the lag operator applied to a vector of controls (including lags of the shock, the dependent variable, and macroeconomic indicators), and $\alpha_i$ captures fixed effects. The coefficient of interest, $\beta_h$, traces out the impulse response function at each horizon $h$. We implement equation~\eqref{eq:lp} at three levels of aggregation. At the monthly national level, we estimate:
\begin{equation}
    y_{t+h} - y_t = \beta_h \cdot \text{shock}_t + \sum_{j=0}^{p} \gamma_j \cdot \text{shock}_{t-j} + \sum_{j=1}^{p} \delta_j \cdot y_{t-j} + \phi \cdot \mathbf{1}_{\text{COVID}} + \alpha_m + \varepsilon_{t+h}
    \label{eq:lp_monthly}
\end{equation}
with $p=2$ lags, month-of-year fixed effects $\alpha_m$ to control for seasonality, and standard errors clustered by date. At the state-year level, we estimate:
\begin{equation}
    y_{s,t+h} - y_{s,t} = \beta_h \cdot \text{shock}_t + \sum_{j=0}^{p} \gamma_j \cdot \text{shock}_{t-j} + \sum_{j=1}^{p} \delta_j \cdot y_{s,t-j} + \theta' W_{s,t} + \alpha_s + \varepsilon_{s,t+h}
    \label{eq:lp_stateyear}
\end{equation}
with $p=1$ lag, state fixed effects $\alpha_s$, controls $W_{s,t}$ for unemployment and inflation, and standard errors two-way clustered by state and year. The state-level specification does not include year fixed effects, as these would absorb all variation in the national monetary policy shock. Identification therefore relies on time-series variation in national monetary policy, with the panel structure providing statistical power through cross-sectional variation in consumption levels.\footnote{An alternative specification with year fixed effects would require interacting the shock with state-level characteristics to study heterogeneous transmission, which we leave for future work.} Finally, at the county-month level, we estimate: 
\begin{equation}
    y_{c,t+h} - y_{c,t} = \beta_h \cdot \text{shock}_t + \sum_{j=0}^{p} \gamma_j \cdot \text{shock}_{t-j} + \sum_{j=1}^{p} \delta_j \cdot y_{c,t-j} + \theta' W_{c,t} + \alpha_c + \alpha_m + \varepsilon_{c,t+h}
    \label{eq:lp_county}
\end{equation}
with county fixed effects $\alpha_c$, month-of-year fixed effects $\alpha_m$, and controls $W_{c,t}$ that include national unemployment, inflation, county-level log house prices, and county-level log income. Because the monetary policy shock varies only at the national level, identification exploits time-series variation in $\text{shock}_t$ across the full panel. Standard errors are two-way clustered by county and date. Time fixed effects are excluded for the same reason as in the state-year specification: they would absorb all identifying variation in the national shock.

Our sample includes the COVID-19 pandemic, which induced extraordinary volatility in consumption patterns. Following \cite{schorfheide2024}, we set observations from March 2020 through October 2020 to missing, thereby excluding this period from estimation while preserving the time-series structure for computing leads and lags. For the monthly specifications, we additionally include a COVID indicator variable spanning March 2020 through June 2022 to absorb residual level shifts associated with fiscal stimulus and the subsequent normalization period.

\subsection{Benchmarking: Monthly National and Annual State-Level}\label{sec:research_application_benchmarking}

We begin by benchmarking the impulse response of credit card spending against those of traditional consumption measures at the national and state level, assessing whether credit card spending tracks not only the levels and growth rates documented in Section~\ref{sec:benchmarking}, but also the dynamic response of consumption to economic shocks.

\begin{figure}[!t]	
    \caption{Monthly National Response to Monetary Policy Shocks}
    \par
    \footnotesize{This figure displays impulse responses of real consumption measures to a 25-basis-point contractionary monetary policy shock, estimated by local projections using monthly national data from 2014 to 2025. Panels (a)–(d) show real PCE, real adjusted PCE, real retail sales, and real credit card spending, respectively. The dependent variable is the cumulative log change from the shock month. All specifications include month-of-year fixed effects, two lags of the shock and dependent variable, and a COVID indicator for March 2020–June 2022, with standard errors clustered by date. Observations from March–October 2020 are set to missing following \cite{schorfheide2024}. The shaded regions represent 68 percent (darker) and 90 percent (lighter) confidence intervals.}
    \label{fig:lp_monthly}
    \begin{center}
        \subfigure[PCE]{\includegraphics[width=.495\textwidth]{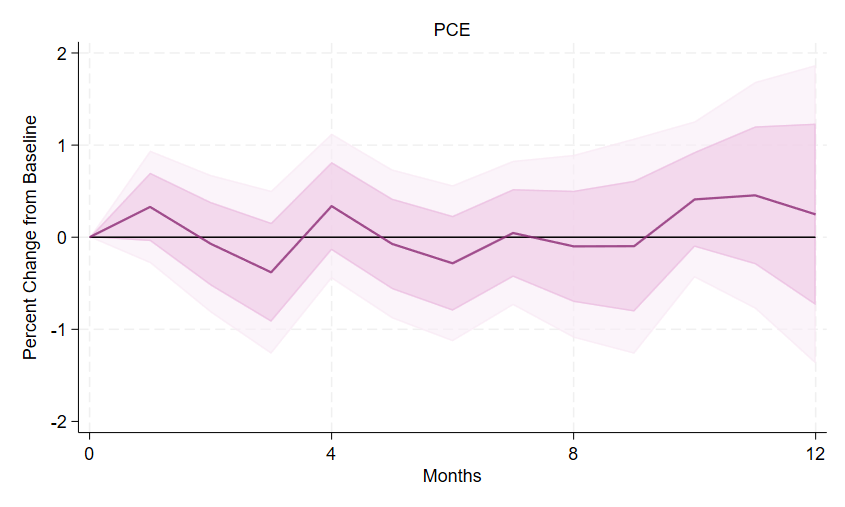}}\label{fig:lp_monthly_pce}
        \subfigure[Adjusted PCE]{\includegraphics[width=.495\textwidth]{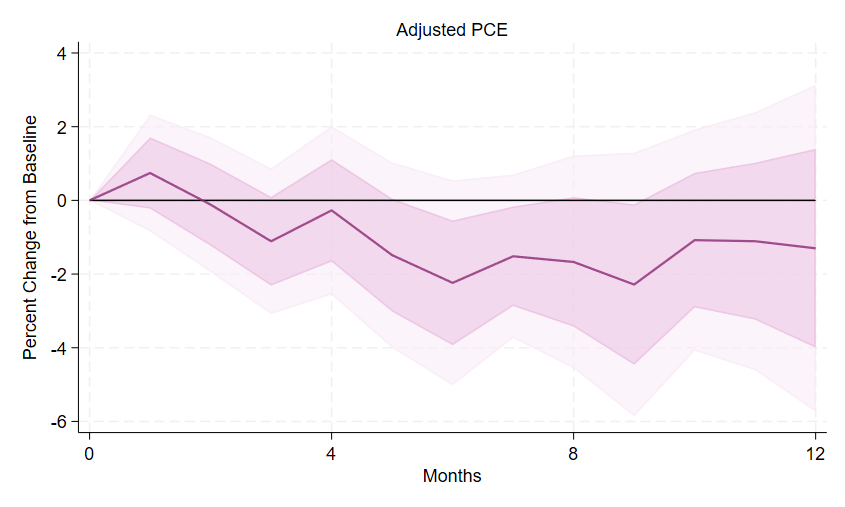}}\label{fig:lp_monthly_pcev1}
        \subfigure[Retail Sales]{\includegraphics[width=.495\textwidth]{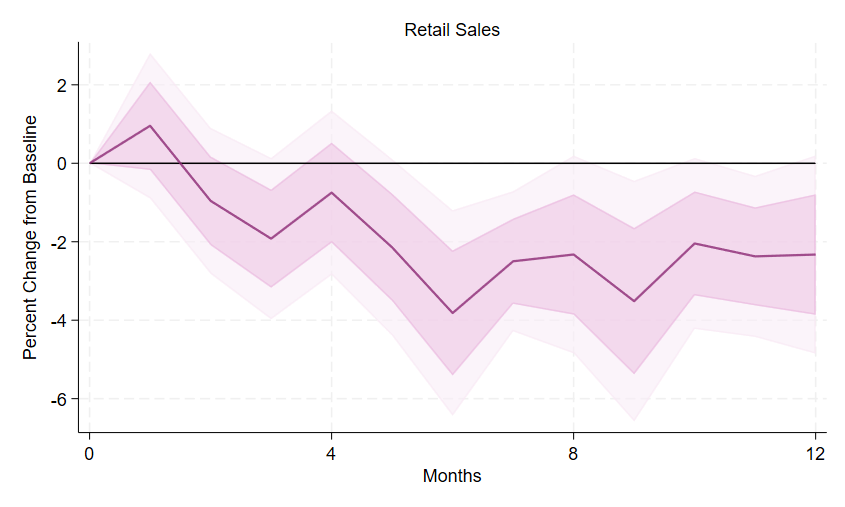}}\label{fig:lp_monthly_retail}
        \subfigure[Credit Card Spending]{\includegraphics[width=.495\textwidth]{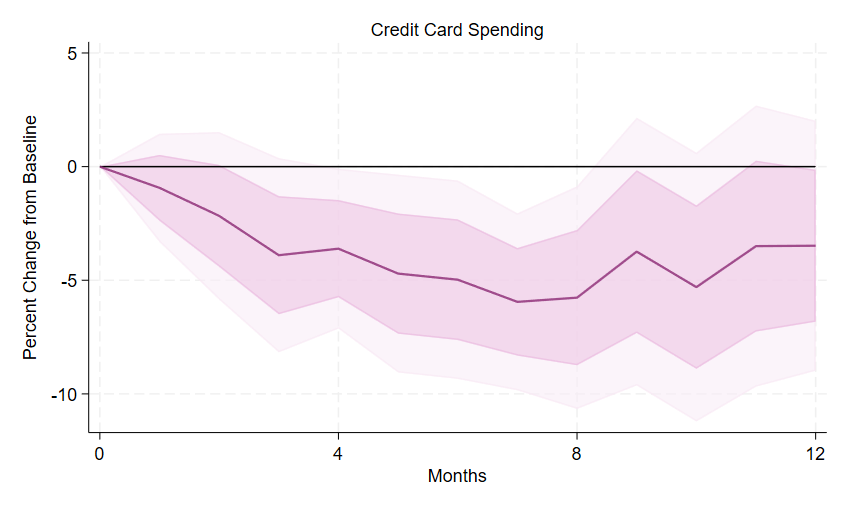}}\label{fig:lp_monthly_cc}
    \end{center}
\end{figure}

\paragraph{Monthly National Results:} Figure~\ref{fig:lp_monthly} displays the impulse responses of the four consumption measures to a 25-basis-point contractionary monetary policy shock, estimated using monthly national data from 2014 to 2025. The specification includes month fixed effects, two lags of the shock and the dependent variable, and a COVID indicator spanning March 2020 to June 2022, with standard errors clustered by date.

Three of the four consumption measures exhibit a clear contractionary response, with credit card spending showing the largest and most precisely estimated decline. Credit card spending falls gradually over the first eight months, reaching a trough of approximately -5\% to -6\%, before partially recovering toward -3\% by month 12. Retail sales display a similar pattern, declining roughly 3\% to 4\% by month 6 and stabilizing thereafter, though with somewhat greater volatility. Adjusted PCE shows a more moderate response, falling approximately 2\% by month 6 before mean-reverting toward -1\% by the end of the horizon.

In contrast, aggregate PCE exhibits a muted response that oscillates around zero throughout the 12-month horizon. This stark difference reflects the composition of total PCE. In particular, total PCE includes housing costs (primarily owners' equivalent rent) which comprise approximately 15-20\% of aggregate consumption and adjust sluggishly to monetary policy due to long-term lease contracts and the mechanical smoothing inherent in imputed rent measures. When this sticky component is excluded (adjusted PCE), the consumption response becomes visible and statistically significant.

The ordering of magnitudes is consistent with economic theory. Credit card purchases disproportionately capture discretionary, interest-sensitive expenditures such as dining, travel, and retail goods, which households can readily postpone or reduce in response to tighter financial conditions. This amplified sensitivity makes credit card spending a particularly useful measure for detecting monetary policy transmission to household consumption, even when aggregate measures appear unresponsive.

\begin{figure}[!h]	
    \caption{Annual State-Level Response to Monetary Policy Shocks}
    \par
    \footnotesize{This figure displays impulse responses of real consumption measures to a 25-basis-point contractionary monetary policy shock, estimated by local projections using annual state-level data from 2014 to 2025. Panels (a)–(c) show real PCE, real adjusted PCE, and real credit card spending, respectively. The dependent variable is the cumulative log change from the shock year. All specifications include state fixed effects, one lag of the shock and dependent variable, and controls for unemployment and inflation, with standard errors two-way clustered by state and year. The shaded regions represent 68 percent (darker) and 90 percent (lighter) confidence intervals.}
    \label{fig:lp_state_year}
    \begin{center}
        \subfigure[PCE]{\includegraphics[width=.495\textwidth]{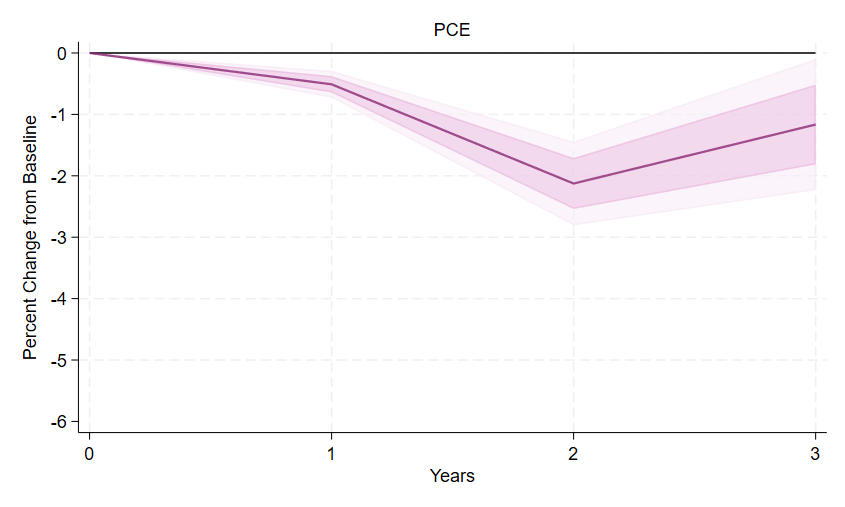}}\label{fig:lp_pce}
        \subfigure[Adjusted PCE]{\includegraphics[width=.495\textwidth]{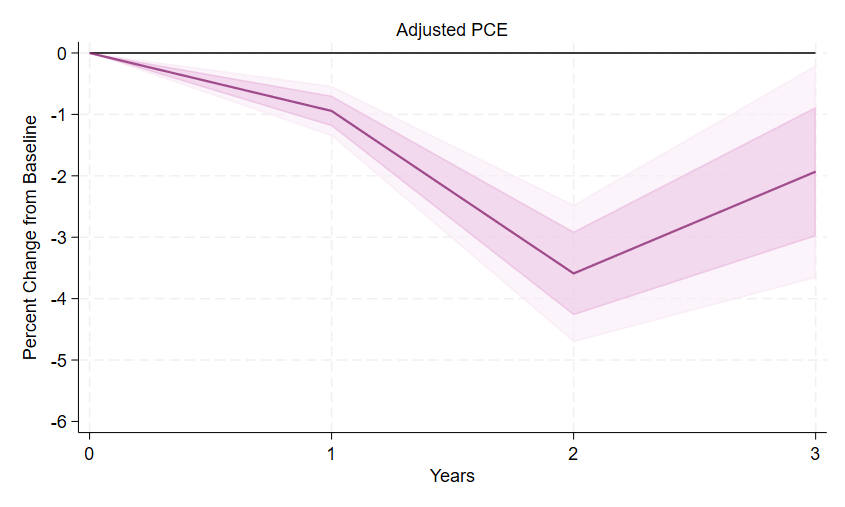}}\label{fig:lp_pce_adj}
        \subfigure[Credit Card Spending]{\includegraphics[width=.495\textwidth]{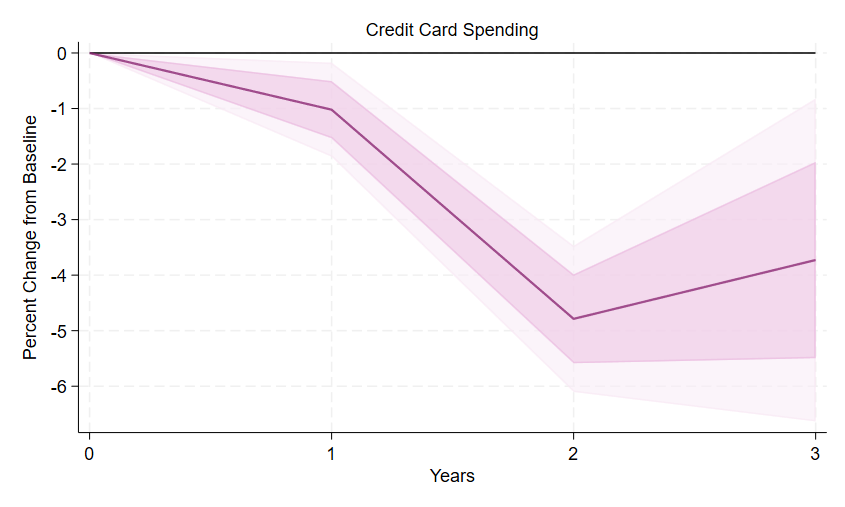}}\label{fig:lp_cc}
    \end{center}
\end{figure}

\paragraph{Annual State-Level Results:} Figure~\ref{fig:lp_state_year} displays the impulse responses of three consumption measures to a 25 basis point contractionary monetary policy shock, estimated using state-year panel data from 2014 to 2025. The specification includes state fixed effects, one lag of the shock and the dependent variable, and controls for unemployment and inflation, with standard errors two-way clustered by state and year. The monetary policy shock is aggregated from monthly high-frequency surprises to annual frequency. The identification exploits time-series variation in the national monetary policy shock, with the state panel structure providing additional statistical power through cross-sectional variation in consumption levels. Because the specification includes state but not year fixed effects, the estimated coefficients reflect the average consumption response across all states to changes in monetary policy over time, controlling for time-invariant state characteristics such as industry composition, demographics, and baseline consumption levels.

All three measures exhibit a strikingly similar response pattern: an initial decline in year 1, a pronounced trough in year 2, followed by partial recovery in year 3. This consistent dynamic across different consumption measures provides strong evidence of a common monetary transmission mechanism operating at horizons beyond the immediate impact period. This consistent dynamic further validates credit card spending as a proxy for consumption in dynamic settings, confirming that it tracks not only the levels and growth rates documented earlier, but also the impulse response to economic shocks.

Notably, unlike at the national monthly level where total PCE was essentially unresponsive, all three measures, including total PCE, exhibit a clear contractionary response at the state-year level, likely reflecting the additional statistical power provided by the panel dimension. The magnitudes, however, differ substantially. Credit card spending shows the largest sensitivity, declining approximately 1\% in year 1 and reaching a peak contraction of 5\% in year 2 before recovering to -4\% by year 3. Adjusted PCE exhibits an intermediate response, falling 1\% initially and bottoming at -3.5\% in year 2. Total PCE displays the most muted reaction, with roughly half the magnitude of adjusted PCE throughout the response path (-0.5\% in year 1, -2\% at the trough, -1\% by year 3).

The amplified response of credit card spending relative to aggregate consumption reflects that credit card purchases capture discretionary, interest-sensitive expenditures by liquidity-constrained households, making them more responsive to monetary policy through both the interest rate and credit channels. The progressively wider confidence bands at longer horizons reflect increased uncertainty about the persistence of monetary policy effects and the relatively short sample period.

\subsection{Beyond Benchmarking: Monthly County-Level}\label{sec:research_application_monthly_county}

We now turn to the county-month level, where the Y-14M data offer a unique advantage: the ability to track consumption at a level of geographic and temporal granularity unavailable in traditional data sources. We exploit this variation to estimate the heterogeneous effects of monetary policy on consumption across the income distribution at the county level. 

\begin{figure}[!t]
    \caption{Monthly County-Level Response to Monetary Policy Shocks}
    \par
    \footnotesize{This figure displays the impulse response of real county-level credit card spending to a 25-basis-point contractionary monetary policy shock, estimated by local projections using monthly county-level data from 2014 to 2025. The dependent variable is the cumulative log change in real credit card spending from the shock month. The specification includes county fixed effects, month-of-year fixed effects, two lags of the shock and dependent variable, controls for unemployment, inflation, county-level house prices, and county-level income, and a COVID indicator for March 2020–June 2022, with standard errors two-way clustered by county and date. Observations from March–October 2020 are set to missing following \cite{schorfheide2024}. The shaded regions represent 68 percent (darker) and 90 percent (lighter) confidence intervals.}
    \label{fig:lp_county}
    \begin{center}
        \subfigure[Credit Card Spending]{\includegraphics[width=.65\textwidth]{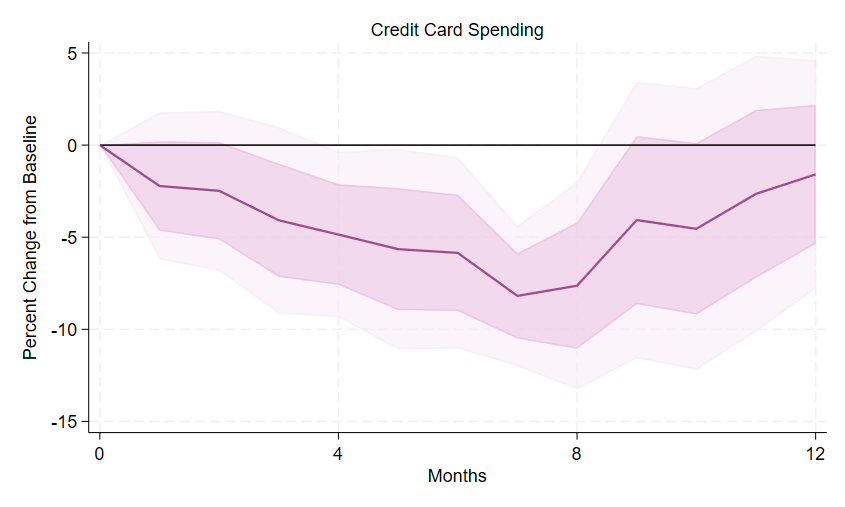}}\label{fig:lp_county_cc}
    \end{center}
\end{figure}

\paragraph{Replication of Aggregate Results:} Figure~\ref{fig:lp_county} displays the impulse response of county-level credit card spending to a 25 basis point contractionary monetary policy shock. The response closely mirrors the national estimate: spending declines monotonically over the first eight months, reaching a trough of approximately $-$7\% to $-$8\% before partially recovering by month 12. The precision of the county estimates is higher than the national counterpart, reflecting the large cross-sectional dimension of the panel.

\paragraph{Heterogeneity by County Income:} We next examine whether the transmission of monetary policy to consumption differs systematically across the income distribution. We split counties into above- and below-median groups based on their time-averaged IRS income, a classification that is highly stable over time: 91\% of counties retain the same classification whether income is measured in 2013 or averaged over the full sample period.\footnote{Figure~\ref{fig:cc_spending_by_county_income} shows raw monthly credit card spending growth for counties in the top and bottom deciles of the income distribution. The two groups exhibit broadly similar aggregate dynamics, with differences concentrated around the pandemic period. The formal local-projection estimates below isolate heterogeneity in the response to monetary policy shocks.} We estimate equation~\eqref{eq:lp_county} augmented with a full interaction between the monetary policy shock and an indicator for above-median income counties:
\begin{align}
    y_{c,t+h} - y_{c,t} &= \beta_h^{\text{low}} \cdot \text{shock}_t + \beta_h^{\Delta} \cdot \text{shock}_t \times \mathbf{1}[\text{High Income}_c]    \notag \\ 
                        &+ \mathbf{1}[\text{High Income}_c] + \Gamma'(L) X_{c,t} + \alpha_c + \alpha_m + \varepsilon_{c,t+h}
    \label{eq:lp_county_het}
\end{align}
where $\beta_h^{\text{low}}$ captures the response of low-income counties and $\beta_h^{\Delta} = \beta_h^{\text{high}} - \beta_h^{\text{low}}$ identifies the differential response, with standard errors for $\beta_h^{\Delta}$ obtained directly from the interaction term rather than from separate regressions, thereby correctly accounting for the covariance between the two group estimators.\footnote{We also examine heterogeneity by county house price levels using time-averaged Zillow HPI. However, 74\% of counties receive the same above/below median classification under income and house prices, reflecting the high correlation between the two measures at the county level. Residualizing house prices on income yields no statistically significant differential response, suggesting that the balance sheet channel operates primarily through the income dimension in our data rather than independently through housing wealth.}

\begin{figure}[!h]
    \caption{Monthly County-Level Response to Monetary Policy Shocks by Income}
    \par
    \footnotesize{This figure displays impulse responses of real county-level credit card spending to a 25-basis-point contractionary monetary policy shock by county income group, estimated from a single interacted local-projection specification using monthly county-level data from 2014 to 2025. Counties are classified as above or below the median based on time-averaged IRS income. Panel (a) shows the impulse responses for high- and low-income counties. Panel (b) shows the difference between high- and low-income counties, with standard errors obtained directly from the interaction term. The dependent variable is the cumulative log change in real credit card spending from the shock month. The specification includes county fixed effects, month-of-year fixed effects, two lags of the shock and dependent variable, controls for unemployment, inflation, county-level house prices, and county-level income, and a COVID indicator for March 2020–June 2022, with standard errors two-way clustered by county and date. Observations from March–October 2020 are set to missing following \cite{schorfheide2024}. The shaded regions represent 68 percent (darker) and 90 percent (lighter) confidence intervals.}
    \label{fig:lp_county_income}
    \begin{center}
        \includegraphics[width=\textwidth]{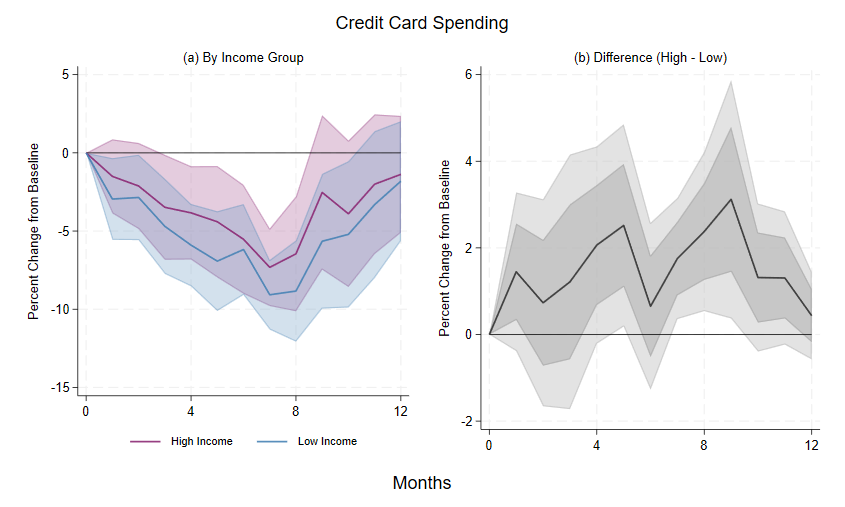}
    \end{center}
\end{figure}

Figure~\ref{fig:lp_county_income} presents the results. Panel (a) shows that both groups exhibit a significant contractionary response, but the magnitude differs substantially: spending in low-income counties declines by approximately 9\% at the trough, compared to $-$7\% for high-income counties. Panel (b) shows the difference $\hat{\beta}_h^{\Delta}$, which is positive throughout---high-income counties are less responsive---and statistically distinguishable from zero at the 68\% level over most of the horizon,  though the 90\% confidence band includes zero at some months.

These findings are consistent with the predictions of heterogeneous-agent New Keynesian (HANK) models \citep{KaplanMollViolante2018}. In these frameworks, monetary policy transmission operates disproportionately through households with low liquid wealth who are unable to smooth consumption in response to income or interest rate shocks. A key insight of \citet{kaplan2014} is that a substantial share of households are ``wealthy hand-to-mouth'' as they hold illiquid wealth but have negligible liquid buffers, making their consumption highly sensitive to current income fluctuations. Low-income counties proxy for areas with a higher concentration of both hand-to-mouth and wealthy hand-to-mouth households, and therefore exhibit amplified responses to contractionary monetary policy through both the direct interest rate channel and the indirect general equilibrium channel operating via labor income \citep{KaplanMollViolante2018}. We note that county-level income is an imperfect proxy for household-level liquidity constraints: there is substantial within-county heterogeneity in wealth and income that our estimates average over, and the true degree of amplification among the most constrained households likely exceeds what we document at the county level. The persistence of the differential response through month 8 is further consistent with buffer-stock saving behavior \citep{carroll1997}: lower-income households maintain smaller precautionary buffers relative to their income, and empirical evidence confirms that their consumption responds more sharply and persistently to adverse income shocks as these buffers are depleted \citep{GanongNoel2019}. Higher-income households, by contrast, can draw down larger liquid reserves to smooth consumption, dampening the impact of tighter monetary conditions over the medium term.

Documenting these patterns requires the geographic and temporal granularity of the Y-14M data. Standard consumption measures are available only at the national or state level and at quarterly or annual frequency, making it impossible to detect whether monetary policy transmission varies systematically across the income distribution. The Y-14M data, covering more than 3,000 counties at monthly frequency over more than a decade, provide precisely this variation.

\subsection{Research Application: Summary}\label{sec:research_application_summary}

Across both monthly and annual specifications, credit card spending exhibits impulse responses that are qualitatively consistent with traditional consumption measures: a gradual decline following contractionary monetary policy, a trough at medium horizons, and partial mean reversion thereafter. The key difference lies in magnitude---credit card spending responds approximately twice as strongly as adjusted PCE and three to four times as strongly as total PCE. Rather than undermining the validity of credit card data, this amplification reflects exactly what theory predicts: credit card transactions disproportionately capture discretionary, interest-sensitive, and postponable expenditures that represent the margin of adjustment for household consumption. 

At the county level, the Y-14M data allow us to go further. Standard consumption measures are not available at both a sufficiently high frequency and sufficiently granular geographic disaggregation, making it impossible to detect whether monetary policy transmission varies systematically across the income distribution. The Y-14M data, covering more than 3,000 counties at monthly frequency over more than a decade, provide precisely this variation. Our finding that low-income counties exhibit a substantially larger consumption decline following contractionary monetary policy is consistent with the predictions of HANK models and represents, to our knowledge, the first such estimate at the county level. Together, these results suggest that credit card spending not only approximates consumption in levels, growth rates, and dynamic responses to shocks, but also opens the door to research designs that existing data cannot support.

\section{A Researcher's Guide: Caveats and Best Practices}\label{sec:researcher_guide}

\subsection{Credit Card Spending and Consumption Subcategories}\label{sec:researcher_guide_subcategories}

The Y-14M data allow us to observe total credit card spending in a given month, but they do not provide information on the types of goods and services purchased. This limitation represents an important drawback relative to transaction-level data, which allow researchers to track spending patterns across merchant categories.

\begin{figure}[!t] 
    \caption{Monthly National Consumption Growth: Major PCE Subcategories} \par 
    \footnotesize{This figure plots monthly year-over-year growth rates in average credit card spending (solid blue line in each panel) and major PCE subcategories (dashed red line in each panel) from June 2014 to March 2026. Panel (a) shows total goods, Panel (b) durable goods, Panel (c) nondurable goods, and Panel (d) services.}  
    \begin{center} 
        \subfigure[Goods]{\includegraphics[width=.495\textwidth]{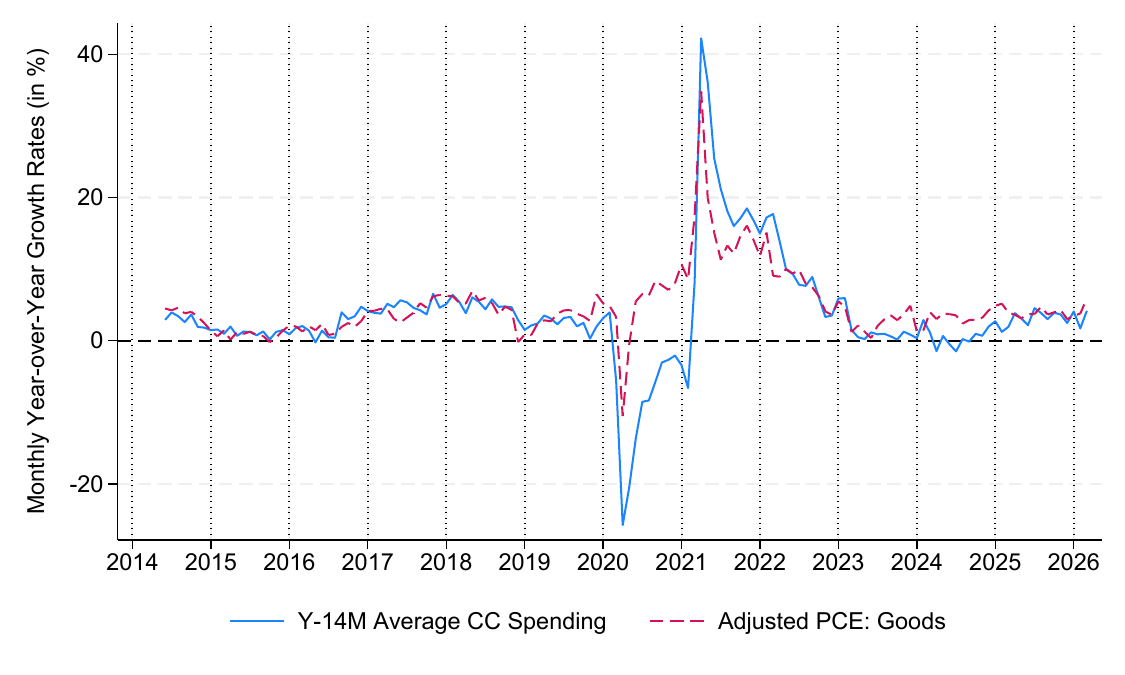}}\label{fig:Fig2a} 
        \subfigure[Durable Goods]{\includegraphics[width=.495\textwidth]{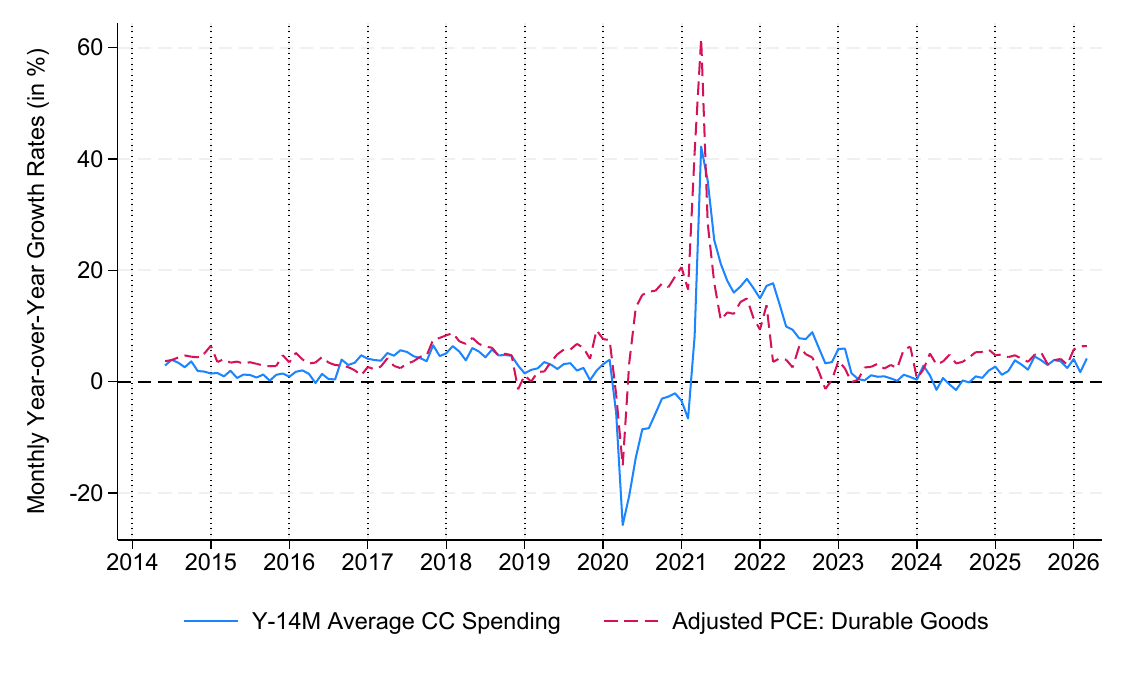}}\label{fig:Fig2b} 
        \subfigure[Nondurable Goods]{\includegraphics[width=.495\textwidth]{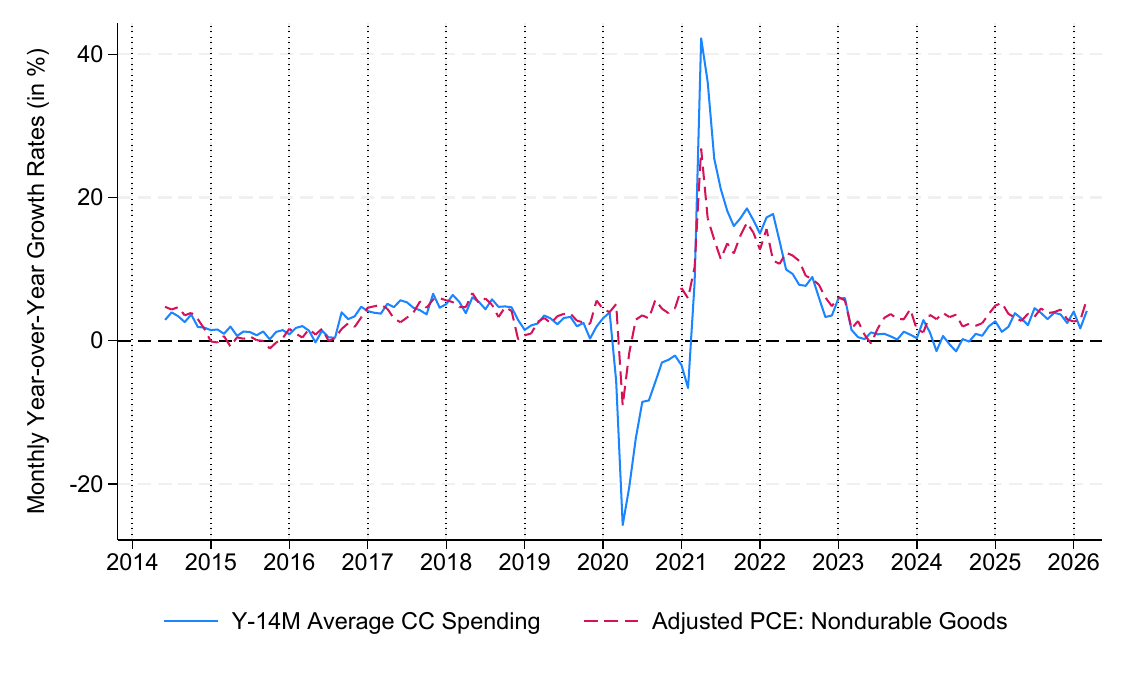}}\label{fig:Fig2c} 
        \subfigure[Services]{\includegraphics[width=.495\textwidth]{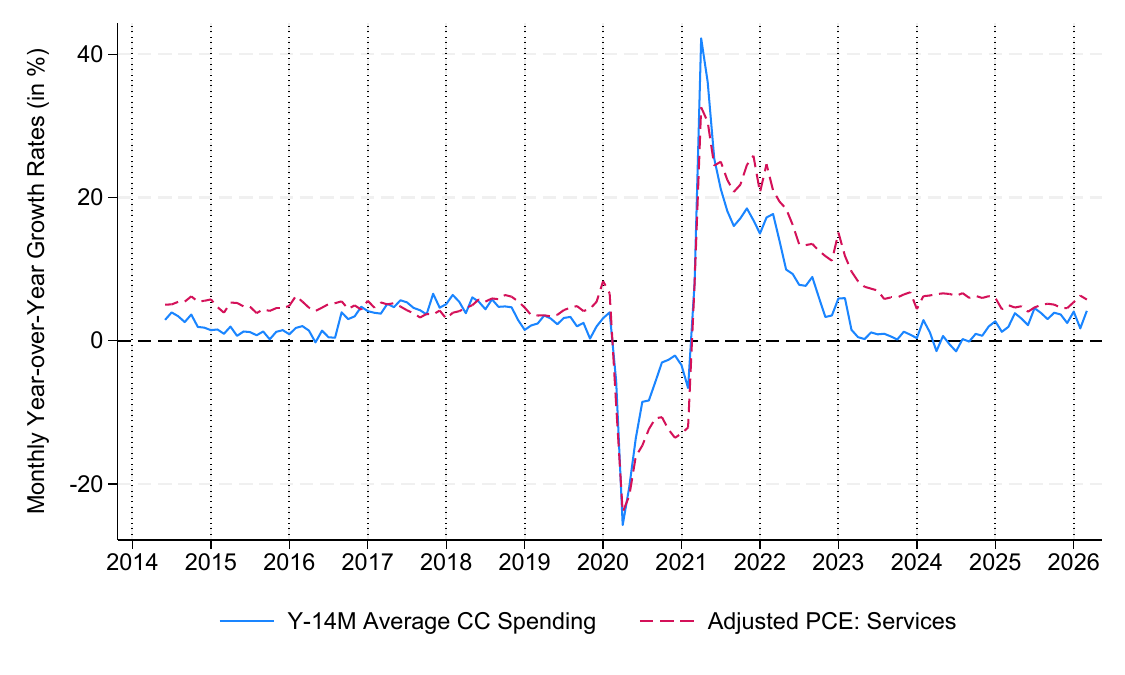}}\label{fig:Fig2d} 
    \end{center}
    \label{fig:cc_spending_pce_subcategories}
\end{figure}

Credit card spending is likely to capture some components of consumption better than others. Figure~\ref{fig:cc_spending_pce_subcategories} compares monthly year-over-year growth rates in credit card spending against growth rates of major PCE subcategories: total goods, durable goods, nondurable goods, and services. In normal times, credit card spending tracks goods consumption reasonably well, both for durable and nondurable goods. During the COVID-19 period, however, credit card spending declined more sharply and recovered less quickly than goods consumption, consistent with pandemic-related shifts in spending composition that cannot be directly decomposed in the Y-14M data. The divergence is particularly visible for services in the post-COVID period, when services PCE growth remained notably higher than credit card spending growth. Thus, as Y-14M spending is not observed at the transaction level, the data are best suited for studying aggregate consumption dynamics rather than shifts across detailed goods and services categories.  

\begin{figure}[!t]	
    \caption{Monthly National Consumption Growth: Excluded PCE Subcategories}
    \par
    \footnotesize{This figure plots monthly year-over-year growth rates in average credit card spending (solid blue line in each panel) and PCE subcategories excluded from our adjusted PCE measure (dashed red line in each panel) from June 2014 to March 2026. Panel (a) shows housing and utilities, Panel (b) health care, Panel (c) financial services and insurance, and Panel (d) final consumption expenditures of nonprofit institutions serving households.}
    \begin{center}
        \subfigure[Housing and Utilities]{\includegraphics[width=.495\textwidth]{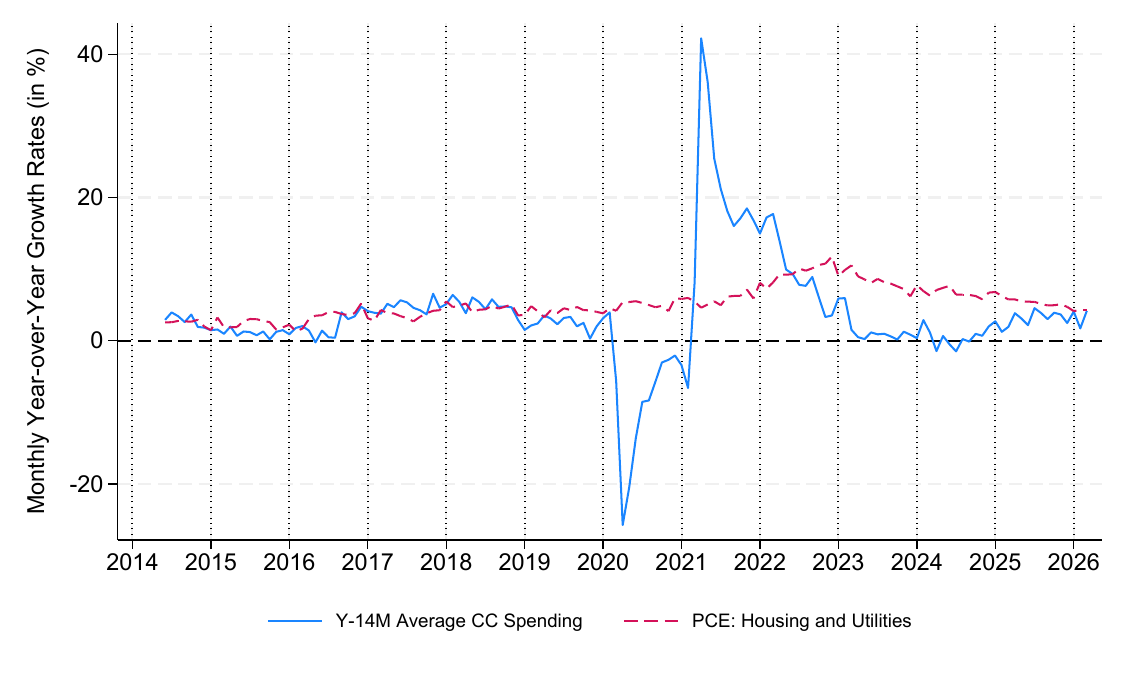}}\label{fig:Fig9a}
        \subfigure[Health Care]{\includegraphics[width=.495\textwidth]{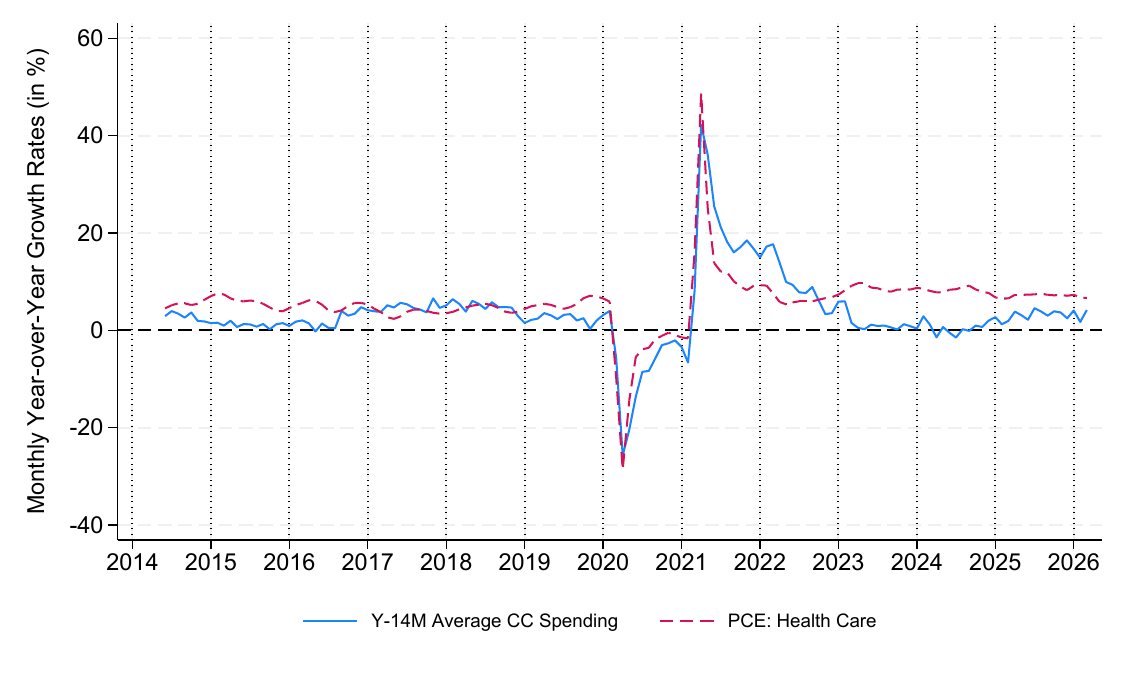}}\label{fig:Fig9b}
        \subfigure[Financial Services and Insurance]{\includegraphics[width=.495\textwidth]{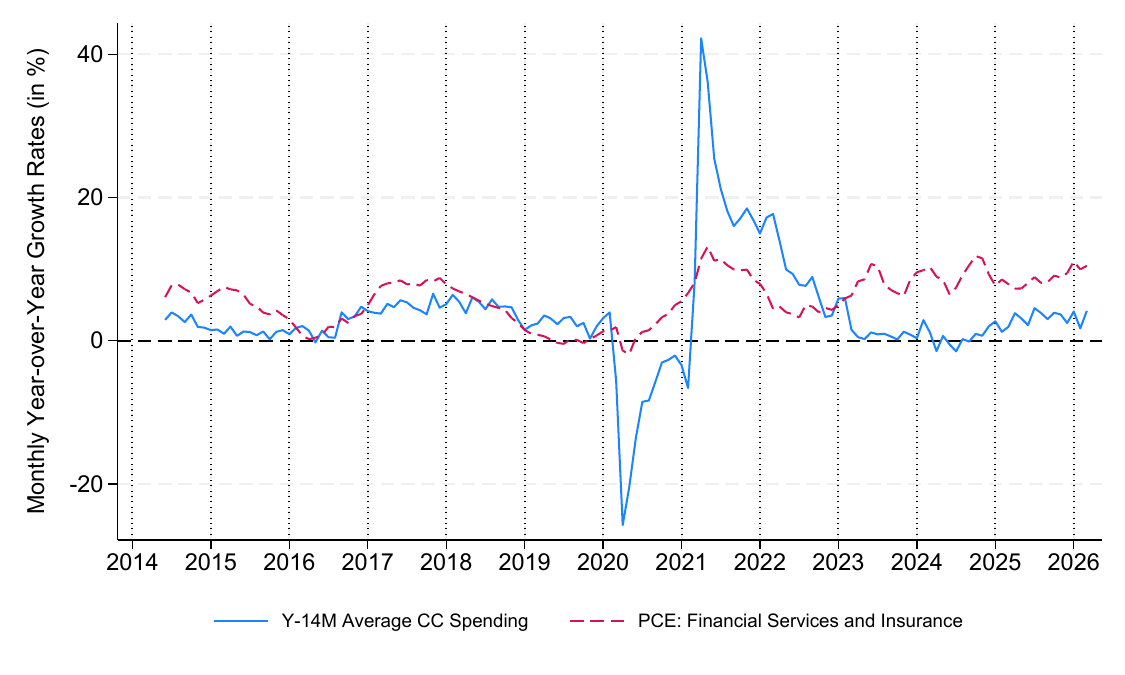}}\label{fig:Fig9c}
        \subfigure[Final CE of NPISHs]{\includegraphics[width=.495\textwidth]{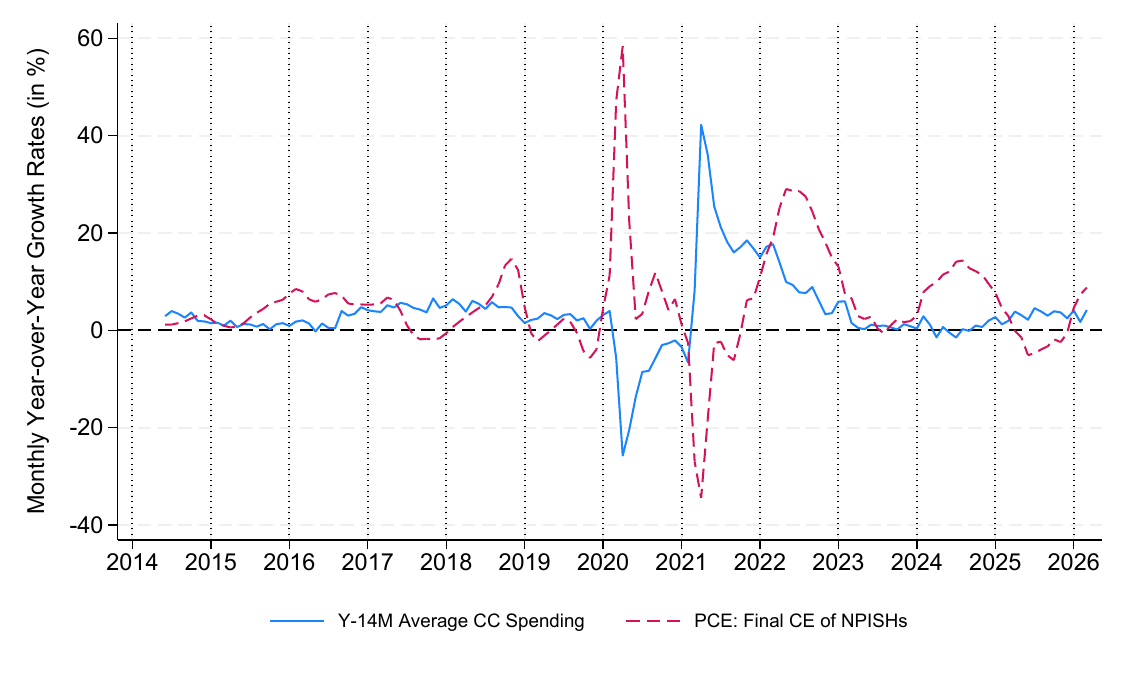}}\label{fig:Fig9d}
    \end{center}
    \label{fig:cc_versuses_excluded_pce_subcategories}
\end{figure}

A related issue is that broad PCE includes several components that are conceptually less comparable to credit card spending as they either do not reflect out-of-pocket household purchases or are rarely paid for using credit cards. These include housing and utilities, health care expenditures paid by insurance, financial services, spending by nonprofit institutions serving households, and some large durable purchases such as motor vehicles. For this reason, our baseline benchmark throughout the paper is adjusted PCE, which excludes categories that are less likely to be captured by credit card spending.\footnote{Figure~\ref{fig:cc_transactions_by_merchant_category} provides complementary evidence from the 2024 Diary of Consumer Payment Choice. Credit card use is highest in categories such as lodging, travel, restaurants, transportation, groceries, gas, and drug stores, and substantially lower for rent, financial services, and transfers.} Figure~\ref{fig:cc_versuses_excluded_pce_subcategories} illustrates this point and shows that the excluded PCE categories exhibit dynamics that differ markedly from credit card spending and from the adjusted consumption measures analyzed earlier. Housing and utilities are relatively smooth, health care and financial services grow more strongly in the post-pandemic period, and nonprofit consumption displays a distinct pattern around the pandemic.

Taken together, Figures~\ref{fig:cc_spending_pce_subcategories} and \ref{fig:cc_versuses_excluded_pce_subcategories} highlight an important caveat for researchers using credit card data to measure consumption. Aggregate card spending is most informative about broad consumption dynamics, especially for out-of-pocket and card-payable expenditures. It is less well suited for measuring categories such as imputed housing services, insurance-financed health care, nonprofit expenditures, or purchases that are rarely made by credit card. Researchers should therefore choose benchmark consumption aggregates that correspond as closely as possible to the spending concept captured in the card data.

\subsection{Payment-Method Shifts and Sample Composition}\label{sec:researcher_guide_payment_shifts}
 
A central challenge in interpreting credit card spending is that aggregate spending growth may reflect not only changes in spending per card, but also changes in the number and composition of cards observed in the sample. Credit card usage has increased over time, both within our sample and in the economy overall \citep{BayehNardoneOBrienPhelps2025}. As a result, growth in \emph{total} credit card spending can overstate growth in underlying consumption if it partly reflects more cards entering the sample rather than higher \emph{average} spending on existing cards.

\begin{figure}[!t]
    \caption{Decomposing Total Credit Card Spending Growth} 
    \footnotesize{This figure plots monthly year-over-year growth in total credit card spending (solid black line) and its decomposition into growth in average credit card spending (blue bars) and growth in the number of credit cards in the sample (red bars) from June 2014 to March 2026, alongside monthly year-over-year growth in adjusted PCE (dashed red line).}
    \begin{center}
        \centerline{\includegraphics[width=1\textwidth]{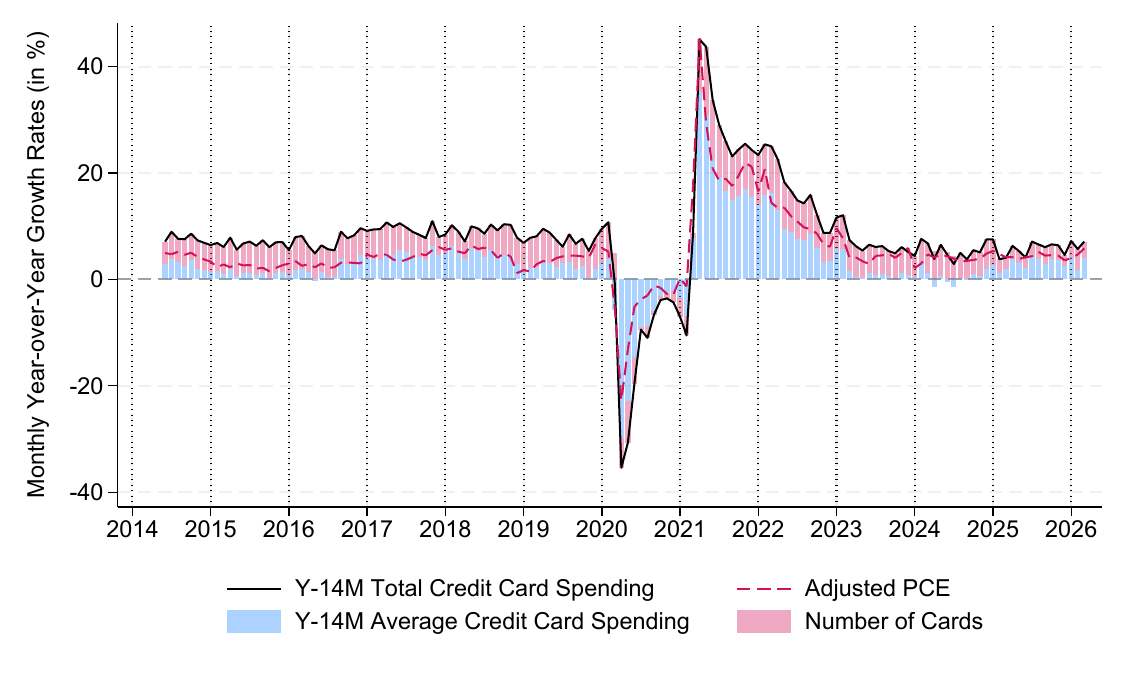}}
    \end{center}  	
    \label{fig:decomposition_of_total_cc_spending_growth}
\end{figure}

Figure~\ref{fig:decomposition_of_total_cc_spending_growth} illustrates why our baseline analysis focuses on average credit card spending rather than total credit card spending. We decompose year-over-year growth in total credit card spending into growth in average spending per card and growth in the number of cards in the sample.\footnote{Let total credit card spending be $T_t=\bar{x}_t N_t$, where $\bar{x}_t$ denotes average spending per card and $N_t$ denotes the number of cards. Taking year-over-year log differences gives $\Delta_{12}\log T_t=\Delta_{12}\log \bar{x}_t+\Delta_{12}\log N_t$.} While growth in average credit card spending tracks adjusted PCE growth well over most of the sample (as also shown in Figure~\ref{fig:monthly_yoy_consumption_growth}), growth in total credit card spending is systematically higher in most periods as the number of cards in the sample increases over time. Thus, total card spending growth is not a pure measure of consumption growth but also reflects increasing card adoption and bank portfolio growth, rather than spending growth by existing cardholders alone.  

\begin{figure}[!t]
    \caption{Decomposing Average Credit Card Spending Growth} 
    \footnotesize{This figure plots monthly year-over-year growth in average credit card spending (solid black line) and its decomposition into an intensive margin among continuing cards (blue bars) and an extensive margin due to net entry and exit (red bars) from June 2014 to March 2026, alongside monthly year-over-year growth in adjusted PCE (dashed red line).}
    \begin{center}
        \centerline{\includegraphics[width=1\textwidth]{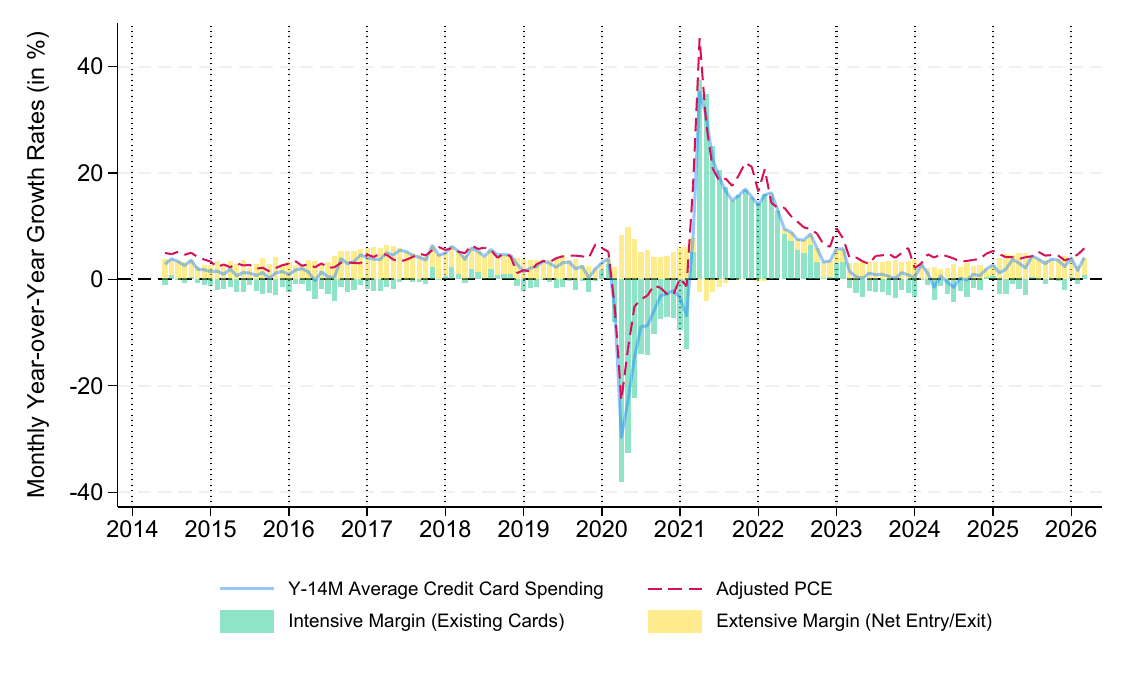}}
    \end{center}  	
    \label{fig:decomposition_of_average_cc_spending_growth}
\end{figure}

While total credit card spending can be decomposed into changes in average spending per card and changes in the number of cards, average per-card spending can be decomposed into an intensive and an extensive margin. The intensive margin captures year-over-year spending growth among cards observed in both the current month and the same month one year earlier. The extensive margin captures the combined net effect of cards entering and exiting the sample. Figure~\ref{fig:decomposition_of_average_cc_spending_growth} illustrates this decomposition. The decomposition shows substantial churn in the card population. In most periods, spending growth among continuing cards is close to zero or negative, with a pronounced exception during the post-COVID recovery. By contrast, the net entry/exit margin is positive in most periods and accounts for much of the observed fit between average credit card spending growth and aggregate consumption growth. This pattern indicates that changes in the composition of cards, rather than changes in spending on a fixed set of cards alone, play an important role in the dynamics of average credit card spending.   

Taken together, Figures~\ref{fig:decomposition_of_total_cc_spending_growth} and \ref{fig:decomposition_of_average_cc_spending_growth} show that researchers should carefully distinguish between total credit card spending growth, average per-card spending growth, and the extent to which these dynamics reflect spending changes among continuing cards versus changes in the number or composition of cards.

\subsection{Bank-Level Heterogeneity and Sample Representativeness}\label{sec:researcher_guide_bank_heterogeneity}

Many credit card datasets available to researchers are drawn from a single bank, fintech platform, or payment processor. Such datasets can offer rich transaction-level detail, but their customer bases may not be representative of the broader population. Differences in customer composition, geographic footprint, product mix, credit quality, and cardholder behavior can all affect measured spending dynamics. The multi-bank structure of the Y-14M data allows us to quantify this concern directly by comparing spending measures constructed from individual banks to the aggregate measure constructed from the full set of reporting banks.    

\begin{figure}[!t]
    \caption{Bank-Level Heterogeneity in Credit Card Spending Growth} 
    \footnotesize{This figure plots monthly year-over-year growth rates in aggregate average credit card spending (solid blue line), alongside the 25th–75th and 10th–90th percentile ranges of bank-level spending growth across banks in our sample, from June 2014 to March 2026.}
    \begin{center}
        \centerline{\includegraphics[width=1\textwidth]{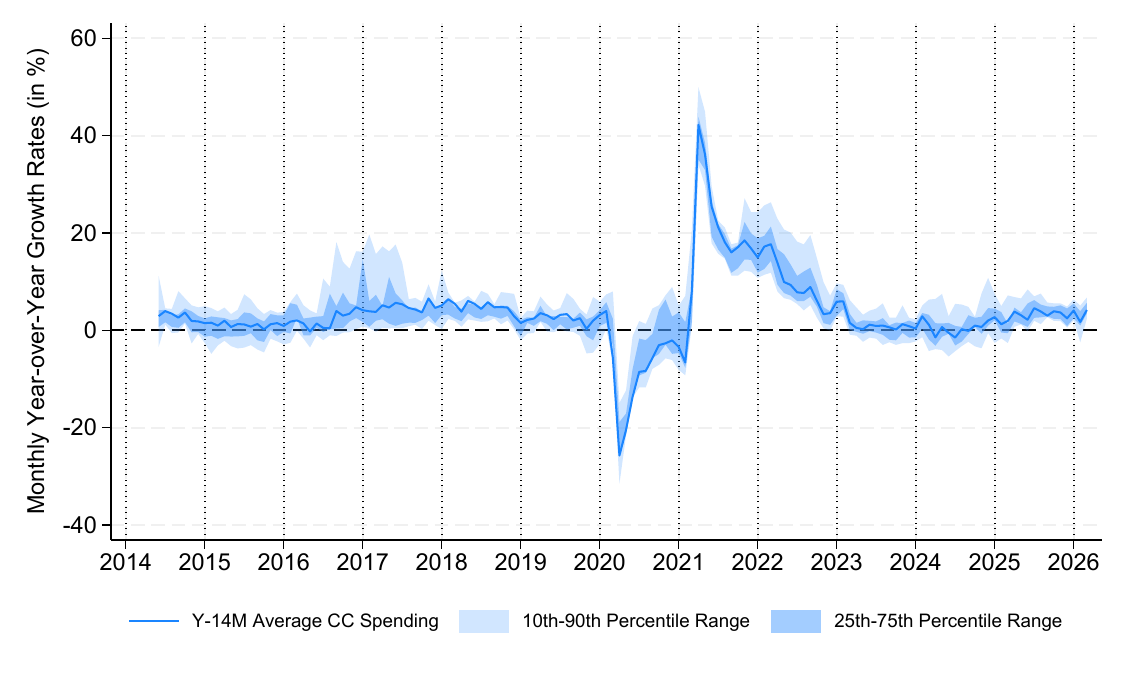}}
    \end{center}  	
    \label{fig:monthly_yoy_consumption_growth_bank_heterogeneity}
\end{figure} 

Figure~\ref{fig:monthly_yoy_consumption_growth_bank_heterogeneity} plots monthly year-over-year growth in average credit card spending for the full Y-14M sample from Figure~\ref{fig:monthly_yoy_consumption_growth}, alongside the $25^{th}-75^{th}$ and $10^{th}-90^{th}$ percentile ranges of bank-level spending growth in our sample. In most periods, bank-level spending growth is distributed tightly around the aggregate Y-14M series, suggesting that the aggregate measure is not driven by a small number of banks and that individual banks often capture similar broad consumption dynamics. At the same time, the dispersion across banks is not negligible. The 10th--90th percentile range widens notably in 2016--2017 and again around 2022, indicating that, in some periods, spending measures constructed from different banks can imply meaningfully different consumption dynamics.  

Researchers using data from individual banks, fintech platforms, or payment processors should therefore validate their spending measures against external benchmarks and, to the extent possible, assess whether their results are driven by particular customer segments or geographic markets.

\subsection{Approximating Credit Card Spending from Credit Bureau Data}\label{sec:researcher_guide_credit_bureau_data}

In the absence of access to regulatory or private-sector credit card data, researchers often infer credit card spending from credit bureau data \citep{gibbs2024consumer}. These data, however, are designed primarily to capture liabilities and repayment behavior rather than consumption. Researchers typically observe credit card balances and, for some furnishers, actual payments, but they do not directly observe the flow of new purchases. Following \citet{GanongNoel2020}, credit card spending on card $i$ in period $t$ can be approximated as  
\begin{align}
    s_{i,t}
    &=
    \left\{
    \begin{array}{c c}
        b_{i,t} - b_{i,t-1} + p_{i,t} & \text{if } b_{i,t} - b_{i,t-1} + p_{i,t} \ge 0 \\[0.4em]
        0                   & \text{otherwise}
        \end{array}
    \right.
    \label{eq:gks_spending_proxy}
\end{align} 
where $b_{i,t}$ and $b_{i,t-1}$ denote current and lagged outstanding balances and $p_{i,t}$ denotes actual payments. This approximation is subject to two important caveats. First, it is inclusive of finance charges and fees, which are generally not separately observed in credit bureau data. Second, it requires actual payment information, which is often unavailable as credit card issuers do not consistently furnish this variable to credit reporting agencies \citep{GuttmanKenneyShahidinejad2025}. The Y-14M data allow us to assess the importance of these limitations as they contain not only outstanding balances, actual payments, finance charges, and fees, but also observed purchase volumes, which we use as the benchmark measure of credit card spending. 

We follow \citet{GuttmanKenneyShahidinejad2025} and estimate three sets of regressions:
\begin{align}
    \text{PV}_{i,t}    &= \alpha + \beta_1 b_{i,t} + \beta_2 b_{i,t-1} + \beta_3 \Delta b_{i,t} + \beta_4 \mathbf{1}\{b_{i,t} > 0\} + \beta_5 \mathbf{1}\{b_{i,t-1} > 0\} + \varepsilon_{i,t}   
    \label{eq:gks_without_payment} \\ 
    \text{PV}_{i,t}    &= \alpha + \beta_1s_{i,t} + \varepsilon_{i,t}  
    \label{eq:gks_with_payment} \\ 
    \text{PV}_{i,t}    &= \alpha + \beta_1s_{i,t} + \beta_2\text{Finance Charges}_{i,t} + \beta_3\text{Fees}_{i,t} + \varepsilon_{i,t}  \label{eq:gks_with_fin_fee}
\end{align}
where $PV_{i,t}$ is actual purchase volume and $s_{i,t}$ is the bureau-style spending proxy from Equation~(\ref{eq:gks_spending_proxy}). Equation~(\ref{eq:gks_without_payment}) evaluates how well spending can be inferred from balances alone, as would be necessary when actual payment data are unavailable. Equation~(\ref{eq:gks_with_payment}) evaluates the approximation when actual payments are observed. Equation~(\ref{eq:gks_with_fin_fee}) augments the bureau-style spending proxy with finance charges and fees to assess how much of the remaining measurement error is attributable to these non-purchase components of credit card balances.

\begin{figure}[!t]
    \caption{Approximating Credit Card Spending from Credit Bureau Data} 
    \footnotesize{This figure plots $R^2$ values from regressions that approximate observed credit card purchase volume using credit-bureau-style measures, based on Equations~(\ref{eq:gks_without_payment}),~(\ref{eq:gks_with_payment}), and~(\ref{eq:gks_with_fin_fee}) and the analysis in \cite{GuttmanKenneyShahidinejad2025}. Each bar reports results from a separate regression for the full sample and for each credit-score segment. The full sample includes 60.82 million credit card accounts. The credit-score segments are subprime, with scores between 300 and 600 (N = 4.06 million); near prime, with scores between 601 and 660 (N = 5.53 million); prime, with scores between 661 and 720 (N = 10.63 million); prime plus, with scores between 721 and 780 (N = 13.77 million); and superprime, with scores of at least 781 (N = 26.84 million).}
    \begin{center}
        \centerline{\includegraphics[width=1\textwidth]{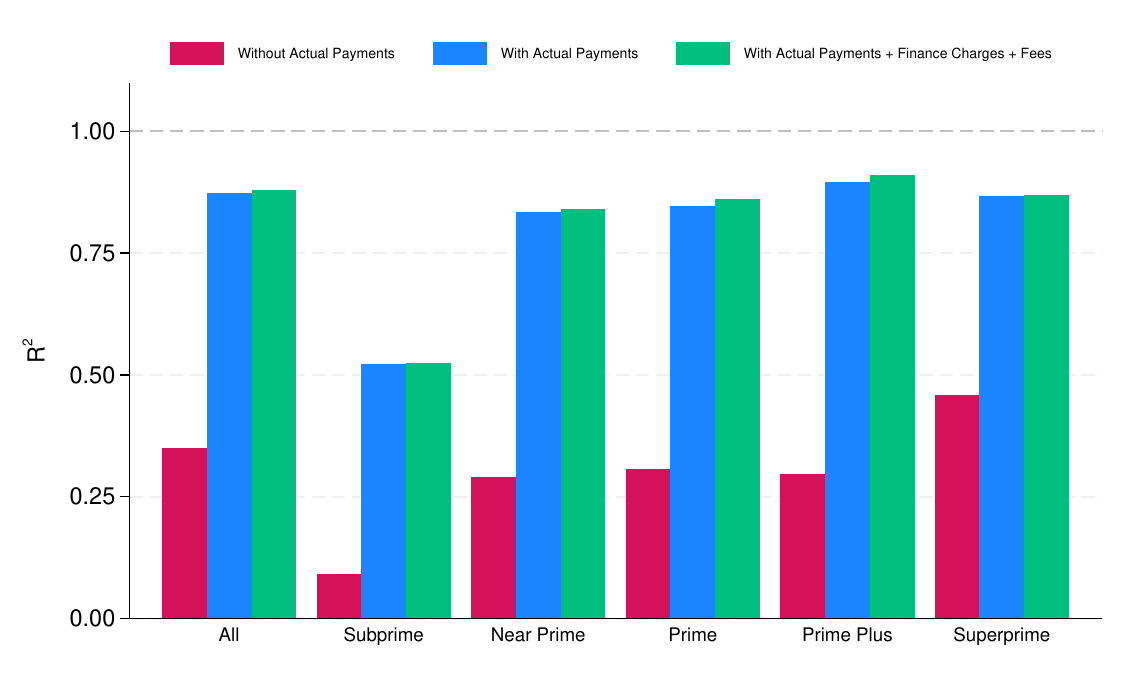}}
    \end{center}  	
    \label{fig:approx_cc_spending_from_cb_data}
\end{figure} 

Figure~\ref{fig:approx_cc_spending_from_cb_data} illustrates how well spending measures based on credit bureau data can approximate actual credit card purchase volumes. The first specification (red bars) uses only balance-based information and mimics the case in which actual payment data are not available. The fit is weak, with an $R^2$ of 0.35 in the full sample, and especially poor for subprime borrowers, where the $R^2$
falls to 0.09. Adding actual payment information substantially improves the approximation. The $R^2$ rises to 0.87 in the full sample and remains reasonably high across most FICO buckets, although the fit continues to be noticeably weaker for subprime borrowers ($R^2$ = 0.52). Finally, adding finance charges and fees to the specification yields almost no additional explanatory power. This suggests that the residual discrepancy between spending inferred from balances and payments and actual purchase volumes is not primarily explained by these components, but instead reflects other non-purchase balance flows, timing differences, or account-level noise.

Taken together, these results show that credit bureau data can provide a reasonable but imperfect proxy for observed credit card purchase volumes, provided that payment information is available.

\section{Conclusion}\label{sec:conclusion}

We show that credit card spending data from the Federal Reserve's Y-14M reports provide a reliable measure for macroeconomic consumption. Monthly credit card spending growth explains 92 percent of the variation in adjusted PCE growth at the national level, and annual credit card spending growth captures meaningful state-level variation in adjusted PCE growth, although the within-year cross-sectional relationship is more mixed. These results hold across a range of economic conditions, including the extraordinary circumstances of the COVID-19 pandemic, thus supporting the use of credit card spending as a credible consumption measure for applied research.

We then introduce a novel monthly county-level credit-card-based consumption dataset, covering more than 3,000 U.S. counties since 2014, which allows us to measure consumption at a frequency and geographic granularity unavailable in traditional data. The dataset reveals granular consumption dynamics that align closely with well-documented economic developments, such as the geographic incidence of the COVID-19 pandemic, and opens the door to research designs that require both high-frequency and granular geographic variation in consumption.

In a proof-of-concept application, we demonstrate the research value of this dataset through an application to monetary policy transmission. We show that credit card spending exhibits impulse responses that are qualitatively similar and quantitatively comparable to those obtained with traditional consumption data. Exploiting the county-level granularity, we find that low-income counties experience substantially larger consumption declines following contractionary monetary policy than high-income counties, consistent with the predictions of HANK models and, to our knowledge, the first such estimate at the county-month level.

Finally, we provide a researcher's guide to the main caveats involved in using credit card data to measure consumption. We show that aggregate card spending is most informative about out-of-pocket, card-payable expenditures; that measured spending growth can reflect changes in the number and composition of cards; that data from different institutions may imply meaningfully different spending dynamics; and that credit bureau data can provide a reasonable but imperfect proxy for observed credit card purchase volume when payment information is available.

\clearpage
\begin{singlespacing}
    \bibliographystyle{jf}    
    \bibliography{References}
\end{singlespacing}

\pagebreak

\begin{center}
    \begin{LARGE}
        Online Appendix for:                            \\
        Measuring Consumption with Credit Card Data:    \\
        Benchmarking and Beyond                         \\
    \end{LARGE}
    \vspace{0.5cm}
    Aditya Aladangady, Ricardo Duque Gabriel, Carlo Wix 
\end{center}

\clearpage
\appendix
\setcounter{section}{0}
\renewcommand{\thesection}{A\arabic{section}}

\clearpage
\setcounter{table}{0}
\renewcommand{\thetable}{A\arabic{table}}
	
\setcounter{equation}{0}
\renewcommand{\theequation}{A\arabic{equation}}

\setcounter{figure}{0}
\renewcommand{\thefigure}{A\arabic{figure}}

\section{Additional Figures}

\begin{figure}[!h]	
    \caption{Annual State-Level Consumption Growth: Time-Series Evidence}
    \par
    \footnotesize{This figure plots annual year-over-year growth rates in average credit card spending (solid blue lines) and adjusted PCE (dashed red lines) for selected states from 2015 to 2024. Panel (a) shows California, Panel (b) New York, Panel (c) Washington, DC, and Panel (d) Maryland.}
    \begin{center}
        \subfigure[California]{\includegraphics[width=.495\textwidth]{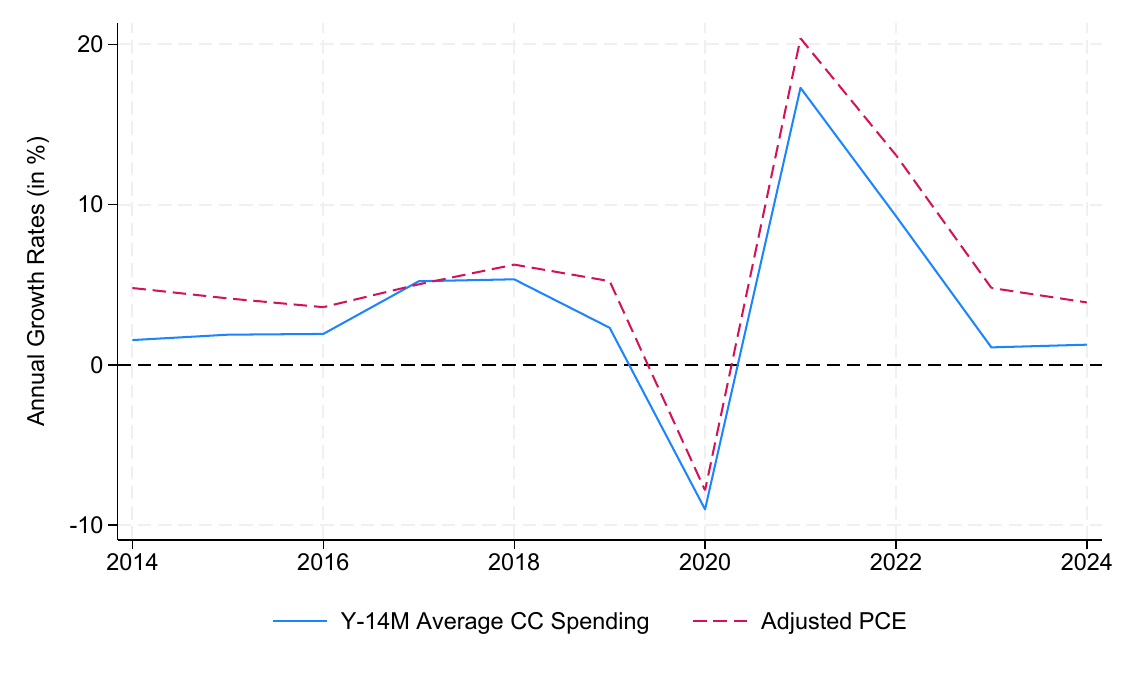}}\label{fig:Fig4a}
        \subfigure[New York]{\includegraphics[width=.495\textwidth]{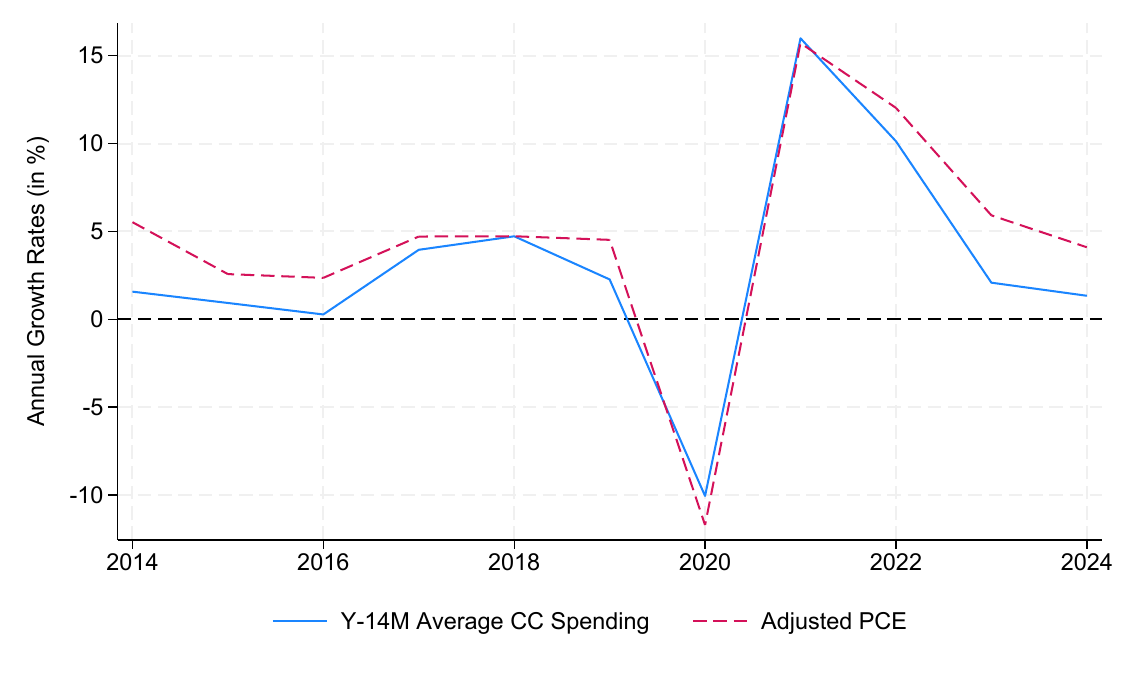}}\label{fig:Fig4b}
        \subfigure[Washington, DC]{\includegraphics[width=.495\textwidth]{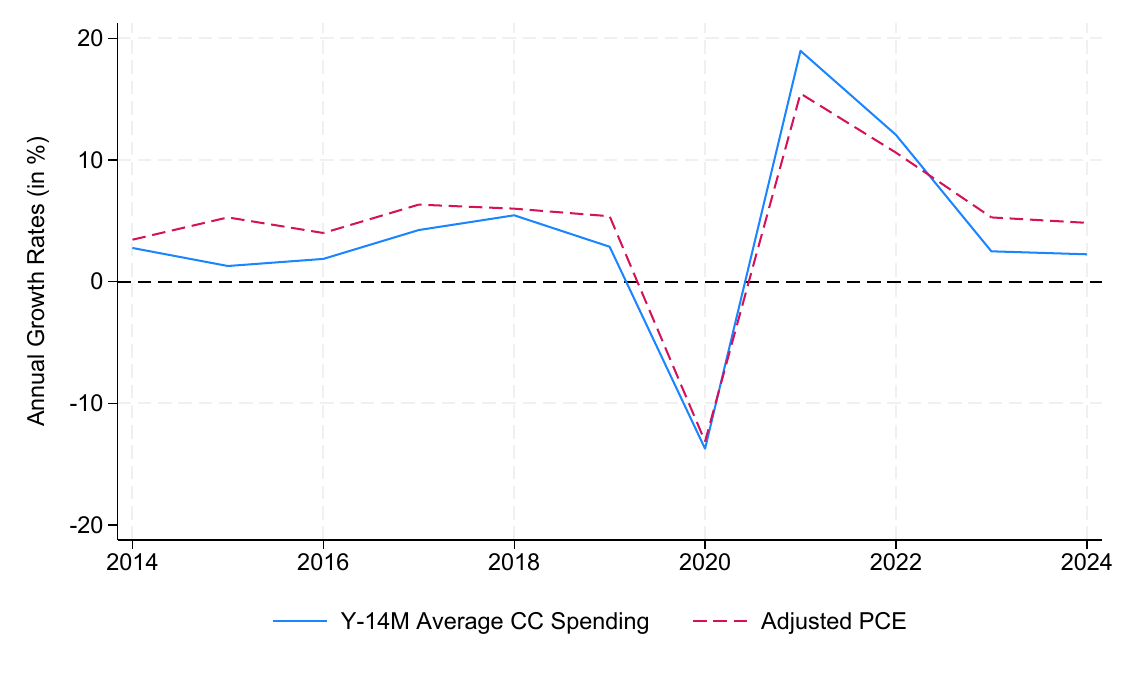}}\label{fig:Fig4c}
        \subfigure[Maryland]{\includegraphics[width=.495\textwidth]{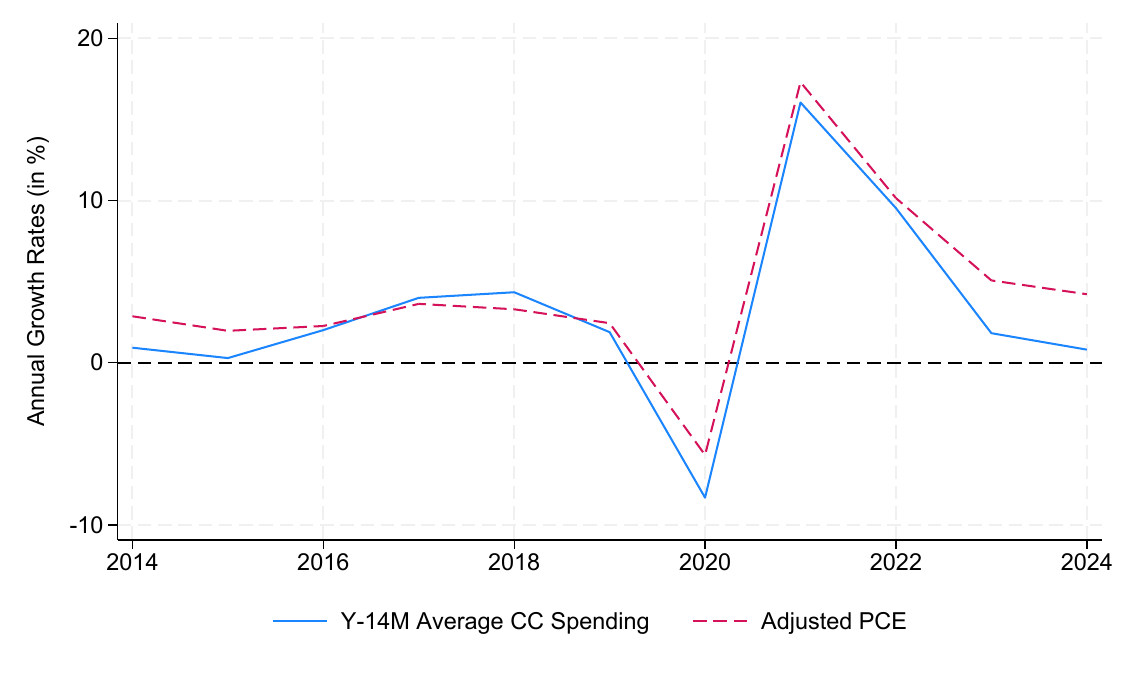}}\label{fig:Fig4d}
    \end{center}
    \label{fig:annual_state_level_consumption_growth}
\end{figure} 

\begin{figure}[!h]	
    \caption{Monthly State-Level Retail Spending Growth: Time-Series Evidence}
    \par
    \footnotesize{This figure plots monthly year-over-year growth rates in average credit card spending (solid blue lines) and Census retail sales (dashed red lines) for selected states from January 2019 to March 2026. Panel (a) shows California, Panel (b) New York, Panel (c) Washington, DC, and Panel (d) Maryland.}
    \begin{center}
        \subfigure[California]{\includegraphics[width=.495\textwidth]{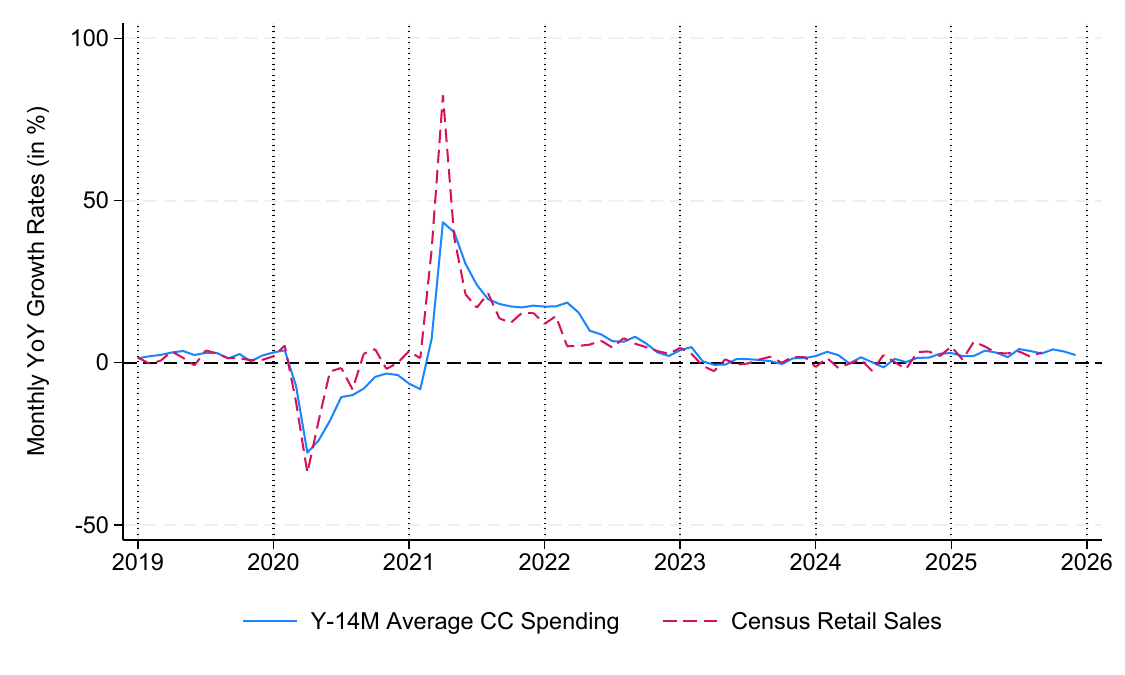}}\label{fig:Fig4a}
        \subfigure[New York]{\includegraphics[width=.495\textwidth]{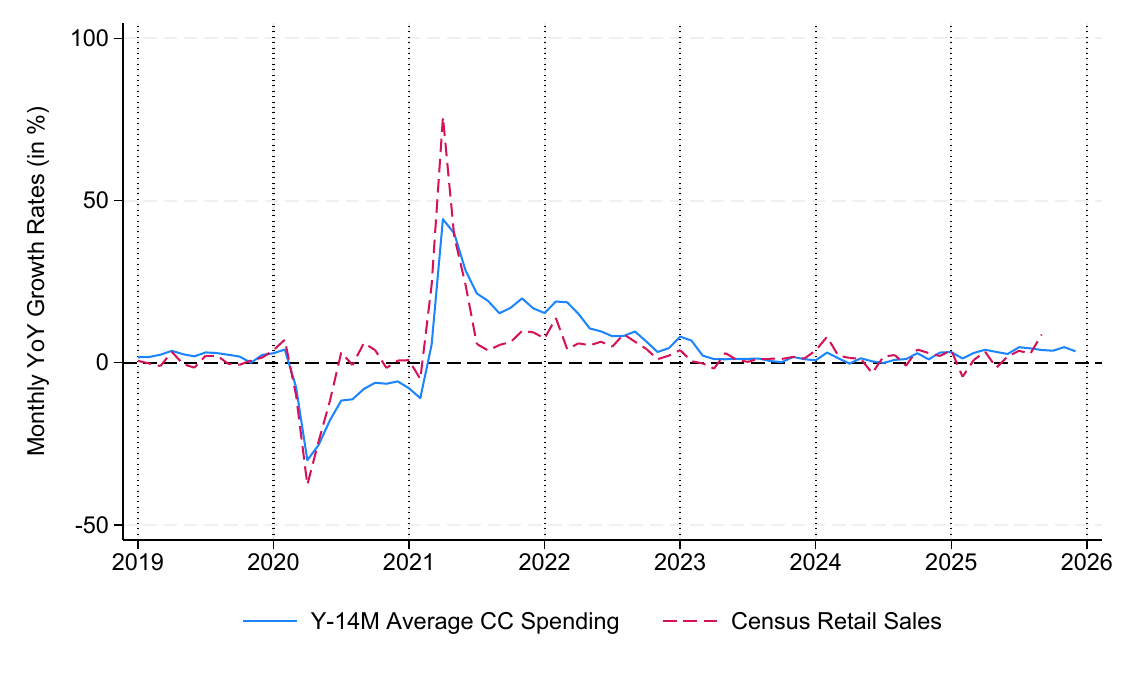}}\label{fig:Fig4b}
        \subfigure[Washington, DC]{\includegraphics[width=.495\textwidth]{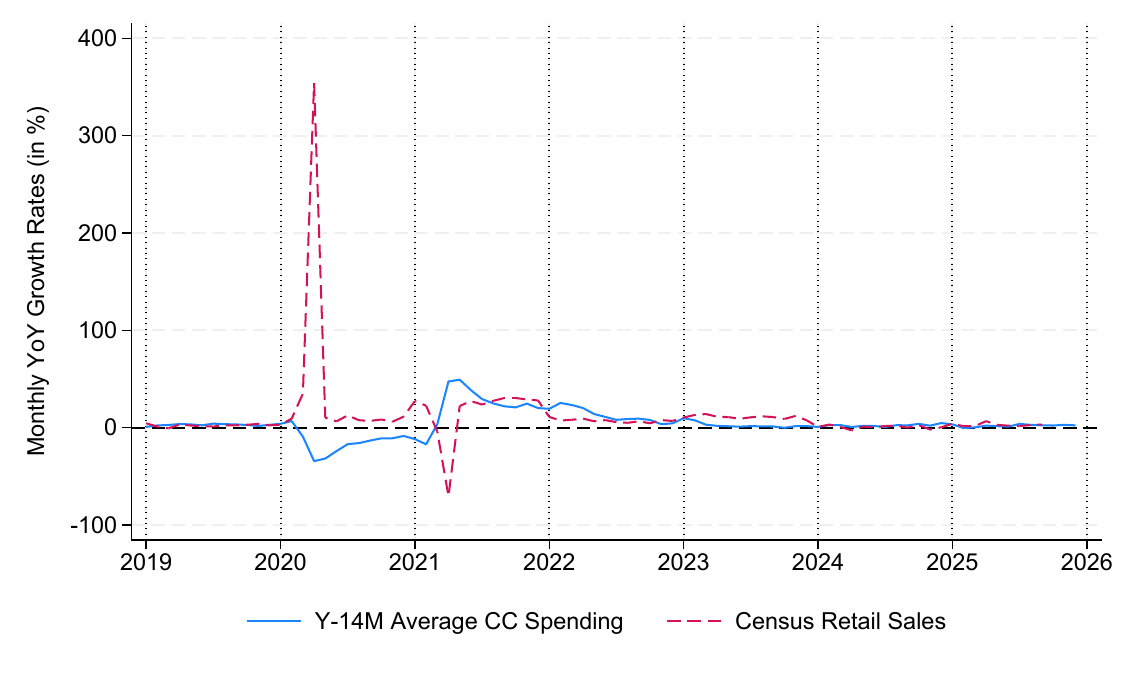}}\label{fig:Fig4c}
        \subfigure[Maryland]{\includegraphics[width=.495\textwidth]{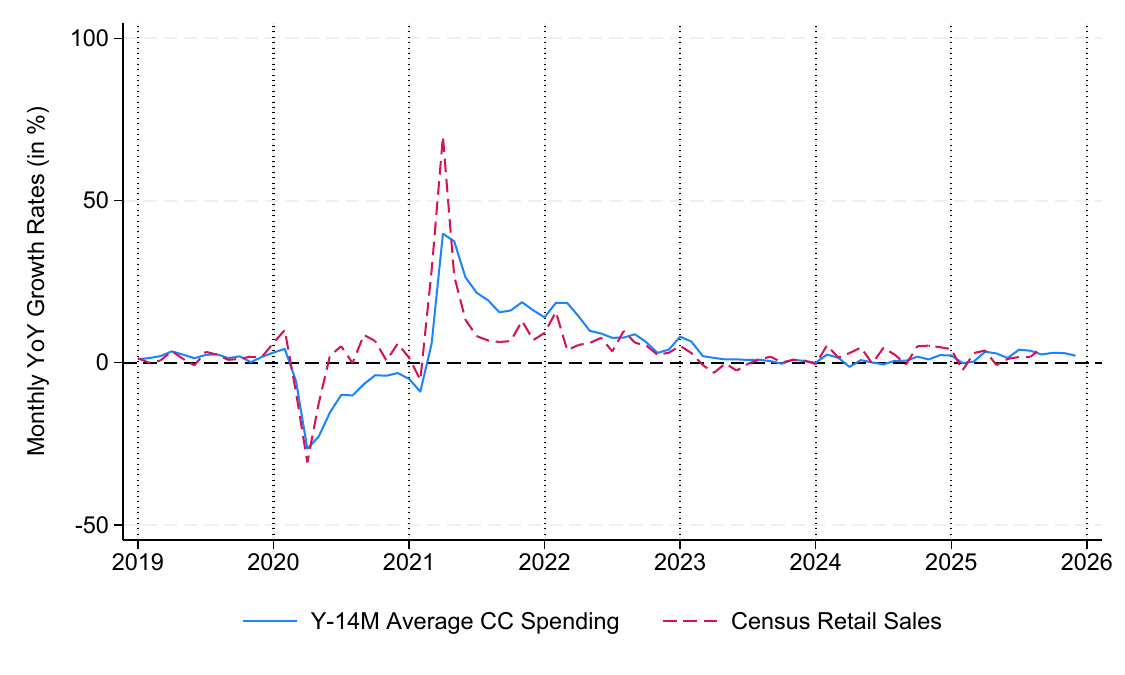}}\label{fig:Fig4d}
    \end{center}
    \label{fig:monthly_state_level_consumption_growth}
\end{figure} 

\begin{figure}[!h]
    \caption{Monthly Spending Relative to January 2020: Affinity Solutions Benchmark} 
    \footnotesize{This figure plots monthly spending in the Y-14M data (solid blue line) and Affinity Solutions data from Opportunity Insights (dashed red line) from January 2020 to March 2026, expressed as seasonally adjusted changes relative to January 2020. The Y-14M series indexes average monthly credit card purchase volume to its January 2020 value and applies an analogous seasonal adjustment based on the 2019 monthly pattern. The Affinity Solutions series is the publicly available Opportunity Insights spending series.}
    \label{fig:monthly_yoy_consumption_growth_affinity}
    \begin{center}
        \centerline{\includegraphics[width=1\textwidth]{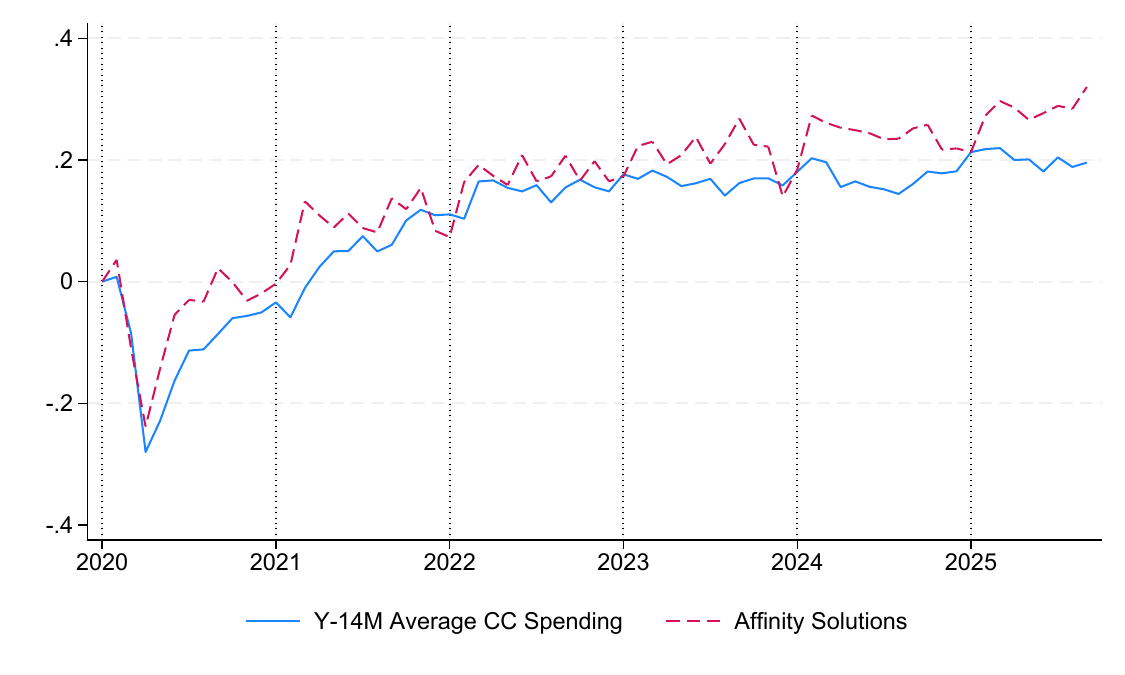}}
    \end{center}  	
\end{figure}

\begin{figure}[!h]
    \caption{County-Level Spending Levels: Per-Capita Credit Card Spending and Economic Census} 
    \footnotesize{This figure plots county-level Economic Census spending per capita against average annual Y-14M credit card spending per capita for the years 2017 and 2022. Each dot represents a county-year observation. The solid red line indicates the fitted regression line, and the dashed black line indicates the 45-degree line.}
    \begin{center}
        \centerline{
            \includegraphics[width=1\textwidth]{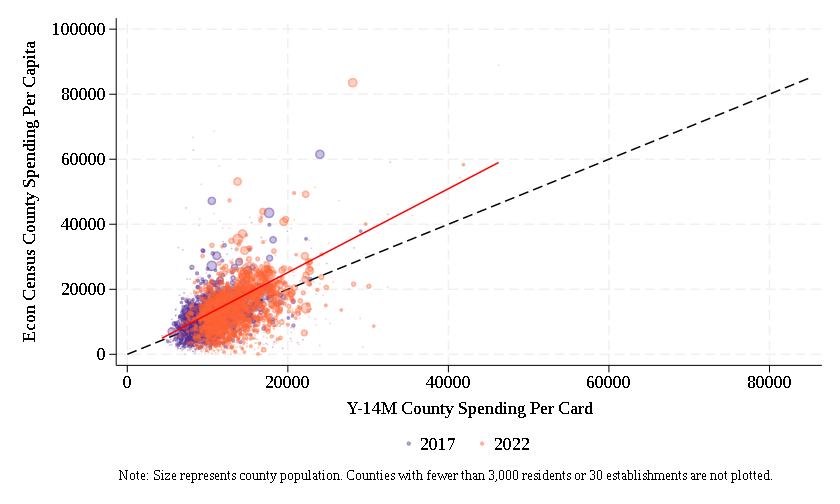}
        }
    \end{center}  	
    \label{fig:y14m_versus_econ_census_spending_level_per_capita}
\end{figure} 

\begin{figure}[!h]
    \caption{Monthly County-Level Consumption Growth by County-Level Income Decile} 
    \footnotesize{This figure plots monthly year-over-year growth rates in average credit card spending for high-income counties (top decile of average income; blue line) and low-income counties (bottom decile of average income; red line) from June 2014 to March 2026.}
    \begin{center}
        \centerline{\includegraphics[width=1\textwidth]{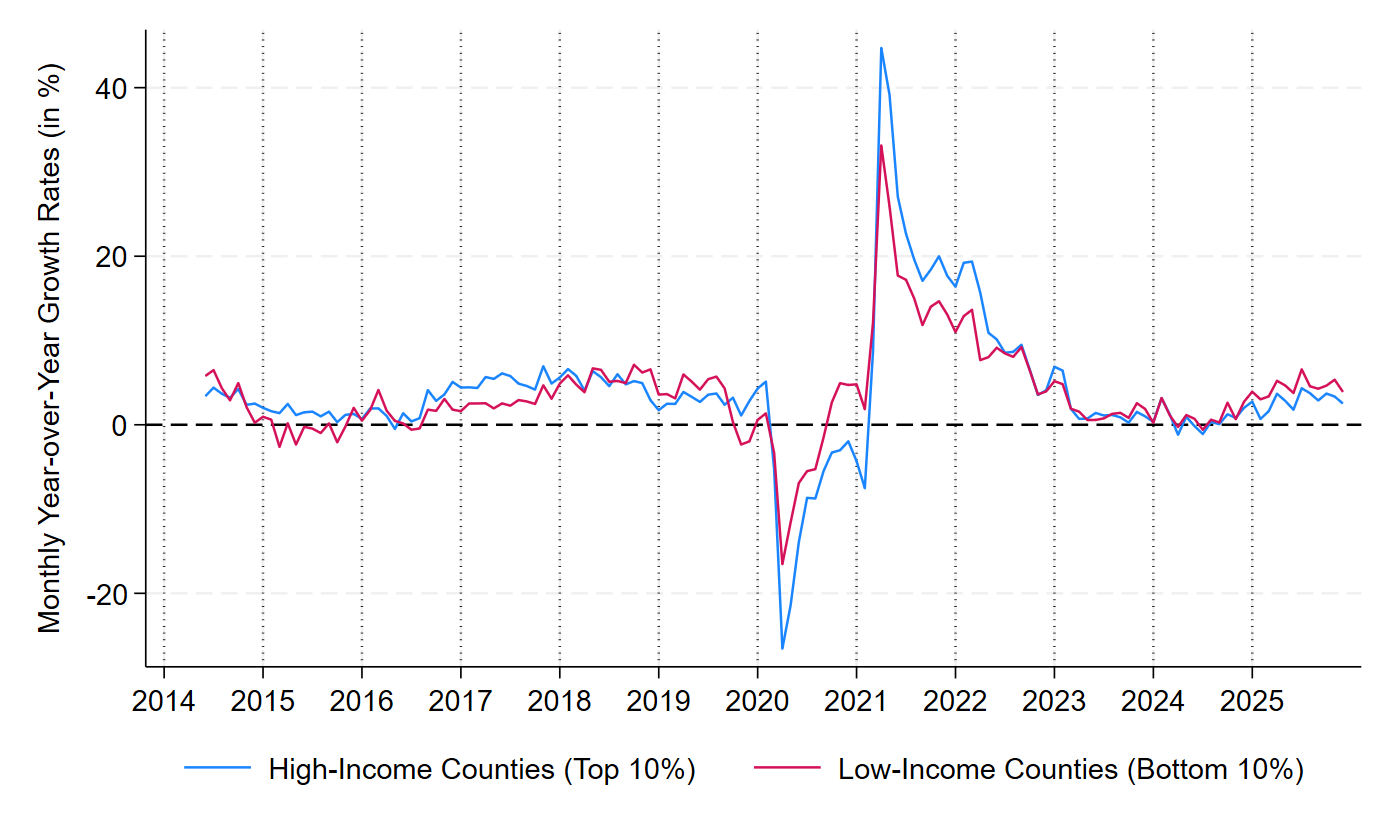}}
    \end{center}  	
    \label{fig:cc_spending_by_county_income}
\end{figure} 

\begin{figure}[!h]
    \caption{Share of Credit Card Transactions by Merchant Category} 
    \footnotesize{This figure plots the weighted share of credit card transactions by merchant category using transaction-level data from the 2024 Diary of Consumer Payment Choice.}
    \begin{center}
        \centerline{\includegraphics[width=1\textwidth]{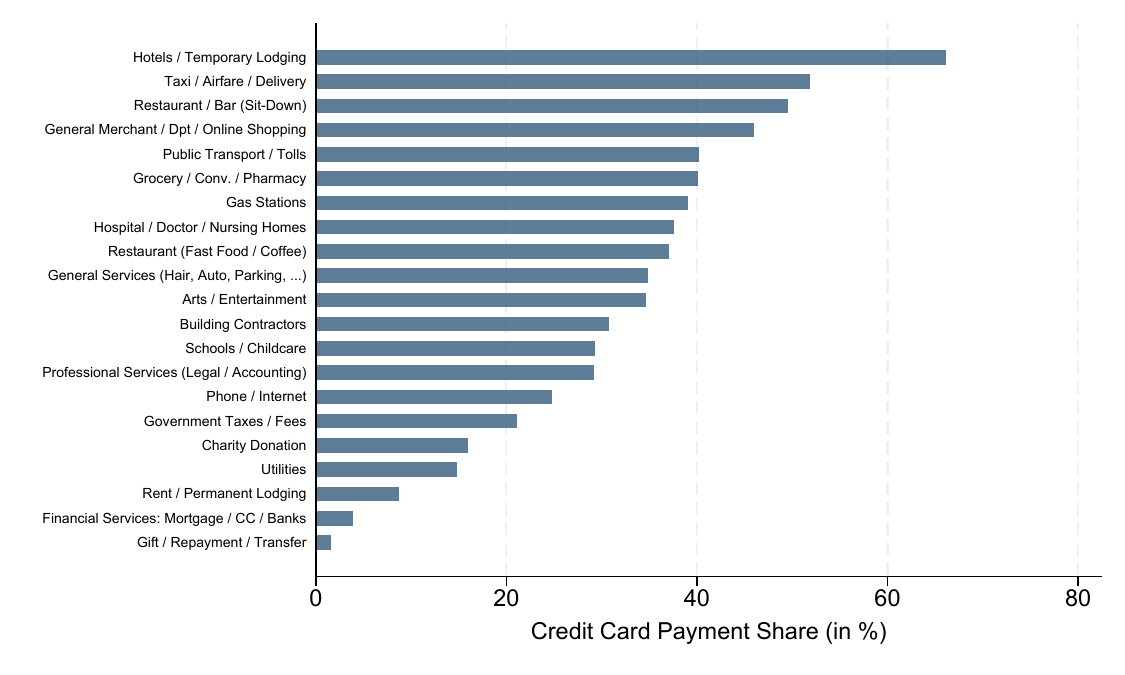}}
    \end{center}  	
    \label{fig:cc_transactions_by_merchant_category}
\end{figure}

\clearpage
\section{Additional Tables}

\begin{table}[!h]
    \caption{Benchmarking Monthly National Consumption: Y-14M and HFIs}
    \footnotesize{\flushleft This table reports monthly national time-series regressions relating private-sector high-frequency indicators (HFIs) to Y-14M credit card spending. The dependent variables are spending measures from Fiserv, Numerator, Verisk Analytics, and Affinity Solutions. For Fiserv, Numerator, and Verisk Analytics, variables are measured as monthly year-over-year growth rates. For Affinity Solutions, both the HFI series and the Y-14M series are expressed as seasonally adjusted changes relative to January 2020. Standard errors are reported in parentheses. $^{***}$, $^{**}$, and $^{*}$ denote statistical significance at the 1, 5, and 10 percent levels, respectively.}
    \begin{center}
        \begin{tabular}{lcccc}
            \toprule
                                    & Fiserv            & Numerator         & Verisk            & Affinity          \\
            \midrule
            $\Delta$ CC Spending    & 0.37$^{***}$      & 0.40$^{***}$      & 0.77$^{***}$      & 0.97$^{***}$      \\
                                    & (0.04)            & (0.04)            & (0.05)            & (0.04)            \\
            \addlinespace
            Adjusted $R^2$          & 0.44              & 0.51              & 0.79              & 0.89              \\
            Sample                  & 2014m6--2024m12   & 2019m1--2025m9    & 2019m2--2024m1    & 2020m1--2025m9    \\
            Observations            & 128               & 84                & 60                & 69                \\
            \bottomrule
        \end{tabular}
    \end{center}
    \label{tab:monthly_yoy_consumption_growth_hfi}
\end{table}

\end{document}